\documentclass[journal]{new-aiaa}
\usepackage[utf8]{inputenc}

\usepackage{graphicx}
\usepackage{amsmath}
\usepackage[version=4]{mhchem}
\usepackage{siunitx}
\usepackage{longtable,tabularx}
\usepackage{subcaption}
\usepackage{hyperref}
\usepackage{textcomp}
\usepackage{multirow}
\usepackage{algorithm}
\usepackage{algpseudocode}

\algnewcommand\algorithmicforeach{\textbf{foreach}}
\algdef{S}[FOR]{ForEach}[1]{\algorithmicforeach\ #1\ \algorithmicdo}

\makeatletter
\newcommand{\multiline}[1]{%
  \parbox[t]{\dimexpr\linewidth-\ALG@thistlm}{%
    \setlength{\parskip}{0pt}
    \linespread{0.9}\selectfont
    #1%
  }\strut 
}
\makeatother

\title{Distributed Semi-Speculative Parallel Anisotropic Mesh Adaptation}

\author{Kevin Garner\footnote{Research Assistant, Center for Real-time Computing, Corresponding Author}, Polykarpos Thomadakis\footnote{Research Assistant, Center for Real-time Computing} and Nikos Chrisochoides\footnote{Richard T. Cheng Chair Professor of Computer Science, Center for Real-time Computing, Corresponding Author, Email: nikos@cs.odu.edu}}
\affil{Center for Real-time Computing, Old Dominion University, Norfolk, VA 23529, USA}

\begin{document}

\maketitle

\begin{abstract}
This paper presents a distributed memory method for anisotropic mesh adaptation that is designed to avoid the use of collective communication and global synchronization techniques. In the presented method, meshing functionality is separated from performance aspects by utilizing a separate entity for each - a multicore cc-NUMA-based (shared memory) mesh generation software and a parallel runtime system that is designed to help applications leverage the concurrency offered by emerging high-performance computing (HPC) architectures. First, an initial mesh is decomposed and its interface elements (subdomain boundaries) are adapted on a single multicore node (shared memory). Subdomains are then distributed among the nodes of an HPC cluster so that their interior elements are adapted while interface elements (already adapted) remain frozen to maintain mesh conformity. Lessons are presented regarding some re-designs of the shared memory software and how its speculative execution model is utilized by the distributed memory method to achieve good performance. The presented method is shown to generate meshes (of up to approximately 1 billion elements) with comparable quality and performance to existing state-of-the-art HPC meshing software.
\end{abstract}

\section{Nomenclature}

{\renewcommand\arraystretch{1.0}
\noindent\begin{longtable*}{@{}l @{\quad=\quad} l@{}}
$S$ & subdomain of a partitioned mesh\\
$I_{1,2}$ & interface boundary between subdomains 1 and 2 \\
$\begin{mathcal}M\end{mathcal}$  & continuous metric field \\
$M$ & discrete metric field defined at the vertices of a mesh \\
$C(\cdot)$& complexity of a metric field \\
$L_a$ & Euclidean edge length evaluated in the metric of vertex \textit{a} \\
$M_{mean}$ & metric tensor interpolated at the centroid of a tetrahedron \\
$|k|$ & volume of a tetrahedron in evaluated metric $M_{mean}$ \\
$Q_k$ & mean ratio shape measure \\
\end{longtable*}}

\section{Introduction} \label{introduction}
Anisotropic mesh adaptation utilizes an anisotropic metric field to modify an existing mesh so that the estimated errors on this mesh, with respect to some simulation, are reduced. This metric field is used to control the orientation and size of elements individually for each direction so that the generated anisotropic mesh accurately captures the features of the underlying simulation. Mesh adaptation is a critical component in the study of Computational Fluid Dynamics (CFD), as CFD simulations in turn propel the design and analysis of aerospace vehicles. NASA's 2030 Vision \cite{CFD2030} identifies mesh generation as a bottleneck in the CFD workflow, underlining the need for leveraging emerging high-performance computing (HPC) architectures to meet the requirements of large-scale, time-dependent CFD problems. This paper presents a distributed memory method for anisotropic mesh adaptation that is designed to meet such needs. One of the challenges related to the development of HPC mesh generation codes is how meshing operations are implemented with regards to performance (such as thread management, load balancing, etc.). If meshing functionality and performance are intertwined within a method's code, the respective method becomes highly complex and difficult to maintain amid the evolving intricacies of HPC hardware. 

Preferably, HPC mesh generation methods should be capable of leveraging the concurrency offered by emerging hardware while requiring minimal updates to their source code. Consequently, the presented method separates performance (scalability) aspects from meshing functionality by utilizing a separate entity for each - PREMA \cite{BarkerPREMA, Thomadakis22PREMA, Thomadakis23TowardRuntime} for parallel runtime support and CDT3D \cite{Drakopoulos19F, EwCCDT3D} for mesh generation. Parallel Runtime Environment for Multicore Applications (PREMA) is a parallel runtime system that is designed to support applications which seek to leverage the concurrency offered by emerging HPC architectures. It provides constructs that enable asynchronous communication between encapsulations of data, work load balancing, and migration capabilities all within a globally addressable namespace. CDT3D is a parallel anisotropic mesh generation application that targets multicore shared memory systems. CDT3D contains meshing operations that are intended to be used within the broader scope of scalable data-parallel and partially coupled methods within a framework we term the Telescopic Approach \cite{Chrisochoides2016TelescopicAF}. 

The Telescopic Approach provides a layout of multiple memory hierarchies within an exascale architecture and describes how different meshing kernels might be utilized at each level to achieve maximum concurrency. This concept poses some challenges however, as recent studies show that methods which utilize complex meshing techniques (such as Delaunay \cite{PODMExperience} or some metric-based edge splitting techniques \cite{TsolakisEvaluation}) do not scale easily when utilizing up to several hundred cores. Although there are isotropic mesh generation methods designed for execution on exascale architectures \cite{ExascaleSubdivisionAIAA2024, t8CodeIMR2023, holke2018scalablealgorithmsparalleltreebased}, they rely on more straightforward, loosely-coupled \cite{ExascaleSubdivisionAIAA2024, t8CodeIMR2023, holke2018scalablealgorithmsparalleltreebased} or fully decoupled \cite{TerminalEdge2006} meshing techniques (such as element subdivision or edge bisection). In addition, the method in \cite{t8CodeIMR2023, holke2018scalablealgorithmsparalleltreebased} does not necessarily guarantee mesh conformity between partitions (subdomains). There are two techniques utilized with regards to mesh generation in an HPC CFD simulation pipeline - the mesh is generated on a single node and then distributed for parallel processing by the solver or mesh generation occurs in distributed memory so that the solver can immediately access a partition once it is generated/adapted. Regardless of challenges related to good performance and scalability on HPC architectures, our results still demonstrate the improved performance that can be achieved when efficiently leveraging the concurrency offered by both shared and distributed memory architectures (as opposed to designing an algorithm that only utilizes either one or the other). The shared memory method, CDT3D, is specifically built to operate within the lowest level of the Telescopic Approach hierarchy, exploiting fine-grain parallelism at the core level. CDT3D was previously tested on a single multicore node and was shown to have excellent scalability and good performance when compared to several other state-of-the-art methods in \cite{TsolakisEvaluation}, encouraging its extension into a distributed memory setting.

To maximize the potential performance of a parallel mesh generation software, there are several attributes that one must consider regarding its design. The following attributes are critical for enhancing a software's adaptability to emerging high-performance computing architectures, thereby ensuring its longevity and reliability when exploiting new methods for concurrency: stability (i.e., good mesh quality), robustness (ability to process any input data), scalability (ratio of the runtime of the best sequential implementation to the runtime of the parallel implementation), code re-use (a modular design that is crucial for long-term maintenance), and reproducibility (producing either identical or same-quality results for the same given input, termed either strong or weak reproducibility, respectively). These attributes are also defined in \cite{Chrisochoides18PDR, TsolakisEvaluation}. As mentioned previously, CDT3D addressed stability, robustness, and scalability (with various test cases, including within the context of a simulation pipeline) in \cite{TsolakisEvaluation, EwCCDT3D}, while addressing code re-use in \cite{Tsolakis21TaskingFramework}. Whereas CDT3D targets the chip level, the presented distributed memory method serves to exploit coarse-grain parallelism at the node level.

In the presented method, an initial mesh is decomposed and its interface elements (subdomain boundaries) are adapted on one multicore shared memory node of an HPC cluster. Subdomains are then distributed among the nodes of the cluster (utilizing the cores within each node) so that their interior elements (those that do not have data dependencies located in other subdomains) are adapted while interface elements (already adapted) remain frozen. Meshing operations within the shared memory method are designed to adopt a speculative execution model, enabling the strict adaptation of either interior or interface elements so that each set of elements can be adapted in separate steps while maintaining mesh conformity. This method offers better performance (shown in our evaluation) compared to an earlier distributed memory CDT3D approach that attempted to adapt and resolve the data dependencies of interface elements after subdomains have been distributed (requiring the movement of element cavities between parallel processes which are needed to permit the adaptation of interface elements) \cite{Garner24EarlyDMCDT3D}. 

One of the reasons why current state-of-the-art HPC mesh generation methods see a deterioration in performance when utilizing large configurations of cores is because many utilize collective communication and global synchronization techniques \cite{LOSEILLEFefloa, avroIMR2022, MassivelyParallel2019, ParkRefine, PragmaticGORMAN2012, OmegahDistributed2016, FMDB2006, FMDB2017, EPIC2012, Garner24EarlyDMCDT3D}. These techniques have been shown to hinder performance and potential scalability \cite{ECPMPIStudy, Klenk2017ECPStudy, TsolakisEvaluation}. The presented method outright avoids the use of collective communication and global synchronization techniques, exhibiting good runtime performance when compared to the state-of-the-art distributed memory mesh generation software \textit{refine} \cite{ParkRefine}. We show that the presented method can generate a model geometry of approximately 1 billion elements in less than 4 hours using 256 cores, compared to the shared memory CDT3D software and \textit{refine} which each take more than a full 8-hour work day to do the same when using 32 cores (CDT3D) and 256 cores (\textit{refine}). The presented method also produces meshes of good quality compared to those generated by the original shared memory CDT3D software and by \textit{refine}.

Several challenges (and their resulting lessons) are also presented given that CDT3D is optimized for a shared memory architecture, and what additional measures were taken to enable the code’s integration into the distributed memory method. Additionally, CDT3D posed a challenge regarding its ability to re-process data that it generated in previous adaptations (a necessity given that adapted interface elements must be re-processed, and subsequently frozen, when adapting interior elements). The original CDT3D method would process previously adapted data successfully (and satisfy the reproducibility requirement) but not efficiently with regards to time. This phenomenon is investigated and resolved in section \ref{how_modifications_affect_cdt3d}. These results echo the recurrent conclusion (also deduced from other distributed memory parallelization efforts of sequential and shared memory mesh generation methods \cite{ZagarisVGRID, APrioriEdgesIso2015, Chrisochoides18PDR, GarnerThesis, PODMExperience}) that a code cannot be simply integrated into a parallel framework as a black box if it was designed and developed independently of that framework. The shared memory CDT3D method underwent some re-design to allow its successful integration into the distributed memory method and to achieve good performance when targeting configurations with large numbers of cores.

The contributions of this paper include a distributed memory anisotropic mesh adaptation method that:
\begin{enumerate}
    \item provides several design abstractions (based on lessons learned) on how a state-of-the-art shared memory mesh generation method (CDT3D) was extended upon from utilizing 40 cores in earlier studies \cite{EwCCDT3D, TsolakisEvaluation} to utilizing 512 cores in a distributed memory setting,
    \item utilizes an a priori interface adaptation approach that enables the method to avoid the use of collective communication and global synchronization techniques, allowing the method to exhibit better runtime performance when compared to an alternate distributed memory (a posteriori interface adaptation) CDT3D approach \cite{Garner24EarlyDMCDT3D} and comparable performance to the state-of-the-art distributed memory mesh generation software \textit{refine} \cite{ParkRefine},
    \item and produces meshes of good quality compared to those generated by the original shared memory CDT3D software and by \textit{refine}.
\end{enumerate}

\section{Background}
\subsection{Shared Memory Mesh Generation} \label{shared_memory_mesh_generation_method}
The shared memory CDT3D implements a tightly-coupled method and exploits fine-grain parallelism at the cavity level using data decomposition, targeting shared memory multicore nodes using multithreaded execution at the chip level. A mesh operation (such as point creation, edge/face swapping, etc.) is performed concurrently on different data by using fast atomic lock instructions to guarantee correctness. These locks are used to acquire the necessary dependencies for the corresponding operation. Failure to do so will result in unlocking any acquired resources (rollback) and attempting to apply an operator on a different set of data. This is the essence of the speculative execution model, which is to exploit parallelism “everywhere possible” from the beginning of refinement when there is no, or very coarse, tessellation (contrary to existing methods that require sequential preprocessing and are in some cases just as expensive as the parallel mesh refinement itself). The speculative execution model is implemented using the \textit{separation of concerns} ideology \cite{DijkstraSeparation} \cite{Tsolakis21TaskingFramework}. Functionality is separated from performance components within CDT3D as well. Meshing operations are abstracted as tasks, and these tasks are only performed when their corresponding dependencies are satisfied (i.e., successfully locked). Such abstractions provide easy interoperability with a low-level runtime system such as PREMA (discussed in more detail in section \ref{parallel_runtime_system}). The presented distributed memory method currently utilizes a combination of POSIX threads (or Pthreads \cite{PThreads}) and OpenMP \cite{OpenMP} for CDT3D \cite{EwCCDT3D} as opposed to other backends offered by its tasking framework (CDT3D's parallel tasking schemes and other execution backends are explored in \cite{Tsolakis21TaskingFramework}).



\begin{figure}[htb]
\begin{center}
\includegraphics[width=1\textwidth]{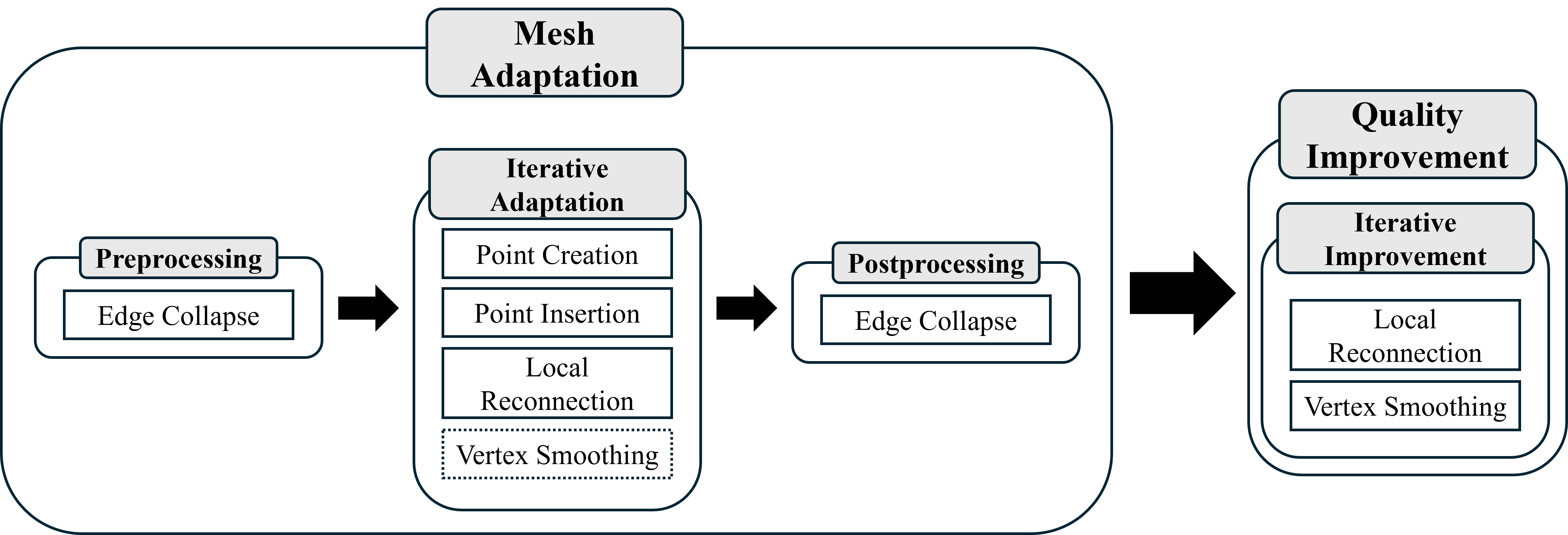}
\end{center}
\caption{The CDT3D pipeline of operations utilized within the presented method for anisotropic mesh adaptation is shown. By default, Vertex Smoothing (dotted) is not utilized during mesh adaptation but is available to potentially further improve mesh quality.}
\label{fig:cdt3d-anisotropic-pipeline}
\end{figure}

CDT3D offers metric-based anisotropic mesh adaptation, accepting an analytic or discrete metric field as input, and can be combined with Computer-Aided Design (CAD)-based information to accomplish adaptation \cite{EwCCDT3D}. The pipeline of operations for CDT3D's anisotropic mesh adaptation can be seen in Fig. \ref{fig:cdt3d-anisotropic-pipeline}. Each module, mesh adaptation and quality improvement, is individually executed over the domain through numerous grid generation passes (i.e., iterations) until the mesh conforms to the target metric field. Details regarding the qualitative measures and criteria that each operation uses to adapt elements can be found in \cite{EwCCDT3D}. While the mesh adaptation phase can improve the overall mesh quality, it should be noted that this phase focuses on satisfying spacing requirements (i.e., edge lengths of elements in the metric space). The quality improvement phase focuses on satisfying element shape requirements (i.e., improving element mean ratio). See section \ref{quality_measure_definitions} for details regarding how mesh quality is measured in the metric space for our evaluation.

As mentioned previously, CDT3D was compared with three other parallel anisotropic mesh adaptation methods used extensively within the industry \cite{TsolakisEvaluation}. The quantitative and qualitative results of each method were compared from testing on benchmarks created by the Unstructured Grid Adaptation Working Group (UGAWG\footnote{\url{https://ugawg.github.io/}}) \cite{UGAWG}. These benchmarks served to evaluate adaptive mesh mechanics for analytic metric fields on planar and simple curved domains. In each case in \cite{TsolakisEvaluation}, CDT3D was shown to maintain stability of metric conformity (with both element shape size and edge length). It also showcased good performance when utilizing up to 40 cores on a single multicore node, exhibiting good weak scaling speedup and almost linear speedup among its strong scaling cases.

\subsection{Parallel Runtime System} \label{parallel_runtime_system}

Message passing and data migration within the presented distributed memory method is handled by utilizing the PREMA runtime system \cite{BarkerPREMA, Thomadakis22PREMA, Thomadakis23TowardRuntime}. This system provides work scheduling and load balancing on both shared and distributed memory architectures, alleviating the application developer of these responsibilities. PREMA introduces constructs called mobile objects, which are encapsulations of data (not necessarily residing in contiguous memory), and mobile pointers which are used to identify mobile objects within a global namespace. Interactions between data can be expressed as remote method invocations (handlers) between mobile objects rather than only between processes or threads. While non-conflicting handlers are executed concurrently across processing elements, PREMA offers the ability to utilize multiple hardware threads to share work within the context of individual handlers \cite{Thomadakis23TowardRuntime}. Due to the nature of adaptive applications, especially in the context of mesh generation, workload disparity is often witnessed among the mobile objects processed by handlers. This fine-grain parallelism allows for more efficient utilization of shared memory resources to help bridge this disparity in workload processing time. 

PREMA offers distributed memory load balancing by monitoring work loads between ranks and performing migrations of mobile objects to available workers without interrupting execution. Communication and execution are separated into different threads to provide asynchronous message reception and instant computation execution at the arrival of new work requests. As stated previously, the mobile pointer construct allows method invocations to be made to mobile objects regardless of their location (potential migration to another rank). PREMA also provides the ability to establish dependencies between mobile objects and to execute user-defined events once all dependencies are satisfied.

\section{Related Work} \label{related_work}
The presented distributed memory method is motivated by past investigations of "black box approaches," where parallelization was attempted for mesh generation programs (most of them were originally sequential) while making the least amount of modifications possible to their source code. This approach is termed "functionality-first," which involves the parallelization of state-of-the-art mesh generation software that are fully functional and optimized for single-core architectures. Several studies addressing the viability of functionality-first black box approaches included VGRID \cite{ZagarisVGRID}, TetGen \cite{Chrisochoides18PDR}, AFLR \cite{GarnerThesis}, Netgen \cite{APrioriEdgesIso2015} and more. The effort regarding VGRID utilized a binary of the sequential software within a distributed memory method \cite{ZagarisVGRID}. Consequently, its performance was hindered by VGRID process creation, file I/O, and data structure initialization. Additionally, its scalability was limited by the sequential design of VGRID and insufficient exploitable concurrency given by the domain decomposition method utilized. With regards to the parallelization of TetGen, TetGen itself failed the reproducibility criterion (it could not initialize its data structures based on a mesh that it itself had already generated) \cite{Chrisochoides18PDR}. Although AFLR's source code was integrated successfully into a distributed method and it satisfied the reproducibility criterion, this parallelization effort was hindered by the mesh quality within each subdomain being constrained by interfaces (subdomain boundaries, as an input boundary was required by AFLR to refine any domain) \cite{GarnerThesis}. The sequential software Netgen was integrated into a parallel framework called PMSH in \cite{APrioriEdgesIso2015}, generating multibillion element meshes on large configurations of cores. While PMSH successfully generated meshes that would otherwise be too large for the memory of a single node, the sequential software required modifications to successfully refine these geometries while some features were not successfully preserved in the parallel implementation (such as specifying mesh sizes through regions of the geometry). The method in \cite{avroIMR2022} exhibited good performance when parallelizing its sequential mesher; however, the method was tested with a low number of processors and is expected to encounter scalability issues (when executed on larger HPC system configurations) given its communication techniques that are known to hinder potential performance (described further below). 

These communication techniques are also seen in the functionality-first approaches in \cite{Pampa2017, MassivelyParallel2019, ParMMG2019}. Another black box approach focused on the integration of a shared memory method, called PODM \cite{PODMFOTEINOS20142}, into a distributed memory framework \cite{PODMExperience}. PODM (a Delaunay-based mesh generation method) was not originally designed for execution within a distributed setting (or integration into the Telescopic Approach). This approach strictly utilized PODM as a black box and involved the migration of large amounts of data (entire subdomains) throughout execution in order to resolve data dependencies that were required to satisfy the Delaunay property. This overhead accounted for more than 50\% of execution time, making this distributed approach 7x slower than the shared memory PODM when utilizing the same numbers of cores \cite{PODMExperience}.  Evaluations of these black box approaches exhibited the recurrent conclusion that if a code is not originally designed for scalability, it cannot be simply integrated into a parallel framework as a black box. Rather than devoting significant amounts of time to redesigning such codes, these studies encourage the development of "scalability-first" approaches (those designed with scalability as the focus and functionality added as needed, such as CDT3D), if one wishes to leverage the maximum potential speedup offered by HPC architectures.

A goal of the presented distributed memory method is to also avoid collective communication, as this is not ideal for distributed computing. The overhead of collective communication was reported in an extensive study involving numerous proxy applications within DoE's Exascale Computing Project (ECP) \cite{ECPMPIStudy, Klenk2017ECPStudy}. The purpose of the ECP study is to understand communication patterns utilized by these applications and to identify where optimization efforts may be focused. It was observed that most of the applications spent more than 50\% of their runtime in communication (as opposed to computation). While the majority of communication calls were primarily point-to-point (messages between individual processes), the amount of runtime spent in communication was dominated by collective calls (e.g. MPI\_Allreduce, MPI\_Alltoall, etc.). As the number of parallel processes are increased, the overhead of collective communication becomes dominant. The parallel adaptive anisotropic meshing strategies utilized by the aforementioned state-of-the-art codes (that were compared to the shared memory CDT3D) all exhibit good speedup in the strong scaling and weak scaling cases presented in \cite{TsolakisEvaluation}. However, there are implicit global synchronization points in these codes that induce increasingly noticeable overhead when utilizing up to several hundred cores. Given the observations presented in the study of DoE's ECP, this overhead is expected to be exacerbated when meshing hundred millions to billions of elements on higher configurations of cores. 

With regards to coarse-grained parallel mesh generation methods, the processing of partition boundaries (subdomain interfaces) is a common problem seen in literature. There are three strategies used to address this problem - solely refining/adapting subdomain interfaces in a separate step before distributing and modifying subdomain interiors \cite{LohnerAPriori1990, Prepartitioning1996, LeonidasDelaunayDecoupling2008, LargeScaleDistributed2024} (a priori approach), simultaneously refining/adapting both interior and interface elements \cite{MITDelaunay1997, NAVE2004191, LargeScaleDistributed2024}, and refining/adapting the interfaces in a separate step after subdomain interior elements have been refined/adapted \cite{LohnerAPriori1990, Shephard1999, BrownLepp1999, ParkRefine, EPIC2012, Lohner2014RecentAdvances, LOSEILLEFefloa, MassivelyParallel2019, Garner24EarlyDMCDT3D} (a posteriori approach). We further explore these methods below.


Some methods share certain techniques within each of these interface approaches. For example, the isotropic mesh generation methods in \cite{Prepartitioning1996, ZagarisVGRID, LargeScaleDistributed2024} each process a surface mesh and first generate elements along bisecting planes that are used to recursively subdivide the domain until a satisfactory decomposition is reached. These interfaces satisfy spacing and qualitative criteria; therefore, they remain fixed while subdomain interiors are then generated and refined. The a priori method in \cite{DistributedDelaunay1999} decomposes an isotropic coarse volume mesh where interface elements are refined before distributing and refining the subdomain interiors. The simultaneous approach (refining interior and interface elements simultaneously rather than separating them into different steps) in \cite{LargeScaleDistributed2024} duplicates interface data across subdomains which are only modified within certain subdomains. Communication follows between processes to then update neighboring subdomains of the modifications to those elements. The isotropic method in \cite{NAVE2004191} migrates the data dependencies of interface elements (i.e., cavities needed for refinement) while also refining subdomain interior elements using a speculative execution model (the model also used by the shared memory CDT3D). This model allows the method to tolerate communication overhead by overlapping computation. Another simultaneous method utilizes similar locking mechanisms specifically for interface element cavities (migrating those data dependencies upon successful locking), and mitigates the time spent waiting for responses by continuing to sequentially refine interior elements \cite{MITDelaunay1997}. 
A posteriori methods typically freeze interface elements to first refine/adapt subdomain interiors. These methods in \cite{Shephard1999, ParkRefine, EPIC2012, Lohner2014RecentAdvances, LOSEILLEFefloa, MassivelyParallel2019} all then utilize some decomposition technique to perform global re-partitioning so that interface elements can be refined/adapted in subsequent steps. Utilizing a domain decomposition technique after each iteration of meshing for re-partitioning was reported to be one of the main parallel overheads in \cite{LOSEILLEFefloa} and \cite{MassivelyParallel2019}. An earlier distributed memory CDT3D implementation \cite{Garner24EarlyDMCDT3D} (discussed briefly in section \ref{aposteriori_DMCDT3D} and shown to be inferior to the presented method in terms of performance) took a similar approach to \cite{Lohner2014RecentAdvances} and \cite{EPIC2012}, where processes would synchronize to migrate layers of elements and turn interface elements into interior elements so that they may be refined/adapted. This shifting of elements was reported to hinder potential scalability in both \cite{Garner24EarlyDMCDT3D} and \cite{LohnerIMR2013, Lohner2014RecentAdvances}. 
While tested with simple isotropic geometries, the method in \cite{BrownLepp1999} has an a posteriori phase where asynchronous messages are exchanged between processes to update neighboring subdomain interfaces. 

The presented distributed memory CDT3D method initially adapts subdomain interfaces on one (shared memory) multicore node, then distributes subdomains to other nodes to adapt the remaining elements within each subdomain (taking an a priori approach). Consequently, it avoids collective communication, as it does not perform global re-partitioning or utilize global synchronization techniques during mesh generation (as opposed to many of the aforementioned methods that do). 
Additionally, many of the aforementioned methods focus on isotropic mesh generation, 
as opposed to the anisotropic 3D mesh adaptation of the presented method. While there are some coarse-grained methods that utilize multi-threading \cite{EPIC2012, Lohner2014RecentAdvances, LOSEILLEFefloa}, no partially-coupled method employs a fine-grained multi-threaded speculative execution model, such as the shared memory CDT3D software, for the adaptation of individual subdomains (to the best of our knowledge). This model is utilized during both phases of adaptation (on one multicore node during interface adaptation and on each individual multicore node during subdomain interior adaptation).

\section{Distributed Memory Method} \label{distributed_memory_method}
The presented distributed memory method is organized into the following sections: a high-level overview of its algorithm (section \ref{high_level_algorithm}), the data decomposition technique utilized (section \ref{data_decomposition}), information pertaining to its data structures (section \ref{data_structures}), modifications made to the shared memory CDT3D code to optimize performance (section \ref{CDT3D_modifications}), how simple connectivity is ensured among subdomains (section \ref{making_subdomains_simply_connected}), a brief overview of an alternate a posteriori interface adaptation approach (section \ref{aposteriori_DMCDT3D}), and a summary of the lessons learned from utilizing the shared memory code within the distributed memory method (section \ref{distributed_requirements}).

\subsection{High-level Algorithm} \label{high_level_algorithm}
Given its design, stability, and performance on a single multicore node, the shared memory CDT3D (SM\_CDT3D) was abstracted as a library to be used in the adaptation of individual subdomains in the distributed memory method. A high-level overview of the distributed memory method (DM\_CDT3D) is shown in Algorithm \ref{alg:high_level_algorithm}. It essentially includes seven steps. First, a coarse mesh is generated at a fraction of the complexity of the target metric (section \ref{quality_measure_definitions} provides a definition of the complexity of a metric field). This step serves to ensure that the initial mesh is dense enough for data decomposition. The remaining steps of the DM\_CDT3D algorithm include: decomposition, interface adaptation preprocessing, interface adaptation, making each subdomain simply connected, subdomain interior adaptation preprocessing, and interior adaptation of all subdomains. Both the interface and interior adaptation preprocessing steps include deactivating elements and preemptively locking points in order to direct CDT3D to adapt specific elements during each phase (interface or interior), discussed in section \ref{CDT3D_modifications}.


\begin{algorithm}
\caption{High-level algorithm of the A Priori Distributed Memory CDT3D implementation}\label{alg:high_level_algorithm}
\scriptsize{
\begin{flushleft}
A Priori DM\_CDT3D($M_i$, \textit{m}, N) \\
\textbf{Input}: $M_i$ is the input mesh \\
\hspace*{\algorithmicindent} \ \ \ \ \ \textit{m} is the target metric \\
\hspace*{\algorithmicindent} \ \ \ \ \ N is the number of subdomains \\
\textbf{Output}: Adapted mesh that conforms to the anisotropic metric field \textit{m}
\end{flushleft}
\begin{algorithmic}[1]
\State $M_C$ = INITIAL\_COARSE\_MESH\_GENERATION($M_i$, \textit{m})
\State Perform data decomposition of $M_C$ to create subdomains $S_1 ... S_N$
\State interface\_points = set of all interface points amongst $S_1 ... S_N$
\State INTERFACE\_ADAPTATION\_PREPROCESSING($M_C$, interface\_points) \Comment{\textit{// Lock and deactivate elements not needed for interface adaptation}}
\State $prelock\_interface\_points_{1...N}$ = set of points preemptively locked by INTERFACE\_ADAPTATION\_PREPROCESSING \Comment{\textit{// $prelock\_interface\_points_{1...N}$ corresponds to $S_1 ... S_N$}}
\State $pseudo\_active\_interior\_elements_{1...N}$ = set of elements deactivated by INTERFACE\_ADAPTATION\_PREPROCESSING \Comment{\textit{// $pseudo\_active\_interior\_elements_{1...N}$ corresponds to $S_1 ... S_N$}}
\State ADAPT\_INTERFACES($M_C$, $S_1$...$S_N$, \textit{m}$_1$...\textit{m}$_N$) \Comment{\textit{// New elements generated from $M_C$ are assigned to $S_1$...$S_N$}}
\State MAKE\_SIMPLY\_CONNECTED($S_1$...$S_N$)
\State Distribute subdomains $S_1$...$S_N$ to processes
\State INTERIOR\_ADAPTATION\_PREPROCESSING($S_1$...$S_N$,$prelock\_interface\_points_{1...N}$, $pseudo\_active\_interior\_elements_{1...N}$) \Comment{\textit{// Lock and deactivate elements not needed for interior adaptation}}
\State ADAPT\_SUBDOMAIN($S_1$...$S_N$, \textit{m}$_1$...\textit{m}$_N$)
\State TERMINATE()
\end{algorithmic}
}
\end{algorithm}

\subsection{Data Decomposition} \label{data_decomposition}
Different methods of decomposition may be applied to the grid. The method of decomposition applied in the current implementation is called PQR, which uses a sorting-based method to partition elements into subdomains based on a boundary-conforming curvilinear coordinate system \cite{CHRISOCHOIDES199475}. A connectivity graph partitioning heuristic is utilized, where the mesh itself is considered to be a Euclidean graph (the elements are vertices and their face-connected neighbors establish edges). This heuristic uses Euclidean metrics and minimizes the diameter of subdomains, delivering quasi-uniform partitions. Once all subdomains have been created, they undergo preprocessing for the interface adaptation phase.

Let a mesh be partitioned into N subdomains \(S_1...S_N\). If \(S_1 \cap S_2 \neq \emptyset\) then \(S_1\) and \(S_2\) are neighbors. Let \(I_{1,2} = S_1 \cap S_2\). All points in \(I_{1,2}\) are interface points. Any edge, face, or tetrahedron defined by an interface point in \(I_{1,2}\) is an interface edge, interface face, or interface tetrahedron, respectively. Any other subdomain which contains elements that are defined by a point in \(I_{1,2}\) is also considered a neighbor of both \(S_1\) and \(S_2\) and will contain a copy of that same interface point. Let \(neighs\) be the set of all neighbor subdomains for \(S_1\). The set of interior points in \(S_1\) is defined as: \(\forall points \in S_1 and \notin I_{1,neighs}\). Elements that are solely defined by interior points are considered to be interior elements.

\subsection{Data Structures} \label{data_structures}
One should identify and encapsulate only data that is essential for jumpstarting the shared memory method in a distributed setting. There is a clean separation between the shared memory code and distributed code. The shared memory code itself is not distributed-aware. It is rather utilized as a library, where a number of adaptation operations can be executed on an input domain (such as the coarse mesh for interface adaptation or each individual subdomain for the interior adaptation phase). Information pertaining to mesh elements within each subdomain must be extracted as input for the shared memory method. Simultaneously, the distributed memory method must maintain global identifiers specifying duplicate data between subdomains. Duplicate data are specified using global identifiers similar to the approach in \cite{NAVE2004191}. These global identifiers are essential when merging subdomains for the final output mesh, as they are used to identify both face and point connections between subdomains. Every vertex in the grid is assigned a subdomain integer id (during decomposition) and an integer id local to that particular subdomain. Decomposition and this assignment of data occur on a single node. Each interface point is assigned to a single subdomain owner, while other subdomains store this point as a duplicate and its global identifier (subdomain id and local id) information. Subdomains are not packed/sent to other nodes of the HPC cluster until interface adaptation and subsequent data assignment (of new elements) is completed. Newly created elements are assigned in a particular manner that maintains simple connectivity amongst subdomains (discussed in section \ref{making_subdomains_simply_connected}). Their respective points (if new) are assigned new global identifiers in the same manner as decomposition (based on the current number of local ids for that particular subdomain). During the subdomain interior adaptation phase, interface points remain untouched throughout adaptation. Any newly created points are assigned new global identifiers at the end of adaptation within their individual subdomains.

The distributed memory CDT3D method is implemented in the C++ programming language and as such, the standard library \textbf{memcpy} function is utilized for the packing/unpacking of data during the distribution of subdomains. It is preferable to organize subdomain data into the simplest data structures possible (arrays, plain old data types, etc.) to make the utilization of this function seamless; otherwise, careful attention must be given to more complex data structures, as they are more likely to induce memory errors if not handled correctly (e.g., container types, dynamically allocated pointers, etc.). These simple distributed memory method data structures are converted into the complex data structures utilized by the shared memory code (linked lists of dynamically allocated pointers to class objects representative of mesh elements, for example) when jump-starting a domain for interface adaptation or a subdomain for interior adaptation.

\subsection{Adapting Specific Elements with CDT3D} \label{CDT3D_modifications}
To efficiently use CDT3D within the distributed method, some preprocessing of the domain is required, both for the interface adaptation phase and the interior adaptation phase. CDT3D denotes elements as "active" if they do not conform to the target metric field. These elements become "inactive" when they do. Active and inactive elements are organized into separate lists, and active elements are further divided into "buckets" to be assigned to threads for CDT3D's point creation and local reconnection operations \cite{Drakopoulos19F}. The interface preprocessing involves deactivating (designating as inactive) elements that should not be modified during the interface adaptation phase (i.e., most of the interior elements). This allows CDT3D to immediately organize interior elements into the inactive list and only assign interface elements (and their respective cavities) to threads. Furthermore, most of the points that define interior elements are preemptively locked (before any mesh operations are executed). Preemptively locked points (locked during interface or interior preprocessing) remain permanently locked during that particular phase of adaptation (interface or interior). There are operations which modify elements regardless of their active status - edge collapse and vertex smoothing. 
The edge collapse and vertex smoothing operations are designed to modify elements only if their respective points are unlocked (i.e., are available for adaptation, as part of the speculative execution model). 
Point creation and local reconnection also adhere to the speculative execution model but additionally check if an element is still active before commencing. It should be noted that points are specifically locked in CDT3D's speculative execution model, not elements. A thread is exclusively given access to modify a particular element provided that element's points are all locked by that thread.

Preemptively locking points also allows CDT3D to focus only on certain elements when applying operations. Preemptively locking points and deactivating elements are two distinct actions that both must be taken to control adaptation (for specific elements) while adhering to the speculative execution model and notion of active elements. Recall that CDT3D assigns elements to threads for the point creation and local reconnection operations. The method does not originally check to see if the points defining these elements are locked. It simply assigns elements (from the list of active elements) to threads. With regards to interface adaptation, deactivating interior elements first allows CDT3D to avoid assigning interior elements (which are implicitly locked due to their preemptively locked points) to threads which would just spin without working (wasting time and computational resources), as these elements' points would never be unlocked for adaptation during the interface adaptation phase. The edge collapse and vertex smoothing operations are designed to iterate through lists of points (rather than elements) when attempting to lock dependencies and commence with adaptation. When executing these operations, CDT3D was modified to only assign points to threads that were not preemptively locked. Otherwise, the method would assign these points to threads and again, those threads would spin unable to complete any work due to these preemptively locked points (negatively impacting performance for the distributed method). It should also be noted that in order to successfully adapt either set of elements (interface or interior), some layers of elements surrounding that set must be active (and their points unlocked) so that the necessary element cavities are available for adaptation to achieve good end quality for the final output mesh. 

As mentioned previously, the original CDT3D method posed a challenge regarding its ability to efficiently process elements that have already undergone adaptation. We have observed that among all the geometry cases with which the original CDT3D has been tested \cite{EwCCDT3D, TsolakisEvaluation}, approximately 20-30\% of elements remain active by the end of adaptation. During adaptation, the method attempts to improve the quality of these elements to reach certain qualitative thresholds. If little improvement is observed after several grid generation passes, the method terminates. Despite this phenomenon, CDT3D was shown to produce meshes with qualitative results that are comparable to those produced by state-of-the-art methods \cite{TsolakisEvaluation}. CDT3D was also tested and shown to function well when coupled with a solver as part of a CFD simulation pipeline, accurately capturing features of underlying simulations and occupying a small fraction of the pipeline's runtime \cite{EwCCDT3D}.

CDT3D was designed with the assumption that the input geometry does not conform to the target metric field. This design assumption is a problem for the distributed method due to the fact that interface elements will have already undergone adaptation and will be connected to interior elements when each subdomain is processed and adapted in distributed memory. Similar preprocessing (i.e., element deactivation and preemptive locking for points) for interior adaptation must first be applied; otherwise, elements that may have still been active by the end of the interface adaptation phase will cause CDT3D to re-apply mesh operations and further attempt to improve these elements during the interior adaptation phase. To maximize potential scalability, we attempt to direct CDT3D to only apply operations to elements upon which the method did not attempt adaptation during the interface adaptation phase.

To address this problem for the distributed method, the shared memory method was modified to acknowledge if elements are classified as "pseudo-active" or "pseudo-inactive." These terms are classifications simply used to help CDT3D determine if elements should be deactivated before adaptation. Essentially, pseudo-active elements should undergo adaptation (through the point creation and local reconnection operations) and pseudo-inactive elements should not undergo adaptation (i.e., should be deactivated before any operation commences). If an element was deactivated (made inactive) by the method during interface adaptation, it is also classified as pseudo-inactive (i.e., the element conforms to the target metric field by the end of interface adaptation and should not be active during interior adaptation). If the point creation operation was applied to an element during interface adaptation, that element is classified as pseudo-inactive (even if the operation failed to create any candidate points that satisfy spacing criteria and the element is still officially active). As specified previously, this is to prevent the method from repeating unnecessary work and attempting to adapt this element again during the interior adaptation phase. If the point creation operation skipped an element (because it was deactivated during interface preprocessing or its points were preemptively locked), then that element is classified as pseudo-active. We use the point creation operation to designate elements as pseudo-active because it is the only operation within CDT3D that is designed to satisfy the metric field's spacing requirements by modifying active elements (the edge collapse and vertex smoothing operations, for example, modify elements regardless of if they are active/inactive). Each element's classification (either pseudo-active or pseudo-inactive) is included when packing and distributing subdomains to processes. Moreover, designations of whether or not points were preemptively locked before interface adaptation is also included (those that were preemptively locked for interface adaptation are designated as $prelock\_interface\_points$ in algorithms \ref{alg:high_level_algorithm} and \ref{alg:interior_preprocessing}). When jumpstarting the shared memory method for a subdomain's interior adaptation, if an element is pseudo-inactive, it is deactivated (made inactive) before any mesh operations are executed. Interface points and their respective cavity points are also preemptively locked (i.e., the opposite of those that were preemptively locked before interface adaptation). With these modifications, we attempt to direct CDT3D to focus on applying mesh operations only to interior elements during interior adaptation. Since the interface elements/points are deactivated/locked during this phase, they are essentially frozen, thus ensuring that all subdomains maintain conformity to one other. The pseudocode for interface and interior preprocessing can be seen in algorithms \ref{alg:interface_preprocessing} and \ref{alg:interior_preprocessing}, respectively. To reduce overhead, both preprocessing algorithms occur while the shared memory CDT3D data structures are created (which preserves the original utilization of OpenMP constructs over the for loops to improve performance \cite{EwCCDT3D}). The final preprocessing step (referred to as UNLOCK\_ACTIVATE\_LAYERS) of unlocking/activating layers of points/elements (with a depth-first traversal from the list of points provided as the first parameter) adds little overhead, as seen in Fig. \ref{fig:apriori_delta_10m_percentage_adaptation_breakdown} and Fig. \ref{fig:apriori_delta_20m_percentage_adaptation_breakdown} of our results. As specified previously, this final step is necessary to ensure that element cavities are available for adaptation to achieve good end quality for the final output mesh (and to satisfy the buffer zone requirement for the point creation operation, discussed in section \ref{point_creation}). While the number of layers unlocked/activated during preprocessing can be specified by the user, we found that at least 3 layers should be specified for each adaptation phase to achieve good final mesh quality in our test cases. 

\begin{algorithm} 
\caption{Interface Adaptation Preprocessing}
\label{alg:interface_preprocessing}
\scriptsize{
\begin{flushleft}
INTERFACE\_ADAPTATION\_PREPROCESSING($M_C$, interface\_points)\\
\textbf{Input}: $M_C$ is the coarse mesh that was decomposed \\
\hspace*{\algorithmicindent} \ \ \ \ \ interface\_points is a set of all interface points from the data decomposition of $M_C$ \\
\textbf{Output}: $M_C$ where interface elements/points (and their respective cavities) are pseudo-active/unlocked while all other elements/points are pseudo-inactive/locked
\end{flushleft}
\begin{algorithmic}[1]
\ForEach{point $\in$ $M_C$}
    \State Create SM\_CDT3D data structure for point
    \State point->layer\_checked = 0
    \State point->attachedTetrahedra = $\emptyset$
    \If{point $\notin$ interface\_points}
        \State Lock(point)
    \EndIf
\EndFor
\ForEach{tetrahedron $\in$ $M_C$}
    \State Create SM\_CDT3D data structure for tetrahedron
    \State Mark tetrahedron as pseudo-inactive
    \ForEach{point $\in$ tetrahedron}
        \State point->attachedTetrahedra.ADD(tetrahedron)
    \EndFor
\EndFor
\ForEach{external boundary face $\in$ $M_C$}
    \State Create SM\_CDT3D data structure for external boundary face
\EndFor
\State num\_layers = user-specified number of layers
\State UNLOCK\_ACTIVATE\_LAYERS(interface\_points, num\_layers) 
\end{algorithmic}
}
\end{algorithm}

\begin{algorithm} 
\caption{Interior Adaptation Preprocessing}
\label{alg:interior_preprocessing}
\scriptsize{
\begin{flushleft}
INTERIOR\_ADAPTATION\_PREPROCESSING($S_1 ... S_N$, $prelock\_interface\_points_{1...N}$, $pseudo\_active\_interior\_elements_{1...N}$) \\
\textbf{Input}: $S_1 ... S_N$ are all the subdomains created from data decomposition \\
\hspace*{\algorithmicindent} \ \ \ \ \ $prelock\_interface\_points_{1...N}$ is a set of points that were preemptively locked for interface adaptation (corresponds to $S_1 ... S_N$) \\
\hspace*{\algorithmicindent} \ \ \ \ \ $pseudo\_active\_interior\_elements_{1...N}$ is a set of elements that are pseudo-active (those not adapted during interface adaptation) (corresponds to $S_1 ... S_N$) \\
\textbf{Output}: $S_1 ... S_N$ where most interior elements/points (and their respective cavities) are pseudo-active/unlocked while other elements/points that were modified during interface adaptation are pseudo-inactive/locked
\end{flushleft}
\begin{algorithmic}[1]
\ForEach{$S_i \in S_1 ... S_N$}
    \ForEach{point $\in$ $S_i$}
        \State Create SM\_CDT3D data structure for point
        \State point->layer\_checked = 0
        \State point->attachedTetrahedra = $\emptyset$
        \If{point $\notin$ $prelock\_interface\_points_i$}
            \State Lock(point)
        \EndIf
    \EndFor
    \ForEach{tetrahedron $\in$ $S_i$}
        \State Create SM\_CDT3D data structure for tetrahedron
        \If{tetrahedron $\in$ $pseudo\_active\_interior\_elements_i$}
            \State Mark tetrahedron as pseudo-active
        \Else
            \State Mark tetrahedron as pseudo-inactive
        \EndIf
        \ForEach{point $\in$ tetrahedron}
            \State point->attachedTetrahedra.ADD(tetrahedron)
        \EndFor
    \EndFor
    \ForEach{external boundary face $\in$ $S_i$}
        \State Create SM\_CDT3D data structure for external boundary face
    \EndFor
    \State pseudo\_active\_inactive\_shared\_points = set of points that define the faces shared by pseudo-active and pseudo-inactive tetrahedra
    \State num\_layers = user-specified number of layers
    \State UNLOCK\_ACTIVATE\_LAYERS(pseudo\_active\_inactive\_shared\_points, num\_layers) 
\EndFor
\end{algorithmic}
}
\end{algorithm}

\begin{figure}[h!]
     \centering
     \begin{subfigure}[htb]{0.45\textwidth}
         \centering
         \includegraphics[width=\textwidth]{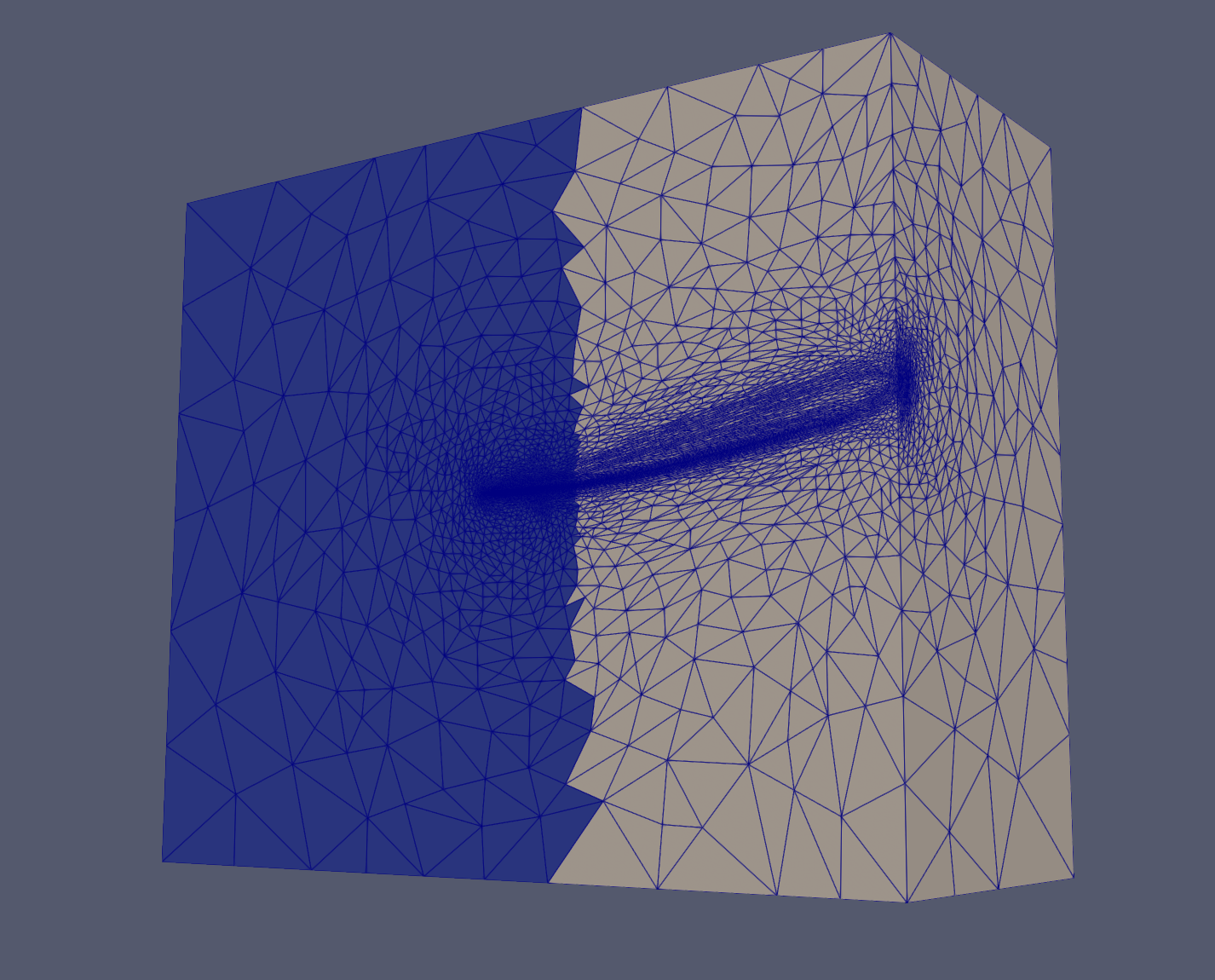}
         \caption{Decomposition (line 2 of Algorithm \ref{alg:high_level_algorithm})}
         \label{visual_decomposition}
     \end{subfigure}
     \hfill
     \begin{subfigure}[htb]{0.45\textwidth}
         \centering
         \includegraphics[width=\textwidth]{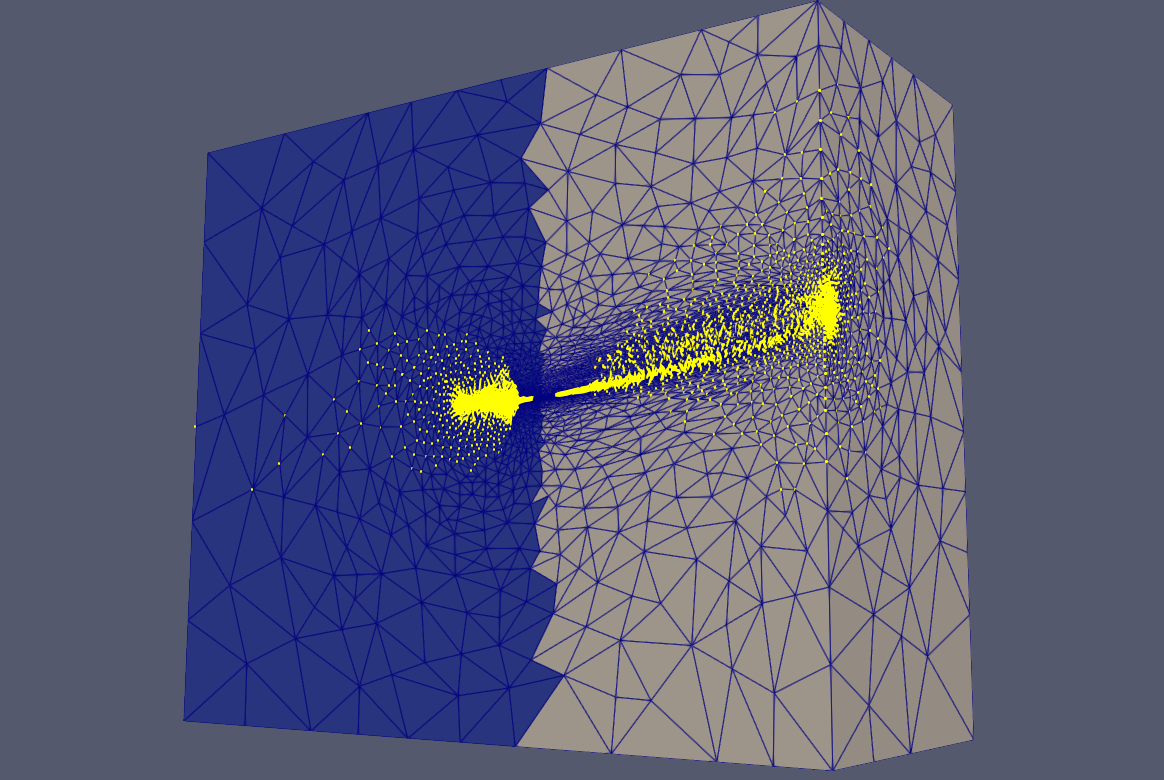}
         \caption{Preemptively Locked Vertices (in Yellow) Before Interface Adaptation (lines 4 and 5 of Algorithm \ref{alg:high_level_algorithm})}
         \label{visual_interface_lock}
     \end{subfigure}
     \hfill
     \begin{subfigure}[htb]{0.45\textwidth}
         \centering
         \includegraphics[width=\textwidth]{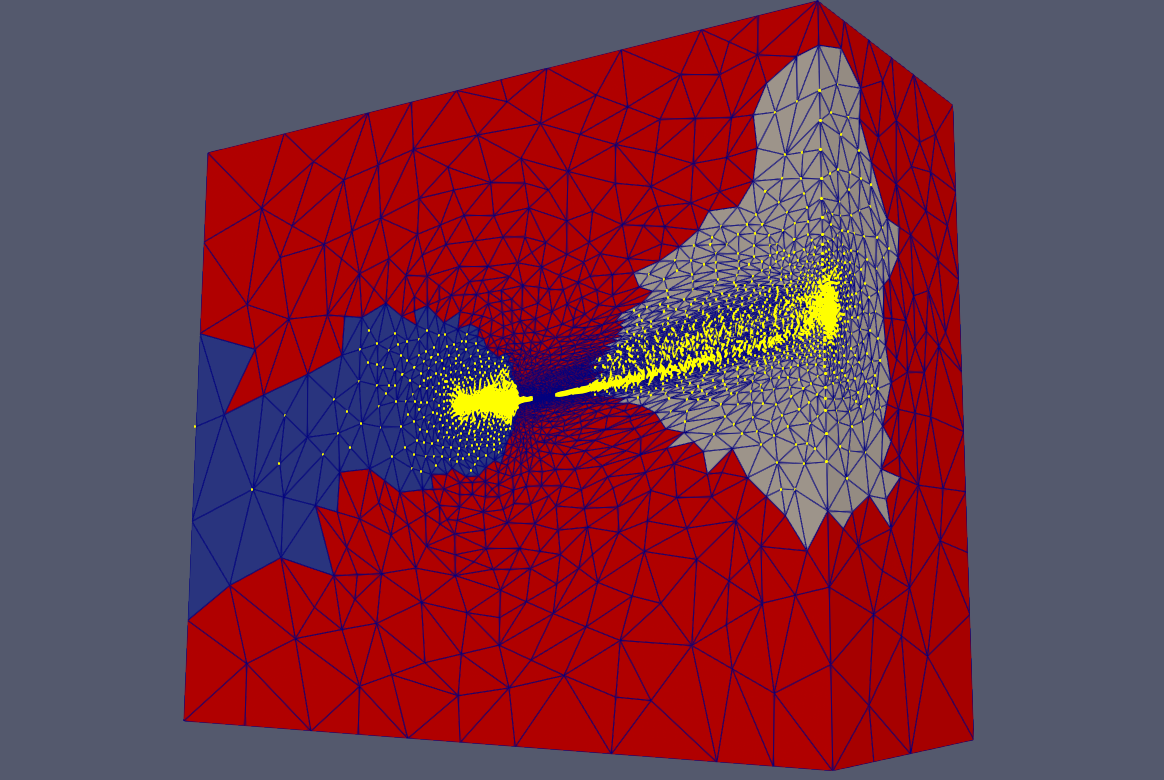}
         \caption{Pseudo-active Elements (in Red) Before Interface Adaptation (lines 4 and 6 of Algorithm \ref{alg:high_level_algorithm})}
         \label{visual_pseudo-active_interface}
     \end{subfigure}
     \hfill
     \begin{subfigure}[htb]{0.45\textwidth}
         \centering
         \includegraphics[width=\textwidth]{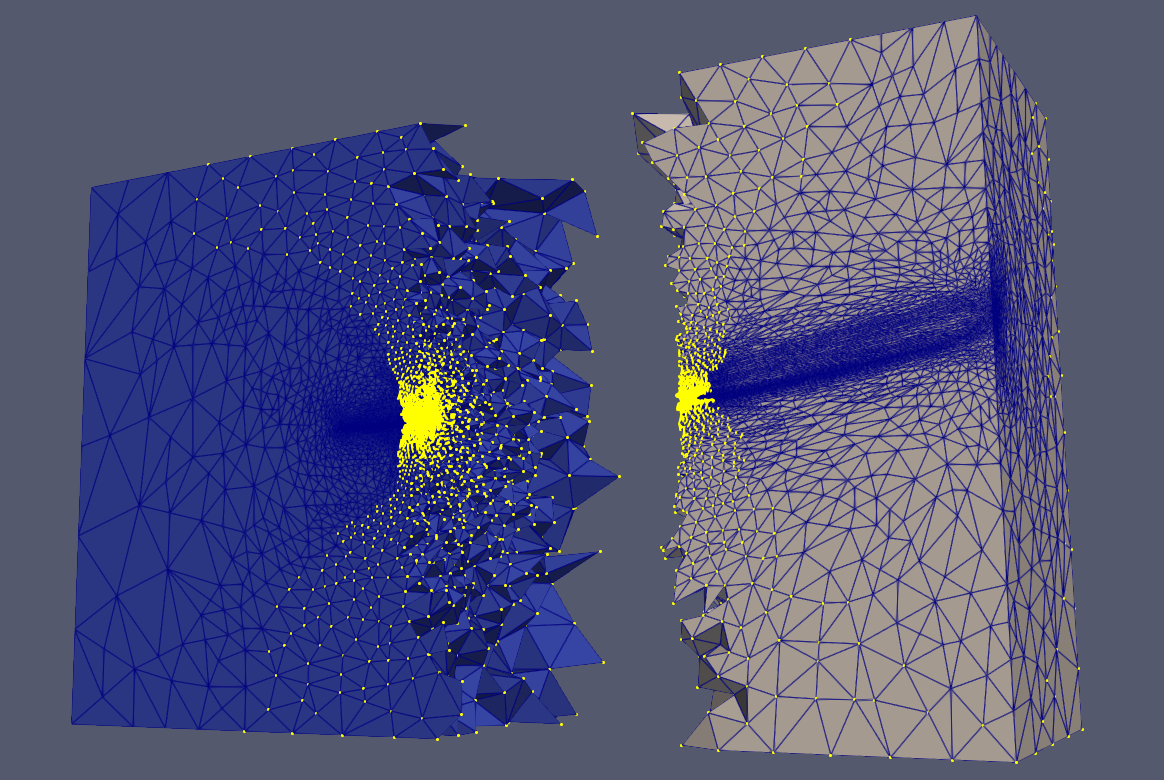}
         \caption{Adapted Interface with New Preemptively Locked Vertices (in Yellow) for Interior Adaptation (line 10 of Algorithm \ref{alg:high_level_algorithm})}
         \label{visual_interior_lock}
     \end{subfigure}
     \hfill
     \begin{subfigure}[htb]{0.45\textwidth}
         \centering
         \includegraphics[width=\textwidth]{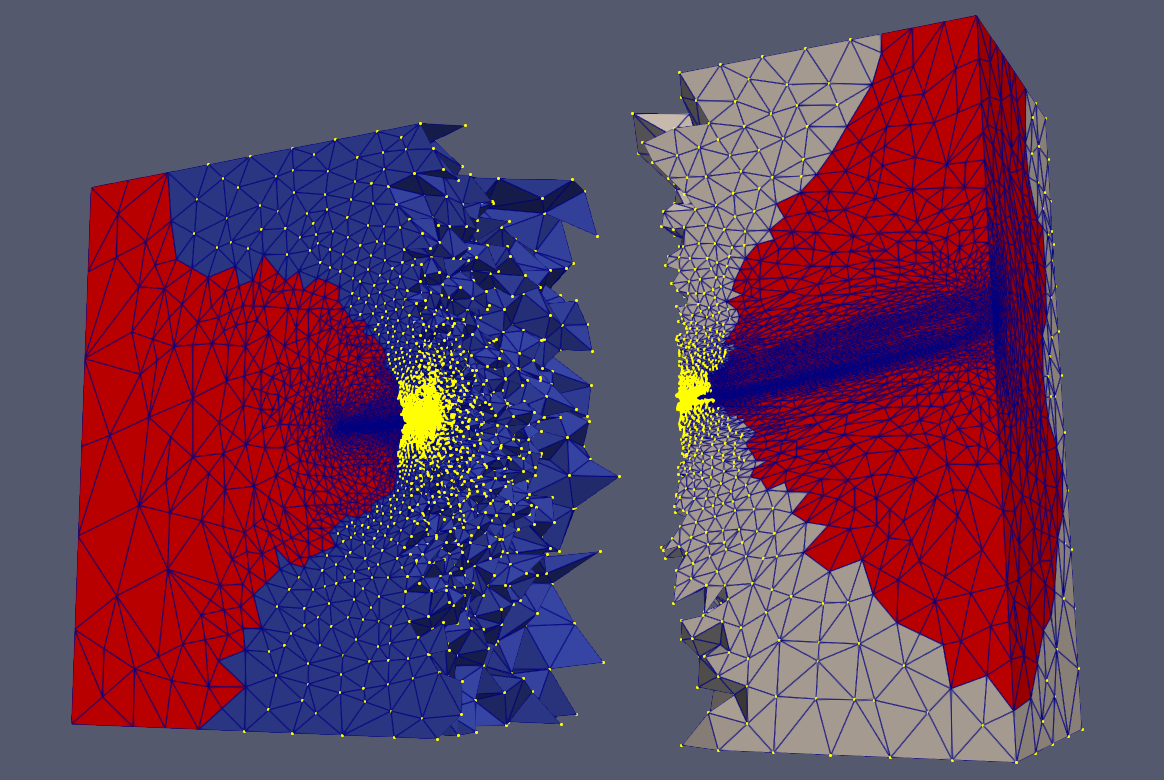}
         \caption{Adapted Interface with Pseudo-active Interior Elements (in Red) Before Interior Adaptation (line 10 of Algorithm \ref{alg:high_level_algorithm})}
         \label{visual_pseudo-active_interior}
     \end{subfigure}
     \hfill
     \begin{subfigure}[htb]{0.45\textwidth}
         \centering
         \includegraphics[width=\textwidth]{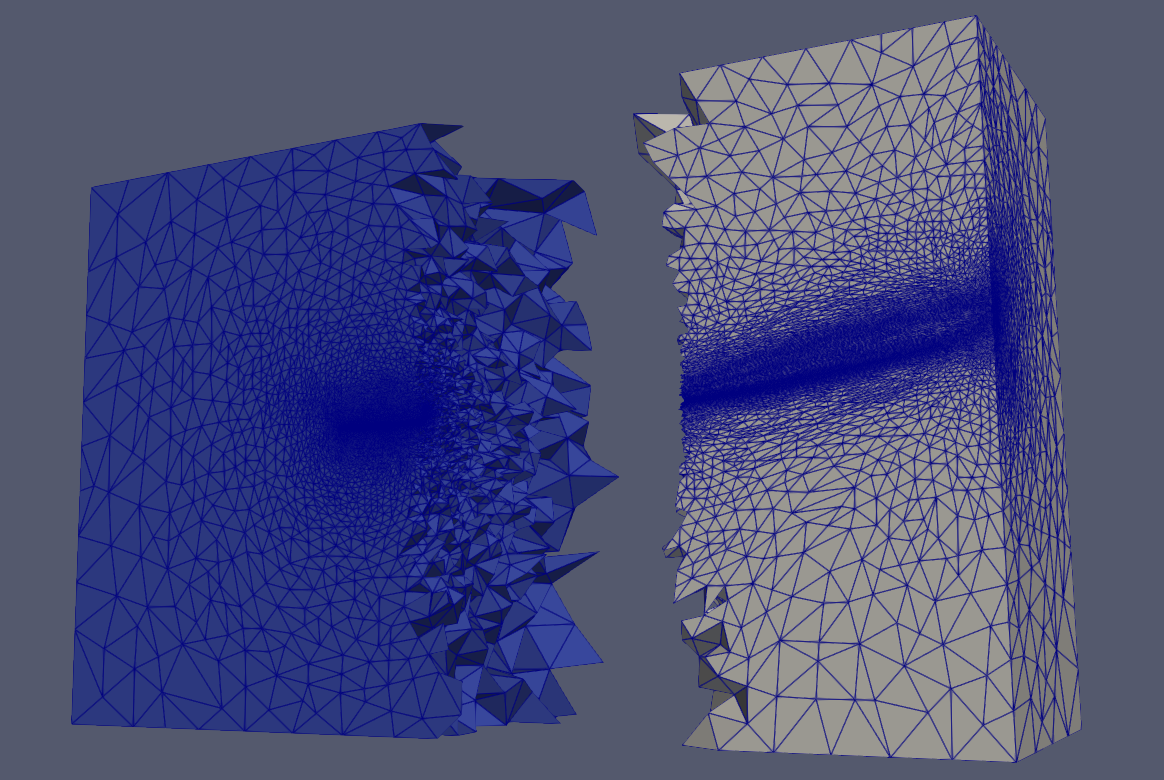}
         \caption{Adapted Subdomain Interiors (line 11 of Algorithm \ref{alg:high_level_algorithm})}
         \label{visual_adapted_interior}
     \end{subfigure}
        \caption{Shown is a visual example of a two-subdomain decomposed delta wing geometry as it is processed by the distributed memory CDT3D method.}
        \label{fig:dmcdt3d_visual_example}
\end{figure}

As mentioned previously, some operations modify elements regardless of their active status (i.e., edge collapse and smoothing). If these operations are permitted to modify inactive elements (deactivated due to being pseudo-inactive), this may cause CDT3D's other meshing operations to propagate over entire regions of the mesh that were pseudo-inactive (rendering the notion of deactivating elements beforehand obsolete). Every newly created element within CDT3D is automatically designated as active, so if a pseudo-inactive element was deleted and replaced with a new active element, a subsequent operation (such as point creation or local reconnection) would be applied to it, affecting surrounding elements, and so on. The preemptive locking of pseudo-inactive element points prevents this phenomenon, establishing control of the edge collapse and vertex smoothing operations within the distributed memory method as a result of the speculative execution model. The combination of the pseudo-active modification and preemptive locking in turn gives control over the point creation and local reconnection operations while maintaining good performance. Fig. \ref{fig:dmcdt3d_visual_example} shows a visual example of a two-subdomain decomposed delta wing geometry as it is processed by the distributed method through both phases of adaptation - interface and interior. Each subdomain is distinguished by a different color. Red elements are pseudo-active elements whose points remain unlocked before adaptation (all other elements are pseudo-inactive and their points are preemptively locked). Preemptively locked points are designated by their yellow color in Fig. \ref{fig:dmcdt3d_visual_example}. 

It should also be noted that the order in which operations are designed to be executed in the shared memory method should also be reflected in their order of execution in the distributed method. As described in section \ref{edge_collapse_vertex_smoothing} and seen in Fig. \ref{fig:cdt3d-anisotropic-pipeline}, the edge collapse operation serves as a pre-refinement and post-refinement operation. The additional layers of elements that will be activated/unlocked during preprocessing may overlap (between those activated/unlocked before interface adaptation and those before interior adaptation). For example, this can be seen in Fig. \ref{visual_pseudo-active_interface} and \ref{visual_pseudo-active_interior} with the fact that there are pseudo-active elements in the corners of the geometry during both phases of adaptation. During interior adaptation, some of these activated layers will have already been adapted during interface adaptation. If CDT3D were to be used as a black box for the interior adaptation of subdomains (meaning that all meshing operations were to be used as default with no preemptively locked or pseudo-inactive data), the method would remove adapted elements and repeat some of the work it already did. More specifically, the edge collapse operation (with regards to its post-refinement design, discussed in section \ref{edge_collapse_vertex_smoothing}) would remove elements from some layers during interface adaptation that would later be re-generated during interior adaptation, wasting time and computational resources. When attempting to use a method like the shared memory CDT3D as a black box, one must be mindful of each operation's design/purpose, how they satisfy the reproducibility requirement described in section \ref{introduction}, and how they can be used appropriately in a distributed setting to maximize potential scalability. Edge collapse is only used as a pre-refinement operation during the interface adaptation phase (line 7 of Algorithm \ref{alg:high_level_algorithm}). Additionally, the point creation and local reconnection operations of CDT3D's mesh adaptation module (seen in Fig. \ref{fig:cdt3d-anisotropic-pipeline}) are utilized during the interface adaptation phase. Edge collapse is not used as a post-refinement operation during the interface adaptation phase. The quality improvement module is executed optionally during this phase if needed to satisfy metric conformity (discussed in sections \ref{results} and \ref{discussion}). The full suite of operations (from both the mesh adaptation and quality improvement modules) are executed during interior adaptation (line 11 of Algorithm \ref{alg:high_level_algorithm}). To satisfy this "order of operations" requirement, the preemptively locked vertices and pseudo-inactive elements from interior preprocessing are unlocked/activated (except the interface elements themselves) before the post-refinement edge collapse operation and quality improvement operations commence during interior adaptation (since those elements/vertices were not processed by the post-refinement edge collapse operation during interface adaptation). Section \ref{how_modifications_affect_cdt3d} in our Results shows an example of how the pseudo-active modification for elements, preemptive locking of points, and respecting the "order of operations" (i.e., utilizing certain operations during each phase of adaptation) affects the shared memory CDT3D method within the distributed method, and why they are necessary. Through the parameters utilized, table \ref{parameters} also shows which operations are applied during each phase of adaptation for different geometry cases (discussed further in sections \ref{results} and \ref{discussion}). More details of each operation and how they function with the pseudo-active modification and preemptive locking are described next.

\subsubsection{Point Creation/Insertion} \label{point_creation}
As stated previously, buckets (or lists) of active tetrahedra are assigned to threads. Each thread iterates through its list of tetrahedra, where a centroid-based point creation technique is utilized to create candidate points. If a candidate point satisfies all proximity checks and encroachment rules (e.g., not too close to an existing point or boundary face, etc.), it is accepted and inserted into the mesh. For more details regarding the implementation of these operations and the spacing criteria used, see \cite{EwCCDT3D}.

\begin{figure}[htbp]
     \centering
     \begin{subfigure}[htb]{0.3\textwidth}
         \centering
         \includegraphics[width=\textwidth]{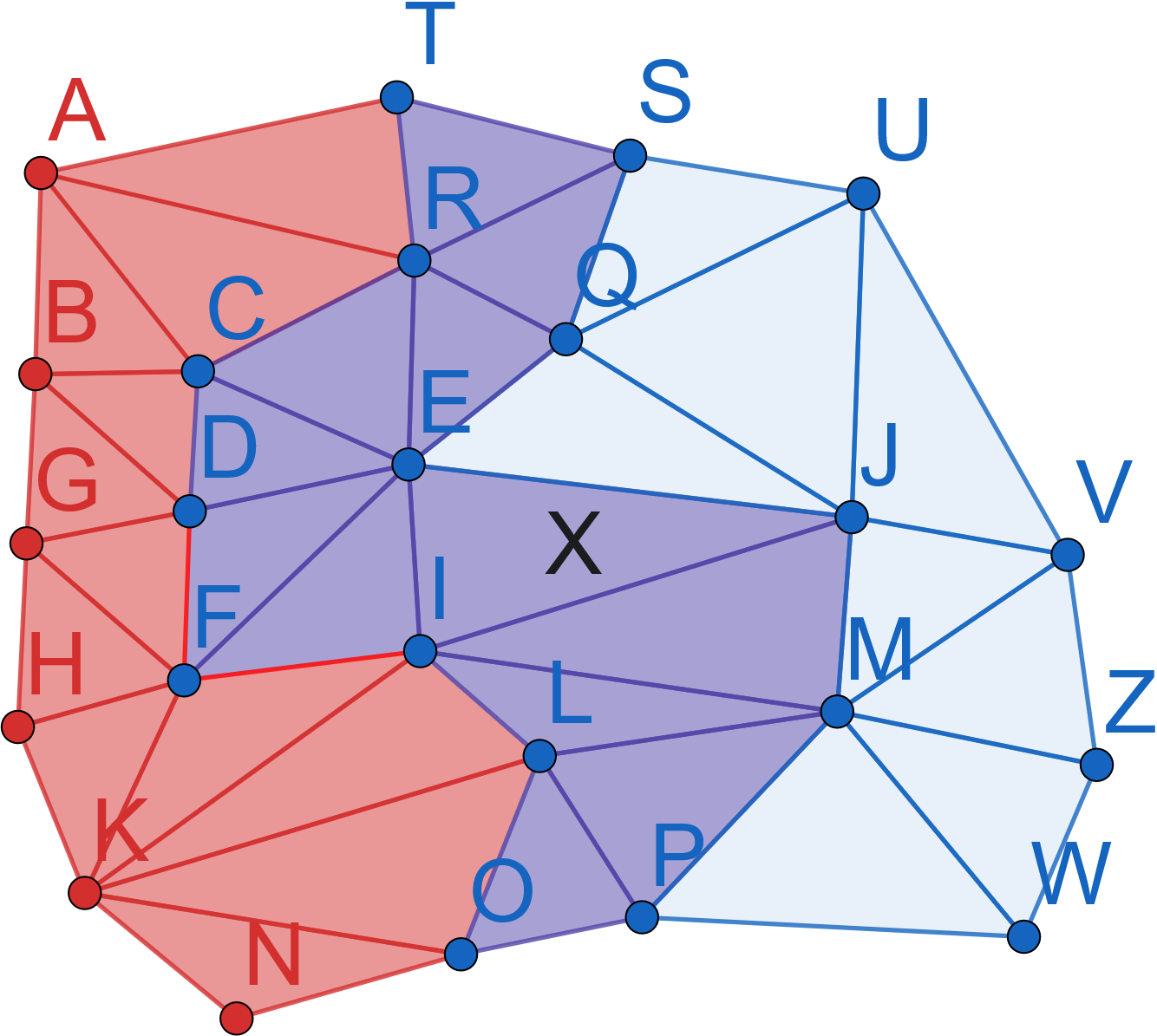}
         \caption{}
     \end{subfigure}
     \hfill
     \begin{subfigure}[htb]{0.3\textwidth}
         \centering
         \includegraphics[width=\textwidth]{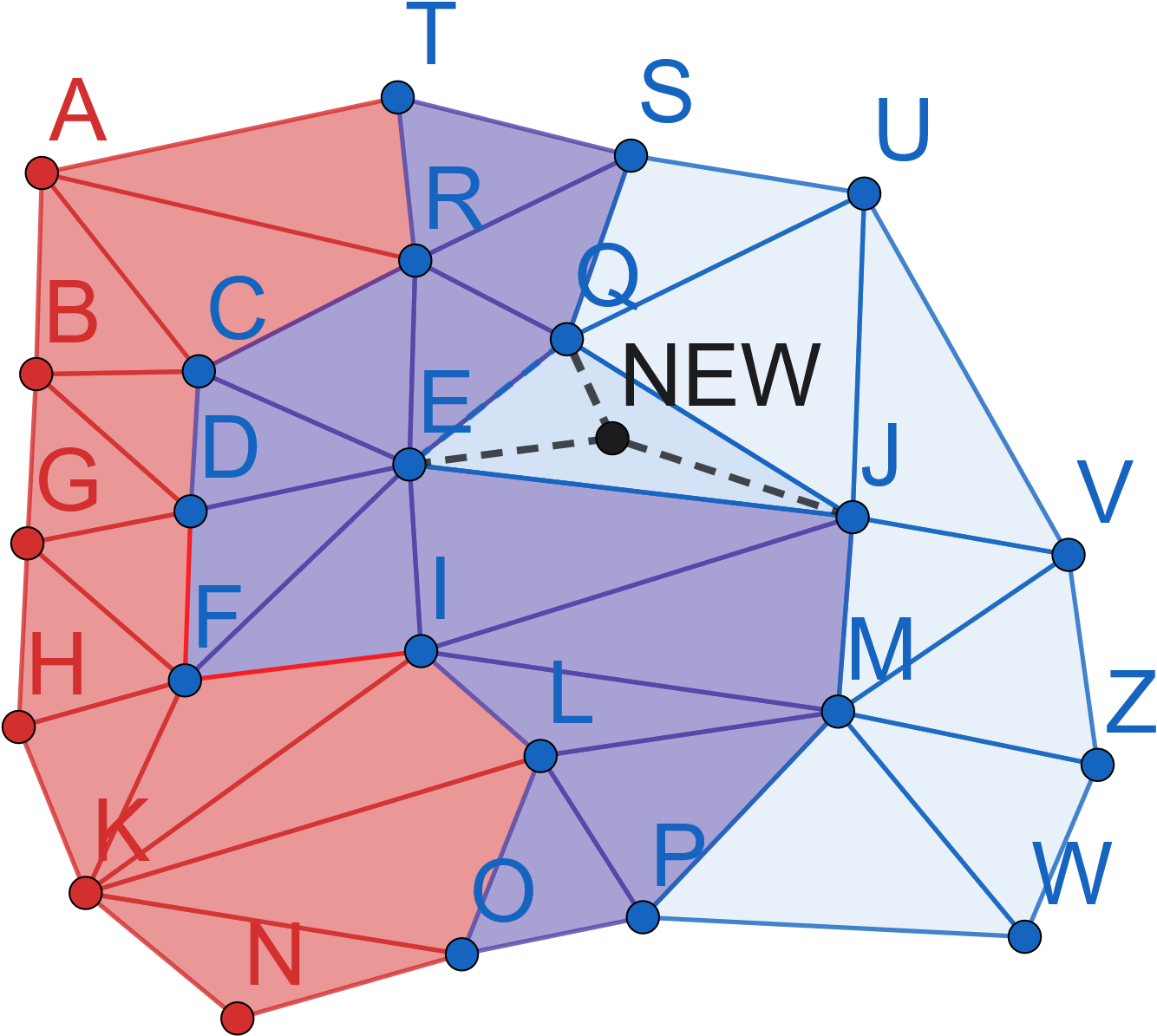}
         \caption{}
     \end{subfigure}
     \hfill
     \begin{subfigure}[htb]{0.3\textwidth}
         \centering
         \includegraphics[width=\textwidth]{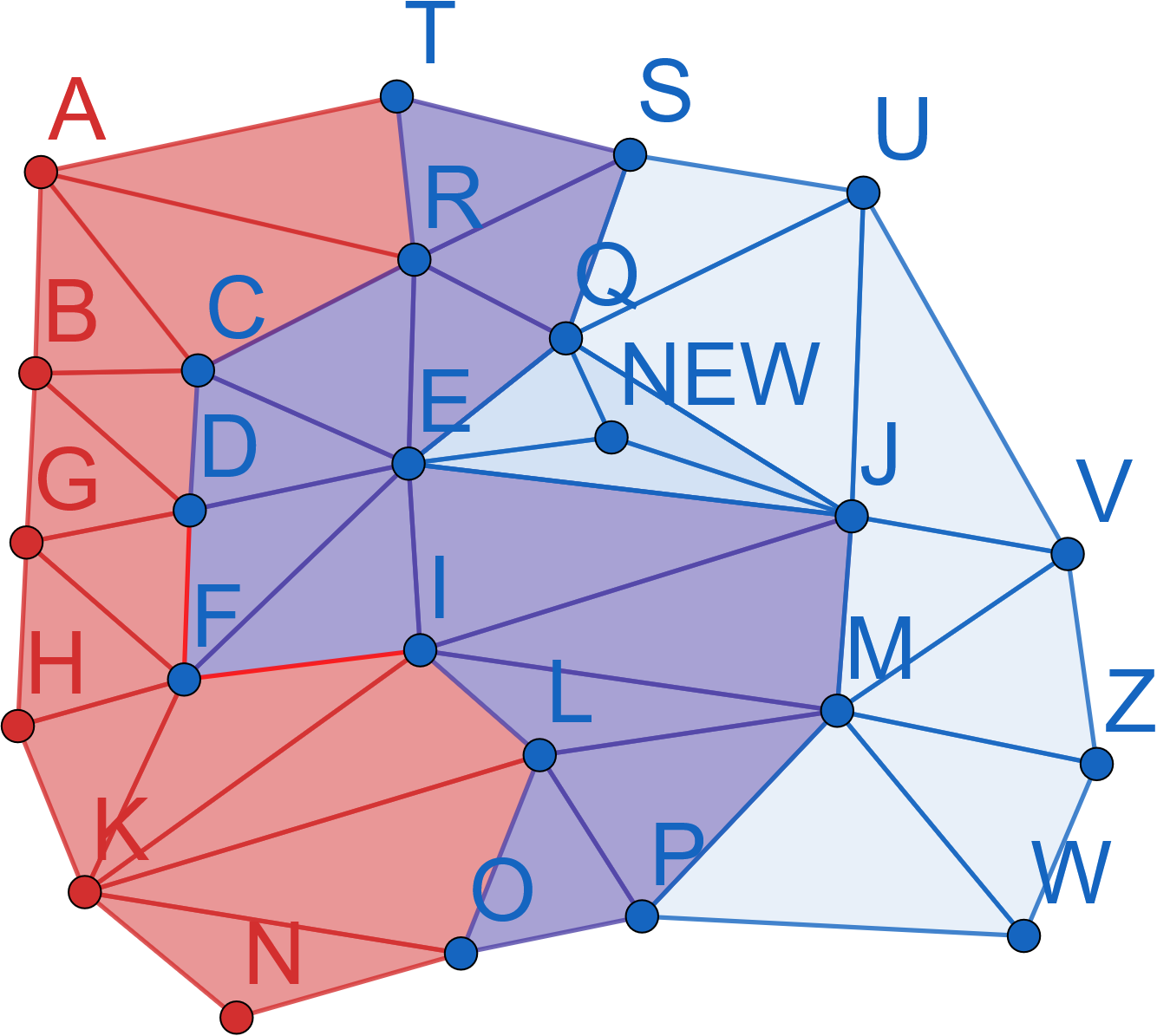}
         \caption{}
     \end{subfigure}
     \begin{subfigure}[htb]{0.3\textwidth}
         \centering
         \includegraphics[width=\textwidth]{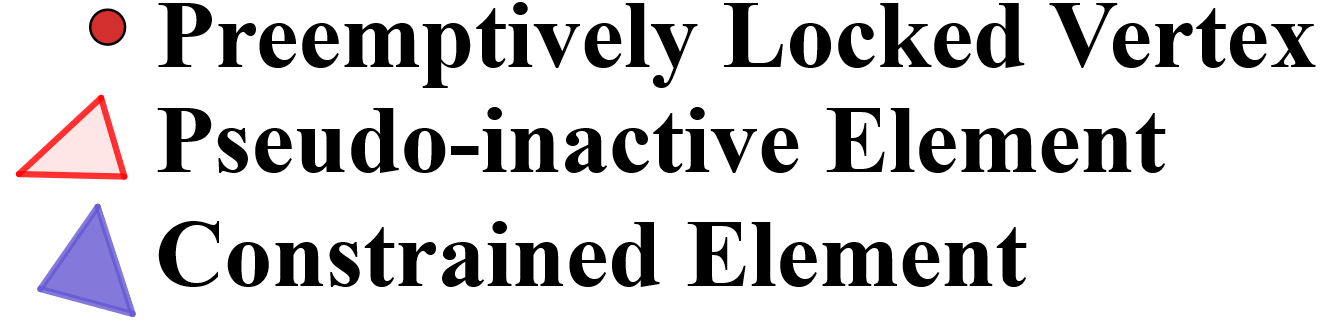}
     \end{subfigure}
        \caption{Shown is a 2D example of constrained elements that cannot undergo point creation/insertion because they are adjacent to pseudo-inactive elements.}
        \label{fig:point_creation}
\end{figure}

It should be noted that point creation/insertion on elements adjacent to pseudo-inactive elements inhibits grid generation convergence. We have found that CDT3D will continuously attempt to improve elements adjacent to these elements by inserting points over numerous iterations until crashing. The method will struggle to truly satisfy spacing requirements within the metric space and will inadvertently create slivers (i.e., elements with a volume near zero). Therefore, a buffer zone must be established around pseudo-inactive elements to reach convergence. Point creation/insertion is prohibited for any element within this buffer zone. Elements within the buffer zone include those defined by a vertex that is also shared by a pseudo-inactive element. Fig. \ref{fig:point_creation} shows a 2D example of constrained elements (those in purple) that cannot undergo point creation/insertion because they are adjacent to (i.e., share a point with) pseudo-inactive elements (those in red, which also cannot undergo point creation/insertion). The term "constrained" is used to designate mesh entities in CDT3D's operations that cannot undergo adaptation simply because they are adjacent to a pseudo-inactive element or preemptively locked point. (a) shows the point creation operation failing to generate any candidate points for element EJI because it is constrained due to point I being shared with elements IKF and ILK. In (b), point creation is permitted for element QJE, resulting in the creation of three new elements in (c). In the context of performing interior adaptation with frozen interface elements, constraining elements that surround interfaces is a practice also seen in the state-of-the-art method \textit{Feflo.a} \cite{LOSEILLEFefloa}, where surrounding cavity elements must be included in the adaptation of interface elements to produce good quality elements. 


\subsubsection{Edge Collapse and Vertex Smoothing} \label{edge_collapse_vertex_smoothing}
\begin{figure}[htbp]
     \centering
     \begin{subfigure}[htb]{0.3\textwidth}
         \centering
         \includegraphics[width=\textwidth]{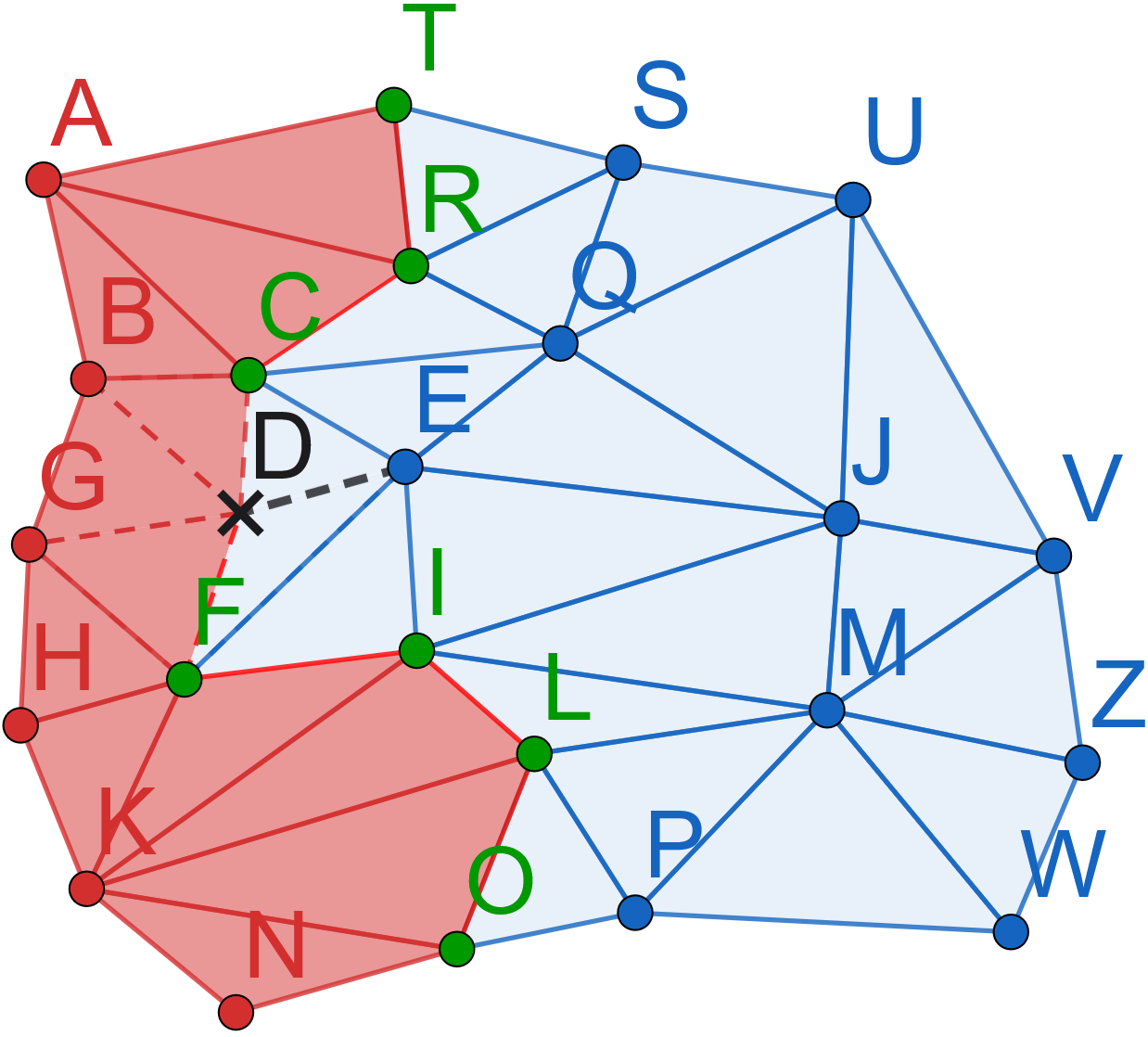}
         \caption{}
     \end{subfigure}
     \hfill
     \begin{subfigure}[htb]{0.3\textwidth}
         \centering
         \includegraphics[width=\textwidth]{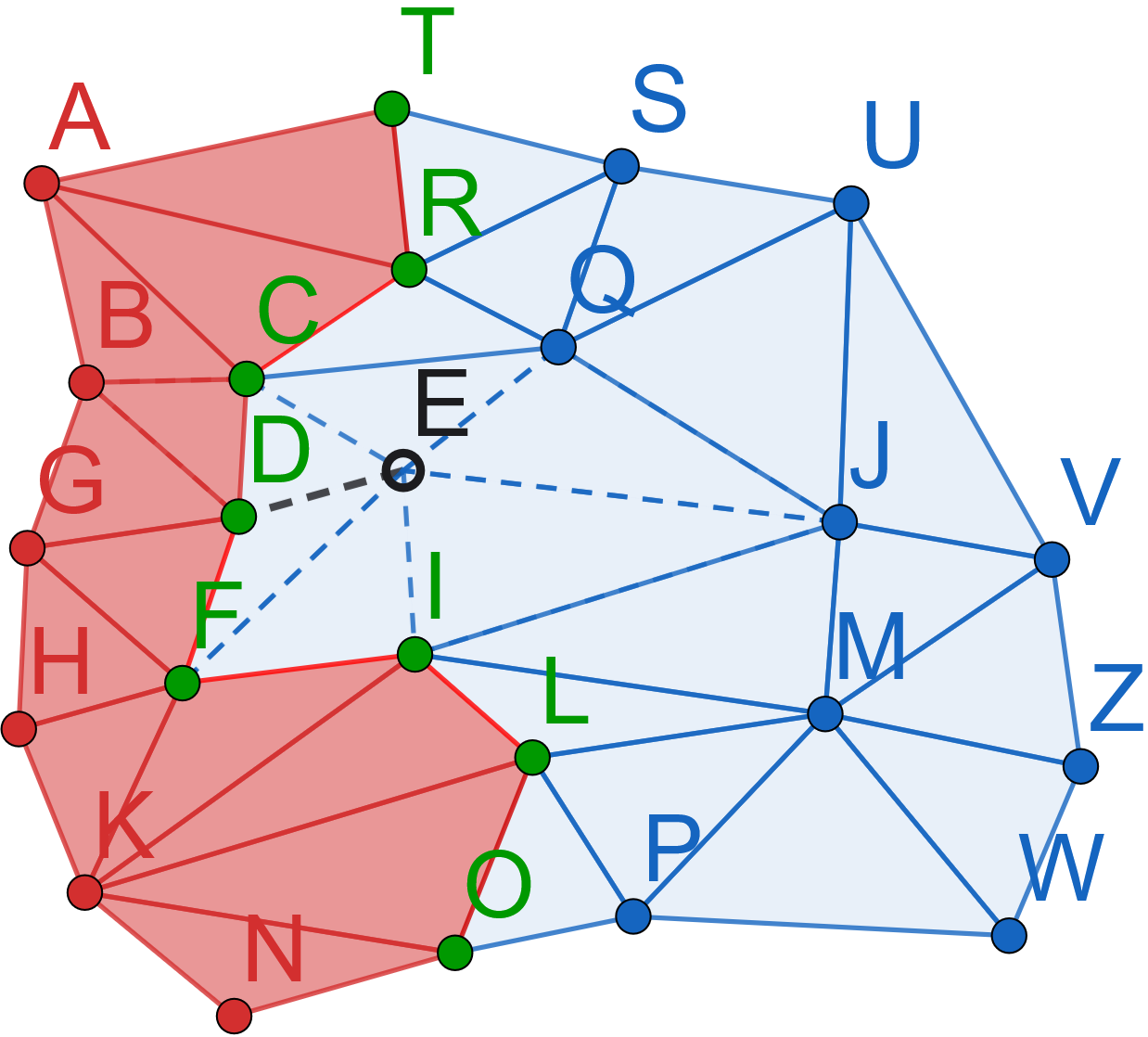}
         \caption{}
     \end{subfigure}
     \hfill
     \begin{subfigure}[htb]{0.3\textwidth}
         \centering
         \includegraphics[width=\textwidth]{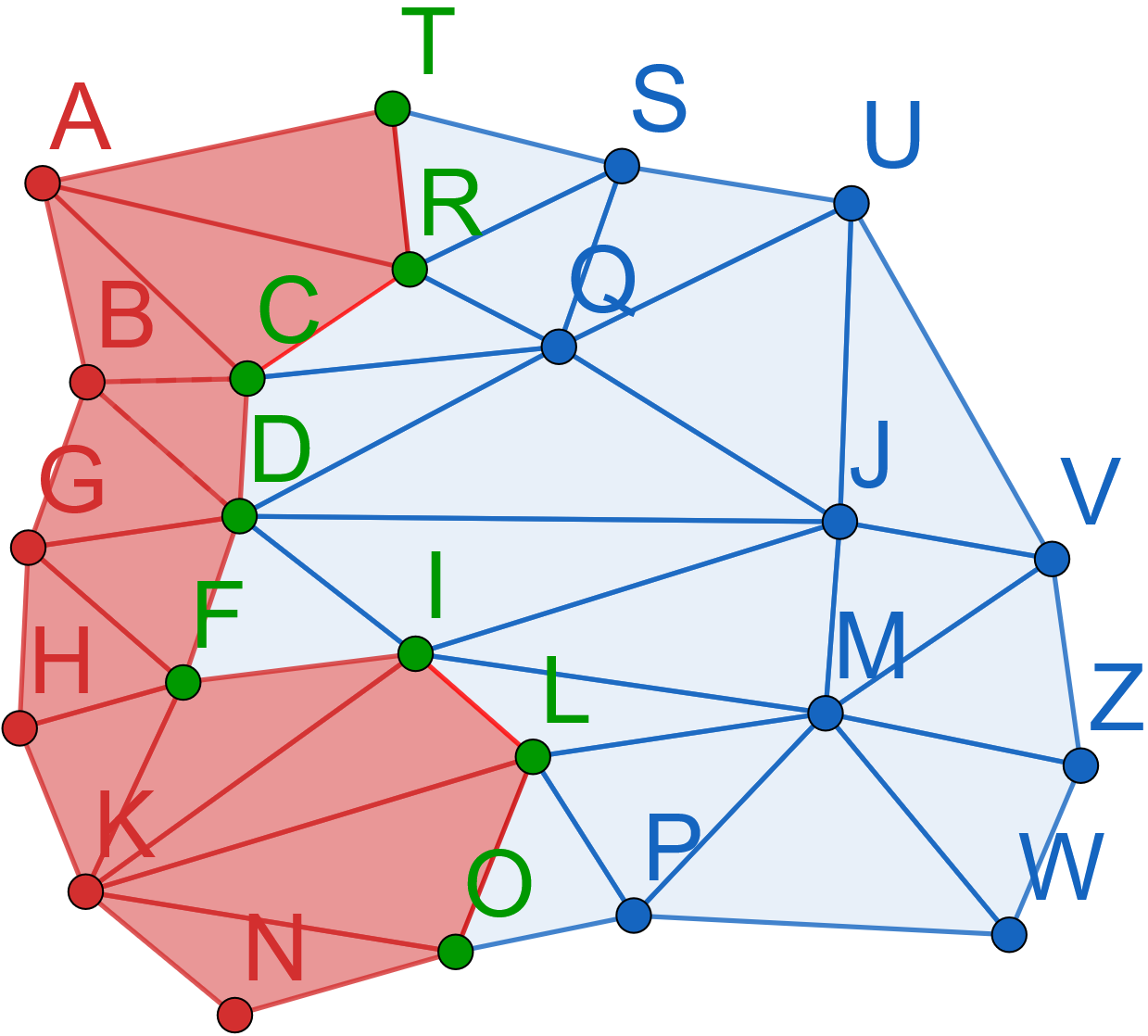}
         \caption{}
     \end{subfigure}
     \begin{subfigure}[htb]{0.5\textwidth}
         \centering
         \includegraphics[width=\textwidth]{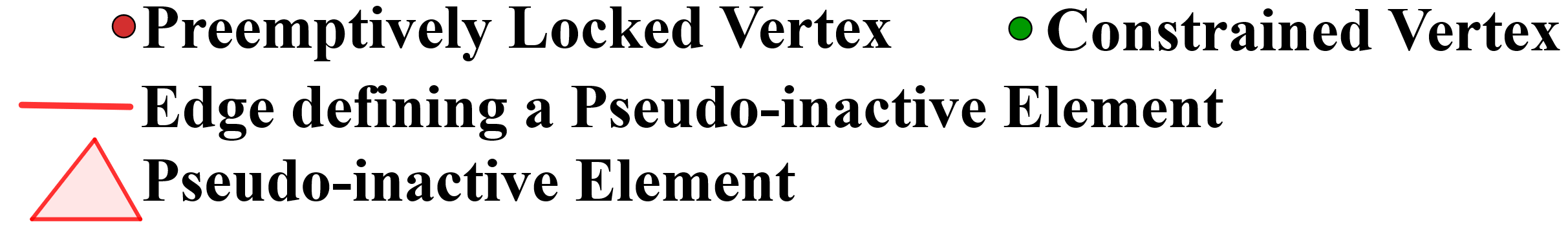}
     \end{subfigure}
        \caption{Shown is a 2D example of an edge and points that would not be allowed to undergo removal or smoothing due to preemptively locked vertices.}
        \label{fig:edge_collapse_vertex_smoothing}
\end{figure}

The edge collapse operation is used within CDT3D as a pre-refinement and post-refinement operation. It serves to remove edges smaller than a target value. As a pre-refinement operation, coarsening a mesh before executing subsequent operations can lead to better quality in the final output mesh (demonstrated in \cite{TsolakisEvaluation}). As a post-refinement operation, it is used to remove short edges created during refinement/adaptation. The vertex smoothing operation moves a vertex incrementally along multiple directions of a search space to determine a position that optimizes the quality of all connected tetrahedra. Both the edge collapse and vertex smoothing operations follow the speculative execution model, described in section \ref{shared_memory_mesh_generation_method}. Each operation iterates through all vertices, attempting to lock each vertex and all its adjacent vertices (implicitly granting exclusive access to all adjacent tetrahedra). If any locks fail, all acquired resources are released and the method proceeds to the next vertex. See \cite{EwCCDT3D} for more details regarding the implementation and edge length criteria used by these operations. Fig. \ref{fig:edge_collapse_vertex_smoothing} shows a 2D example of an edge and points that would not be allowed to undergo removal or smoothing due to preemptively locked vertices. If an edge contains such a vertex, it cannot be removed. If a vertex is adjacent to a preemptively locked vertex, it cannot be smoothed (i.e., it is constrained). Any edge that does not define a pseudo-inactive element, or contain a preemptively locked vertex, may be removed (i.e., any edge that is blue in the figure). However, the primary vertex that a thread attempts to lock must be considered. The dotted lines are edge connections to points that are dependencies of the primary vertex (those that a thread will attempt to lock). In (a), the black dotted edge DE cannot be removed because the primary vertex is D, which is constrained because it is adjacent to preemptively locked vertices B and G. In (b), the primary vertex is E, and all adjacent vertices are successfully locked by the thread. Edge ED can therefore be removed, resulting in (c). Because vertex smoothing uses the same locking mechanisms, a similar principle applies. No preemptively locked vertices may undergo smoothing because they are already locked. No constrained vertex (those in green) may undergo smoothing because a thread would fail to lock all their adjacent vertices (i.e., those that are preemptively locked vertices). Edge collapse is only used as a pre-refinement operation during the interface adaptation phase (line 7 of Algorithm \ref{alg:high_level_algorithm}). It is used as a pre-refinement and post-refinement operation during the interior adaptation phase (line 11 of Algorithm \ref{alg:high_level_algorithm}). Vertex smoothing can be utilized during both phases of the distributed memory method and is not restricted to either. This usage of the operations was shown to yield the optimal qualitative results in the fastest time during our testing of DM\_CDT3D (section \ref{results}).

\subsubsection{Local Reconnection}
\begin{figure}[htbp]
     \centering
     \begin{subfigure}[htb]{0.3\textwidth}
         \centering
         \includegraphics[width=\textwidth]{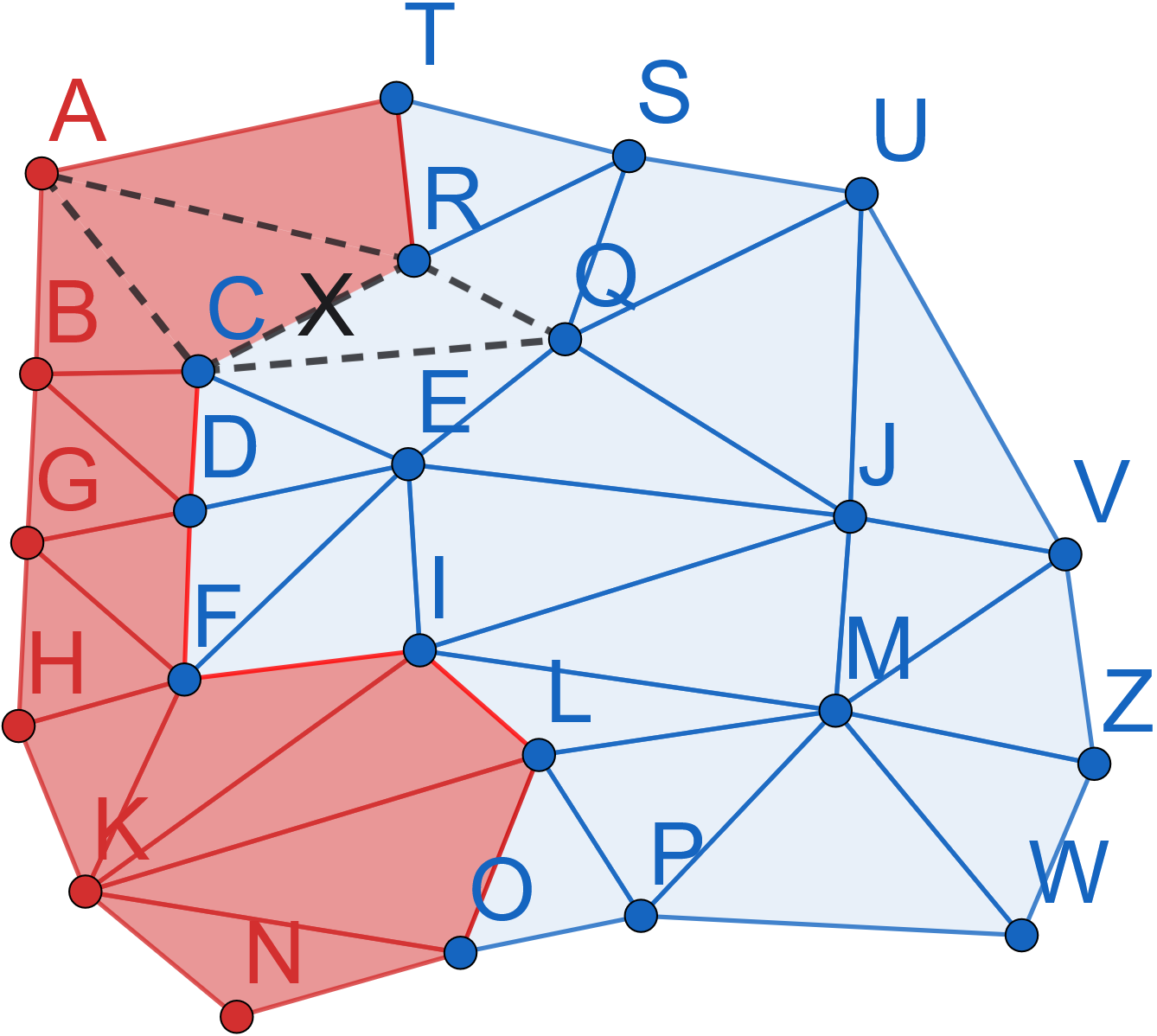}
         \caption{}
     \end{subfigure}
     \hfill
     \begin{subfigure}[htb]{0.3\textwidth}
         \centering
         \includegraphics[width=\textwidth]{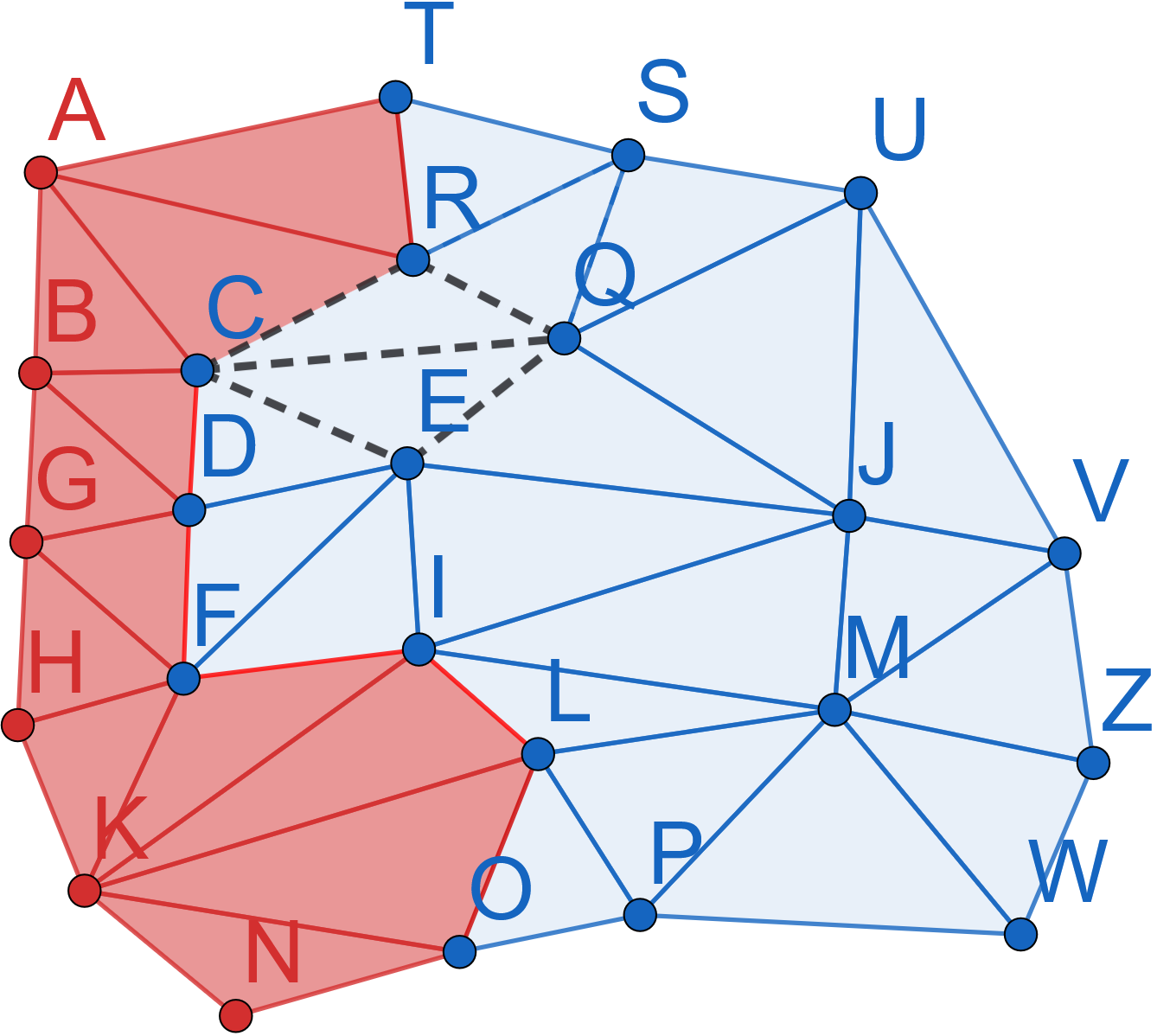}
         \caption{}
     \end{subfigure}
     \hfill
     \begin{subfigure}[htb]{0.3\textwidth}
         \centering
         \includegraphics[width=\textwidth]{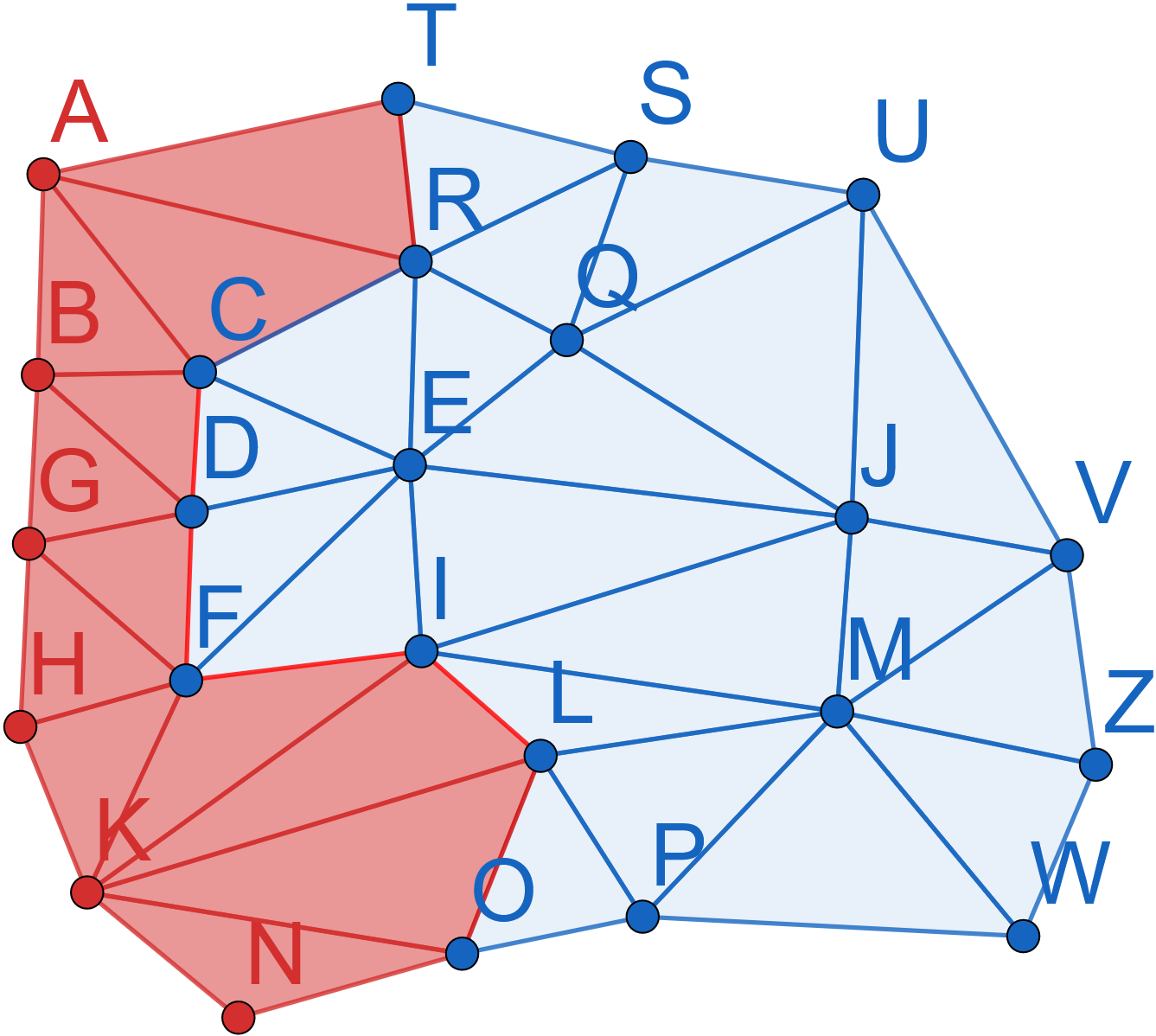}
         \caption{}
     \end{subfigure}
     \begin{subfigure}[htb]{0.3\textwidth}
         \centering
         \includegraphics[width=\textwidth]{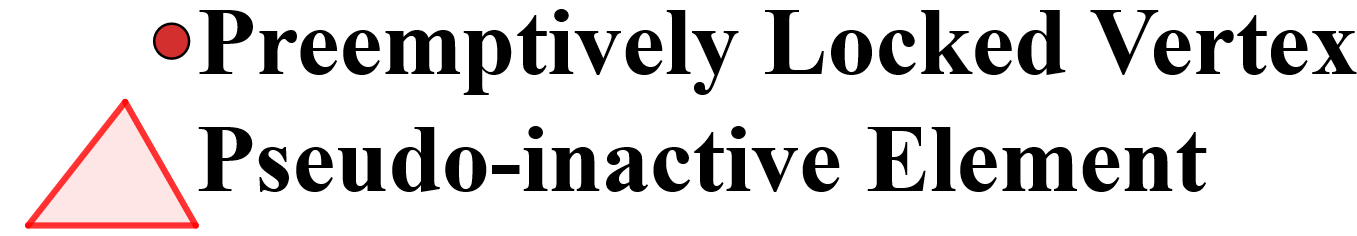}
     \end{subfigure}
        \caption{Shown is a 2D example of elements that cannot undergo flips if they are pseudo-inactive or defined by preemptively locked vertices.}
        \label{fig:local_reconnection}
\end{figure}

Local Reconnection includes four types of topological transformations, or flips, that include metric-based information to improve element quality. This operation also adheres to the speculative execution model. When each thread iterates through its list(s) of assigned tetrahedra, the thread attempts to lock the vertices of a tetrahedron. For each of the tetrahedron's faces that have a neighboring tetrahedron, the thread attempts to lock the opposite vertex of that neighbor tetrahedron and attempts flips. If any locking fails, the thread either moves on to the next neighbor or releases any acquired resources and moves on to the next tetrahedron. For more details regarding this operation's implementation, see \cite{EwCCDT3D}. Fig. \ref{fig:local_reconnection} shows a 2D example of elements that would be permitted to undergo flips. Any combination of elements are permitted to undergo local reconnection iff none of those elements are pseudo-inactive elements (i.e., those in red). As an example, the black dotted lines represent elements that the method attempts to lock in order to perform a flip. In (a), elements ARC and RQC cannot undergo swapping. Element ARC cannot be locked by a thread due to point A already being locked (since it is a preemptively locked vertex). In (b), elements RQC and CQE are permitted to undergo swapping since neither are defined by preemptively locked points. (c) is the result of the flip, creating elements CRE and RQE. 

\subsection{Making Subdomains Simply Connected} \label{making_subdomains_simply_connected}
The shared memory CDT3D software is designed to process only manifold geometries. Consequently, every subdomain must be simply connected before allowing CDT3D to adapt them. This means that no tetrahedron, or partition of tetrahedra, are loosely-connected, sharing only a point or an edge, to other tetrahedra or partitions within that subdomain. Every tetrahedron must share a face with at least one other tetrahedron. All tetrahedra within a subdomain can be considered as an undirected graph, where a tetrahedron is a vertex and an edge denotes that two tetrahedra share a face. Through edge traversal, if every vertex of the undirected graph can reach every other vertex, then the subdomain is simply connected.

A challenge with data decomposition methods (such as PQR and other graph-based methods \cite{CHRISOCHOIDES199475}) is that they do not necessarily guarantee a decomposition with simply connected subdomains. During interface adaptation in the distributed method, elements are assigned a subdomain id based on the dominant subdomain to which their face-connected neighbor tetrahedra are already assigned (this however does not guarantee that subdomains will be simply connected). After interface adaptation is complete, algorithm \ref{alg:simply_connected} is executed. Elements are reassigned between subdomains, ensuring that all subdomains are simply connected before they are packed and distributed. The largest partition of face-connected tetrahedra within each subdomain are designated to remain in that subdomain. Loosely-connected partitions are identified and re-assigned to another subdomain based on how many tetrahedra share faces with neighboring tetrahedra within that other subdomain. It should be noted that this routine of making subdomains simply connected is sequential, with a runtime complexity of $O(nm^2)$ (where $n$ is the number of subdomains and $m$ is the number of tetrahedra generated thus far in each subdomain). This routine remains to be parallelized in future work. It should also be noted that after this routine is complete and the corresponding shared memory data structures are converted into simple distributed memory data structures (as specified in section \ref{data_structures}), subdomains are packed in parallel (utilizing OpenMP threads) and are then unpacked sequentially as they are received by the respective nodes to which they were sent.

\begin{algorithm} 
\caption{Making Subdomains Simply Connected}
\label{alg:simply_connected}
\normalsize{
\begin{flushleft}
MAKE\_SIMPLY\_CONNECTED($S_1 ... S_N$) \\
\textbf{Input}: $S_1 ... S_N$ are data decomposed subdomains \\
\hspace*{\algorithmicindent} \ \ \ \ \ Every tetrahedron in $S_1 ... S_N$ should be marked as "NOT checked" \\
\textbf{Output}: Each subdomain in $S_1 ... S_N$ will be simply connected (i.e., manifold)
\end{flushleft}
\begin{algorithmic}[1]
\ForEach{subdomain $\in S_1 ... S_N$}
    \State simplyConnected = true
    \State partitions = $\emptyset$
    \State tetCounter = 0
    \State numTetsInSubdomain = number of tetrahedra currently assigned to subdomain
    \State subdomainTets = list of tetrahedra assigned to subdomain
    \State nextTetrahedron = first tetrahedron in subdomainTets
    \While{tetCounter $<$ numTetsInSubdomain}
        \State partition = new empty list
        \State FACE\_TRAVERSAL(nextTetrahedron, tetCounter, partition)
        \State Partitions.ADD(partition)
        \ForEach{tetrahedron $\in$ subdomainTets}
            \If{tetrahedron is NOT checked}
                \State nextTetrahedron = tetrahedron
                \State simplyConnected = false
                \State Exit For Loop \Comment{// Exit loop \textbf{foreach} tetrahedron $\in$ subdomainTets}
            \EndIf
        \EndFor
    \EndWhile
    \If{NOT simplyConnected}
        \State largest\_partition = partition with greatest number of tetrahedra \Comment{// randomly selected if same size}
        \ForEach{partition $\in$ partitions}
            \If{partition is NOT largest\_partition}
                \State \multiline{dominant\_subdomain = another subdomain to which the majority of face-connected neighbor tetrahedra (of all tetrahedra in partition) have been assigned that is NOT this subdomain}
                \State subdomain.REMOVE(partition)
                \State dominant\_subdomain.ADD(partition)
            \EndIf
        \EndFor
    \EndIf
\EndFor
\end{algorithmic}

\begin{flushleft}
FACE\_TRAVERSAL(tetrahedron, tetCounter, partition) \\
\textbf{Input}: tetrahedron is the tetrahedron from which faces will be traversed \\
\hspace*{\algorithmicindent} \ \ \ \ \ tetCounter keeps track of the number of tetrahedra traversed \\
\hspace*{\algorithmicindent} \ \ \ \ \ partition stores each tetrahedron traversed \\
\textbf{Output}: Traversed tetrahedra will have been counted, marked as "checked," and stored in partition
\end{flushleft}
\begin{algorithmic}[1]
\State tetCounter = tetCounter + 1
\State Mark tetrahedron as checked
\State partition.ADD(tetrahedron)
\ForEach{neighbor\_tetrahedron that is face-connected to tetrahedron}
    \If{neighbor\_tetrahedron is NOT checked}
        \State FACE\_TRAVERSAL(neighbor\_tetrahedron, tetCounter, partition)
    \EndIf
\EndFor
\end{algorithmic}
}
\end{algorithm}

\subsection{Alternate A Posteriori Interface Adaptation Approach} \label{aposteriori_DMCDT3D}
Algorithm \ref{alg:aposteriori_algorithm} shows a distributed memory a posteriori interface adaptation approach with CDT3D that was previously implemented in \cite{Garner24EarlyDMCDT3D}. This early approach focused on adapting the interior of all subdomains first. Layers of elements would then be shifted between subdomains to satisfy the dependencies of interface elements and those subdomains would undergo adaptation again to focus on these interface elements. This would occur over numerous iterations. The runtime system PREMA was particularly useful for this approach. It allowed for subdomains (i.e., mobile objects) to communicate with one another within neighborhoods when shifting data. Each neighborhood would perform interface shifts and adaptation independently of each other while updating duplicate data between subdomains, avoiding the use of all-to-all communication techniques. However, this approach had several drawbacks that deteriorated its performance. The pseudo-active modification was utilized but there was no preemptive locking to efficiently control the edge collapse and vertex smoothing operations during interface adaptation. There is also an implicit global synchronization point in each iteration of shifting data (in the while loop, lines 5 - 10 of algorithm \ref{alg:aposteriori_algorithm}), due to the fact that a master/worker model was used. The master would determine which subdomains would send or receive data during these interface shifts. Although a decentralized model would likely improve the method's performance, the following drawback renders this a posteriori approach inferior to the a priori method presented. The overhead of shifting data between subdomains (i.e., the migration of data between parallel processes) quickly becomes a bottleneck after several iterations. When generating large meshes, the method eventually moves large amounts of data between processes just to satisfy the dependencies of a small number of interface elements that have not yet undergone adaptation. This communication overhead occupies approximately 35\% of end-to-end runtime compared to computation (i.e., meshing), as shown in our results. If the preemptive locking mechanism were to be implemented in this a posteriori approach, meshing time would likely decrease and this communication overhead would occupy an even larger percentage of runtime. More details regarding the method and this performance bottleneck can be seen in \cite{Garner24EarlyDMCDT3D}. In section \ref{results}, we compare the end-to-end performance of this a posteriori approach with the improved a priori approach presented.

\begin{algorithm}
\caption{A Posteriori Interface Adaptation-based Distributed Memory CDT3D Approach}\label{alg:aposteriori_algorithm}
\footnotesize{
\begin{flushleft}
A Posteriori DM\_CDT3D($M_i$, \textit{m}, N) \\
\textbf{Input}: $M_i$ is the initial mesh \\
\ \ \ \ \ \ \ \ \ \ \ \ \textit{m} is the target metric \\
\ \ \ \ \ \ \ \ \ \ \ \ N is the number of subdomains \\
\textbf{Output}: Adapted mesh that conforms to the anisotropic metric field \textit{m}
\end{flushleft}
\begin{algorithmic}[1]
\State $M_C$ = INITIAL\_COARSE\_MESH\_GENERATION($M_i$, \textit{m})
\State Perform data decomposition of $M_C$ to create subdomains $S_1 ... S_N$
\State Distribute subdomains $S_1$...$S_N$ to processes
\State ADAPT\_SUBDOMAIN($S_1$...$S_N$, \textit{m}$_1$...\textit{m}$_N$) \Comment{// Adapt interior elements amongst subdomains}
\While{($\exists$ elements $\in$ $S_1$...$S_N$ that have NOT undergone mesh adaptation)} \Comment{// Adapt interface elements}
\State $S_{S_1}$...$S_{S_n}$, $S_{R_1}$...$S_{R_k}$ = COLOR\_SUBDOMAINS($S_1$...$S_N$) \Comment{// Color subdomains to designate which will send or receive interface data}
\State INTERFACE\_SHIFT($S_1$...$S_N$)
\State ADAPT\_SUBDOMAIN($S_{R_1}$...$S_{R_k}$, \textit{m}$_{R_1}$...\textit{m}$_{R_k}$)
\State UPDATE\_TOPOLOGY($S_1$...$S_N$)
\EndWhile
\State QUALITY\_IMPROVEMENT($S_1$...$S_N$, \textit{m}$_1$...\textit{m}$_N$) \Comment{// Commence quality improvement over all subdomains' interior elements}
\State TERMINATE()
\end{algorithmic}
}
\end{algorithm}

\subsection{Requirements for the Distributed Method's Design} \label{distributed_requirements}
The following requirements summarize the lessons learned in needing to re-design the shared memory CDT3D method and successfully integrate it into the distributed memory method:

\begin{enumerate}
    \item Operations should be designed to execute successfully even if only subdomain input data is provided. Operations should not assume access to the entire input domain. Some data structures within the shared memory method were originally initialized based on this assumption. Their initialization has been modified to only account for what is needed for a particular subdomain.
    \item All operations should check if points are locked, adhering to the mechanisms used by the speculative execution model, thus allowing the precursory freezing of specific data without needing to further modify each operation to process new input types (interface vs. interior data). For example, a routine within CDT3D called "boundary sliver removal" \cite{Drakopoulos19F} iterates over all tetrahedra of the input domain and attempts to remove slivers from the external boundary of the domain. Originally, this method assumed that if a tetrahedron contained a face with no neighbor tetrahedron attached, then that face is a boundary face. When applied to a subdomain, this face could be an interface face (of which the shared memory method is not aware). Despite being sequential, this routine has been modified to simply check if any of the face's points are locked. If so, it moves on to the next tetrahedron.
    \item The shared memory method should be designed to assign work to threads only if the respective data are expected to become accessible for processing. This prevents preemptively locked points from being assigned to threads for certain meshing operations within CDT3D, thereby ensuring that threads won't waste time/resources attempting to process data that will never become accessible (unlocked) during a particular phase of adaptation.
    \item Point creation/insertion on elements adjacent to pseudo-inactive elements inhibits grid generation convergence; therefore, a buffer zone must be established around pseudo-inactive elements to reach convergence.
    \item The order in which operations are designed to be executed in the shared memory method should also be reflected in their overall order of execution in the distributed method. The impact of the order of operations is observed in our evaluation, in combination with the pseudo-active modification and preemptive locking technique, which applies operations only to specific elements. This reduces the time spent processing elements that have already undergone adaptation.
\end{enumerate}

\section{Results} \label{results}
The results of our experimental evaluation is organized into the following sections: details and definitions of the measures used to evaluate the grid adaptation method (section \ref{quality_measure_definitions}), a brief overview of the state-of-the-art method \textit{refine} to which the presented method is compared (section \ref{refine_overview}), our experimental setup (section \ref{experimental_setup}), results pertaining to the adaptation of a delta wing geometry (generating approximately 100-200 million elements) (section \ref{delta_wing_geometry}), and results for the adaptation of a cube geometry (generating approximately 1 billion elements) (section \ref{cube_geometry}).
\subsection{Grid Adaptation Method} \label{quality_measure_definitions}
The same grid adaptation method utilized in \cite{TsolakisEvaluation}, for the shared memory CDT3D's evaluation and comparison to other parallel meshing strategies, is utilized here to evaluate the presented distributed memory CDT3D method. The goal is to adapt a given grid so that it conforms to an anisotropic metric field \emph{M}. A comprehensive introduction to the definition and properties of the metric tensor field is provided in \cite{LoseilleAlauzetCMFI}. The complexity \emph{C} of a continuous metric field \begin{math}\begin{mathcal}M\end{mathcal}\end{math} is defined as:

\begin{equation}
C(\begin{mathcal}M\end{mathcal}) = \int_\Omega \sqrt{det(\begin{mathcal}M\end{mathcal}(x))}dx.
\end{equation}

Complexity on the discrete grid is computed by sampling \(\begin{mathcal}M\end{mathcal}\) at each vertex \emph{i} as the discrete metric field \emph{M},

\begin{equation}
C(M) = \sum_{i=1}^{N} \sqrt{det(M_i)}V_i,
\end{equation}

\noindent where \(V_i\) is the volume of the Voronoi dual surrounding each node. The complexity of a grid is known to have a linear dependency with respect to the number of points and tetrahedra, shown theoretically in \cite{LoseilleAlauzetCMFI} and experimentally verified in \cite{LoseilleAlauzetCMFII, Park2015ComparingAO}. The number of vertices are approximately 2\emph{C} while the number of tetrahedra are approximately 12\emph{C}. As shown in \cite{LoseilleAlauzetCMFI, TsolakisEvaluation}, scaling the complexity of a metric can generate the same relative distribution of element density and shape over a uniformly refined grid compared to the original complexity. The metric tensor \(M_{C_r}\) that corresponds to the target complexity \(C_r\) is evaluated by \cite{LoseilleAlauzetCMFI}:

\begin{equation}
    M_{C_r} = \left(\frac{C_r}{C(M)}\right)^\frac{2}{3} M,
\end{equation}

\noindent where \emph{M} is the metric tensor before scaling and C(M) is the complexity of the discrete metric before scaling.

Quantitative results are examined with respect to overhead incurred by the a priori interface adaptation approach of the presented method, as opposed to alternative methods which use global synchronization and re-partitioning techniques to process interface elements during mesh generation. Because there is no sequential method included in our evaluation, we consider scalability for each method to be the ratio of the runtime when it is executed using 1 core (1 processing element) to its runtime when executed with more processing elements. Qualitative results are examined with respect to metric conformity of the adapted mesh. These qualitative measures described below are the same as those used by the Unstructured Grid Adaptation Working Group (UGAWG\footnote{\url{https://ugawg.github.io/}}). The adapted meshes of the presented method are compared to those of the shared memory CDT3D method (and state-of-the-art method \textit{refine} \cite{ParkRefine}) in order to verify the stability of the distributed memory CDT3D method. 

The aim of metric conformity is the creation of a unit grid, where edges are unit-length and elements are unit-volume with respect to the target metric. For calculating edge length, we adopted the same definition that appears in \cite{ALAUZETSizeGradation}. For two vertices \emph{a} and \emph{b}, an edge length in the metric \(L_e\) can be evaluated using:

\begin{equation}
\begin{split}
& L_e = 
\begin{cases}
    \frac{L_a - L_b}{log(L_a/L_b)} & |L_a - L_b| > 0.001 \\
    \frac{L_a+L_b}{2} & otherwise
\end{cases} \\
& L_a = (v_e^TM_av_e)^{\frac{1}{2}},L_b = (v_e^TM_bv_e)^{\frac{1}{2}}
\end{split}
\end{equation}

and an element mean ratio shape measure can be approximated in the discrete metric as:

\begin{equation}
    Q_k = \frac{36}{3^{1/3}} \frac{\left(|k|\sqrt{det(M_{mean})}\right)^\frac{2}{3}}{\sum_{e\epsilon L} v_e^TM_{mean}v_e},
\end{equation}

where \emph{v} is a vertex of element \emph{k} and \(M_{mean}\) is the interpolated metric tensor evaluated at the centroid of element \emph{k}. Since the goal is to create edges that are unit-length, edges with length above or below one are considered to be sub-optimal. The measure for mean ratio is bounded between zero and one since it is normalized by the volume of an equilateral element. One is the optimal quality for an element's mean ratio shape. 

\subsection{\textit{refine}} \label{refine_overview}
\textit{refine} is an open-source\footnote{Available at \url{https://github.com/NASA/refine}}, state-of-the-art parallel adaptive anisotropic mesh generation method \cite{ParkRefine} to which we compare the presented method. \textit{refine} uses a partially-coupled, coarse-grained approach that exploits parallelism at the subdomain level using domain decomposition. A combination of edge split, edge collapse, and point smoothing operations is utilized to generate a unit grid for a given metric field. Subdomain interior elements are first adapted in parallel while their interfaces are frozen, taking an a posteriori approach. The method then focuses on adapting interface elements where communication occurs between neighboring subdomains to provide updates of the changes to these shared elements. A combined load balancing and migration technique is utilized during re-partitioning at the end of each adaptation pass to equalize the number of points within subdomains. Moreover, all-to-all communication occurs at the end of each adaptation pass to ensure that newly inserted grid points each have unique global identifiers (to denote duplicate points between subdomains). In our evaluation, the default parameters for \textit{refine} are utilized with the exception of the domain decomposition method. We use ParMETIS, as this resulted in the best runtimes for \textit{refine} among its domain decomposition options.

\subsection{Experimental Setup} \label{experimental_setup}
PREMA, the original shared memory CDT3D, \textit{refine}, and the distributed memory CDT3D codes were all compiled using the GNU GCC 11.4.1 and Intel MPI compilers. Data were collected on two supercomputers - Old Dominion University's Wahab cluster (utilizing up to 512 cores in our evaluation) and Purdue University's Anvil cluster \cite{AnvilCommunity2014, Anvil2022} (utilizing up to 1,536 cores in our evaluation). Wahab contains dual socket nodes that each feature two Intel Xeon Gold 6148 CPUs @ 2.40 GHz (20 slots) and 384GB of memory. Anvil contains dual socket nodes that each feature two AMD EPYC 7763 CPUs @ 2.45 GHz (64 slots each) and 256 GB of memory. With regards to data reported for the execution of DM\_CDT3D on Wahab, when using configurations of cores between 1 and 32, a single multicore node is allocated and the corresponding number of cores are allocated on that node. When utilizing more than 32 cores on Wahab, nodes are allocated with 32 cores each. For example, a configuration of 256 cores means that 8 nodes are allocated with 32 cores each. With regards to execution on Anvil, up to 96 CPU cores are allocated per node. Thus, a configuration of 768 cores on Anvil means that 8 nodes are allocated with 96 cores each. Within the context of the distributed method, the shared memory method is utilized on each node, leveraging the allocated cores on that node for either interface or a subdomain's interior adaptation. With each configuration (except that of 1 core), 1 core is also utilized for PREMA's communication (to allow for asynchronous communication and computation).

\subsection{Delta Wing Geometry} \label{delta_wing_geometry}
\begin{figure}[htbp]
\begin{center}
\includegraphics[width=0.65\textwidth]{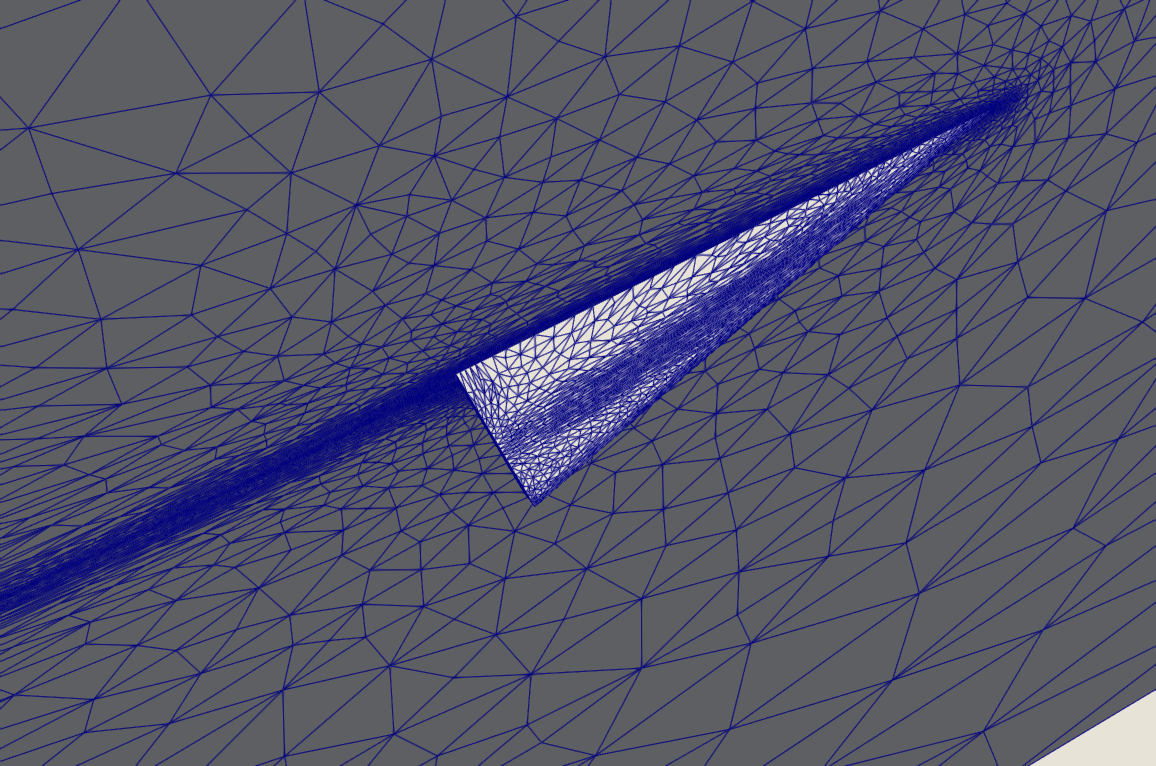}
\end{center}
\caption{Shown is a delta wing geometry with a multiscale metric derived from a laminar flow and a complexity of 50,000.}
\label{fig:delta_wing}
\end{figure}

Fig. \ref{fig:delta_wing} shows a delta wing geometry made of planar facets. Its multiscale metric \cite{ALAUZET2010561} is constructed based on the Mach field of a subsonic laminar flow. The input grid is adapted from a complexity of 50,000 and is scaled to two complexities in the below test cases for both a qualitative and quantitative evaluation. 
This case was also utilized in the aforementioned shared memory CDT3D evaluation study \cite{TsolakisEvaluation}. Details of the verification of the delta wing/grid adaptation process is provided in \cite{VerificationUGAComponents}.

\subsubsection{How The Shared Memory Method Modifications Affect The Distributed Method's Performance} \label{how_modifications_affect_cdt3d}

Table \ref{table_redundancy} shows an example of how the pseudo-active modification, preemptive locking, and respecting the "order of operations" affect CDT3D within the a priori distributed memory method. We first show results using a two-subdomain decomposition (across the x-axis) of the delta wing geometry. While we show end-to-end performance statistics with other decomposition configurations further below, we initially focus on the interface adaptation and interior adaptation statistics for a simple decomposition to better understand why these modifications are necessary. In these first experiments, a coarse mesh has already been generated which contains about 12 million tetrahedra. The target metric complexity is 10 million (approximately 100 million elements will be generated). Two nodes were allocated with 32 cores each on Wahab. Interface adaptation occurs on one node and each subdomain's interior adaptation occurs on separate nodes. While the usage and parameters for the meshing operations within Fig. \ref{fig:cdt3d-anisotropic-pipeline} are optional, we adhere to their usage (and respective parameters) from \cite{TsolakisEvaluation}. This is because the original shared memory CDT3D method generated meshes of good quality (including for this delta wing geometry) compared to several state-of-the-art methods. There are two experiments presented in table \ref{table_redundancy}. Each experiment uses the pseudo-active modification and preemptive locking for interface adaptation. However, one only utilizes the previously specified subset of operations from the mesh adaptation module in Fig. \ref{fig:cdt3d-anisotropic-pipeline} for interface adaptation - the pre-refinement edge collapse, point creation, and local reconnection operations. This experiment is referred to as the "Modified" experiment. The other experiment ignores the importance of order of operations (referred to as the "Naive" experiment) and utilizes the full suite of operations from both the mesh adaptation and quality improvement modules for interface adaptation (except the optional 
smoothing operation of the mesh adaptation module, as this additional operation is not needed to satisfy spacing requirements in the metric space for this geometry case). With regards to interior adaptation, the "Modified" experiment uses the pseudo-active modification and the preemptive locking mechanism while the "Naive" experiment does not use these modifications (i.e., it uses CDT3D as a black box). 

\begin{table}[htbp]
\caption{Quantitative statistics that show the affect of the pseudo-active modification and preemptive locking on CDT3D within DM\_CDT3D are shown. The Modified column is the experiment that includes these modifications for each subdomain's interior adaptation while respecting the order of operations throughout execution. The Naive column is the experiment that does not respect the order of operations and does not include the pseudo-active modification or preemptive locking for interior adaptation (default settings as a black box during both adaptation phases). `M' is million, `K' is thousand, and `Tets' means tetrahedra.}
\centering
\begin{tabular}{cccc}
\hline\hline
Adaptation Phase & Statistics & Naive & \textbf{Modified (Reduced By)} \\\hline
\multirow{3}{*}{Interface} & Adaptation Time (sec) & 321.09 & \textbf{75.4 (4.25x)} \\
& Start \# Tets & 12.1M & 12.1M \\
& End \# Tets & 18.1M & 18.1M \\\hline
\multirow{3}{*}{Subdomain 1 Interior} & Adaptation Time (sec) & 1949.23 & \textbf{1220.99 (1.59x)} \\
& Start \# Tets & 9.04M & 9.06M \\
& End \# Tets & 48.7M & 48.8M \\\hline
\multirow{3}{*}{Subdomain 2 Interior} & Adaptation Time (sec) & 1937.07 & \textbf{1185.03 (1.63x)} \\
& Start \# Tets & 9.1M & 9.11M \\
& End \# Tets & 48.9M & 48.9M \\
\hline\hline
\end{tabular}
\label{table_redundancy}
\end{table}

While the interface adaptation time in the Naive experiment is expectedly longer due to the additional operations utilized, the interior adaptation times in the Modified experiment are shorter, despite the fact that the same operations are utilized in both experiments for interior adaptation. Despite the extra runtime in the Naive experiment, CDT3D generates approximately the same number of elements as in the Modified experiment. An expanded version of this table (table \ref{table_redundancy_full}) can be seen in the appendix. 
Table \ref{table_redundancy_cdt3d_breakdown} shows a further breakdown of time spent by each CDT3D operation in each experiment. `Deactivation' refers to the time spent deactivating pseudo-inactive elements and checking the quality of active elements (to potentially deactivate if found to be satisfactory). `Partitioning Tets' refers to the partitioning of tetrahedra into the aforementioned buckets to be assigned to threads. Some minor routines are not included in the table given that they occupy a small fraction of runtime (activating all elements before the quality improvement module, data structure cleanup, etc.). There are several key takeaways from this data. When the Modified version of the method respects the order of operations by skipping the post-refinement edge collapse for interface adaptation, the time spent in deactivation, creating vertices, and local reconnection are significantly reduced in each subdomain's interior adaptation. When the Naive approach executes the full suite of operations for the interface adaptation, it indeed forces CDT3D to repeat work during interior adaptation because mesh adaptation operations are executed over elements that have already been processed by the post-refinement edge collapse operation. This post-refinement operation is designed to only be executed after those in the mesh adaptation module are complete. This repetition of work in the Naive approach causes the method to spend more time in every operation. The bottleneck operation is local reconnection (both the Mesh Adaptation, MA, and Quality Improvement, QI, operations summed), which is approximately doubled in the Naive approach. Meanwhile, the `Insert Vertices' operation has approximately the same runtime in each experiment. This underscores the observation made in table \ref{table_redundancy} that the Naive (black box) approach takes almost double the amount of time to essentially generate the same size mesh.

\begin{table}[htbp]
\caption{Breakdown of time spent (seconds) in CDT3D operations for each adaptation phase of the subdomains in Table \ref{table_redundancy} - The Modified experiment includes the order of operations, pseudo-active modification, and preemptive locking while the naive experiment does not. The bottleneck operations are the mesh adaptation and quality improvement local reconnection operations (in bold font), which refer to the respective modules in Fig. \ref{fig:cdt3d-anisotropic-pipeline} in which the local reconnection operation is utilized.}
\centering
\begin{tabular}{cccccc}
\hline\hline
Experiment & Operations & Interface & Subdomain 1 Interior & Subdomain 2 Interior & Maximum \\\hline
\multirow{11}{*}{Naive} & Pre- Edge Collapse & 0.3 & 3.48 & 3.37 & 3.78 \\
& Deactivation & 13.33 & 88.77 & 82.8 & 102.1 \\
& Partitioning Tets & 7.02 & 18.6 & 19.8 & 26.82 \\
& Create Vertices & 17.2 & 190.73 & 180.01 & 207.93 \\
& Insert Vertices & 2.15 & 15.34 & 14.25 & 17.49 \\
& \textbf{Mesh Adaptation} & \multirow{2}{*}{\textbf{64.84}} & \multirow{2}{*}{\textbf{667.16}} & \multirow{2}{*}{\textbf{700.96}} & \multirow{2}{*}{\textbf{765.8}} \\
& \textbf{Local Reconnection} & & & \\
& Post- Edge Collapse & 2.87 & 25.74 & 25.17 & 28.61 \\
& \textbf{Quality Improvement} & \multirow{2}{*}{\textbf{171.86}} & \multirow{2}{*}{\textbf{572.02}} & \multirow{2}{*}{\textbf{544.2}} & \multirow{2}{*}{\textbf{743.88}} \\
& \textbf{Local Reconnection} & & & \\
& Smoothing & 36.08 & 340.67 & 338.78 & 376.75 \\
\hline
\multirow{11}{*}{Modified} & Pre- Edge Collapse & 0.31 & 2.47 & 2.64 & 2.95 \\
& Deactivation & 2.82 & 41.02 & 39.89 & 43.84 \\
& Partitioning Tets & 2.23 & 19.89 & 17.23 & 22.12 \\
& Create Vertices & 11.43 & 119.79 & 97 & 131.22 \\
& Insert Vertices & 2.11 & 14.25 & 14.08 & 16.36 \\
& \textbf{Mesh Adaptation} & \multirow{2}{*}{\textbf{53.97}} & \multirow{2}{*}{\textbf{404.3}} & \multirow{2}{*}{\textbf{405.81}} & \multirow{2}{*}{\textbf{459.78}} \\
& \textbf{Local Reconnection} & & & \\
& Post- Edge Collapse & $-$ & 24.29 & 26.58 & 26.58 \\
& \textbf{Quality Improvement} & \multirow{2}{*}{$-$} & \multirow{2}{*}{\textbf{227.27}} & \multirow{2}{*}{\textbf{228.37}} & \multirow{2}{*}{\textbf{228.37}} \\
& \textbf{Local Reconnection} & & & \\
& Smoothing & $-$ & 341.35 & 339.15 & 341.35 \\
\hline\hline
\end{tabular}
\label{table_redundancy_cdt3d_breakdown}
\end{table}

\subsubsection{End-to-end Performance \& Qualitative Results Adapted at 10 Million Complexity}
\begin{figure}[htbp]
\begin{center}
\includegraphics[width=0.75\textwidth]{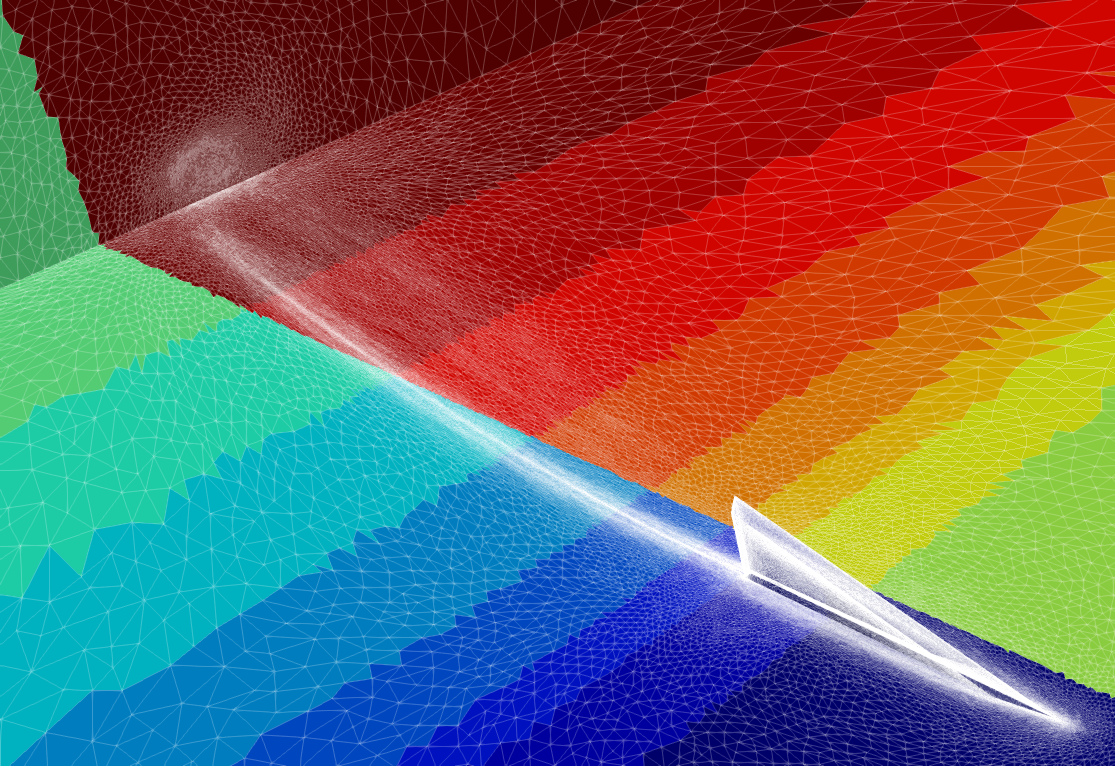}
\end{center}
\caption{Shown is a delta wing geometry decomposed into 16 subdomains using PQR.}
\label{fig:delta_wing_decomposition}
\end{figure}

There are several factors that affect the distributed memory CDT3D methods' output grid's metric conformity and performance. Regarding both the a posteriori and a priori DM\_CDT3D approaches, these include the method of decomposition and the metric complexity of the coarse mesh generated before decomposition. A unique parameter that impacts the a posteriori method's results is the number of interface shift iterations. For the a priori DM\_CDT3D method, the number of extra layers activated/unlocked before adaptation is important. The optimal settings of these heuristics vary between different geometries and require careful consideration when meshing large geometries. After extensive testing, the settings for adapting the delta wing geometry at 10 million complexity with the a posteriori distributed memory CDT3D method was determined to be a PQR decomposition of 16 subdomains (8 split across the x-axis and 2 split across the z-axis), the coarse mesh generated to a complexity of 1,250,000, and 12 interface shift iterations. The parameters used for the a priori DM\_CDT3D method are: a PQR decomposition of 16 subdomains (split into 8 across the x-axis and 2 across the z-axis), the coarse mesh generated to a complexity of 1,250,000, and 5 extra layers unlocked for interface adaptation with 3 unlocked for interior adaptation. As in section \ref{how_modifications_affect_cdt3d}, the interface adaptation phase of the a priori DM\_CDT3D approach utilizes only the mesh adaptation module (of Fig. \ref{fig:cdt3d-anisotropic-pipeline}). The interior adaptation phase utilizes both. The decomposition for the DM\_CDT3D approaches is shown in Fig. \ref{fig:delta_wing_decomposition}, where each color represents an individual subdomain to which elements are assigned.

\begin{figure}[htbp]
\begin{center}
\includegraphics[width=0.9\textwidth]{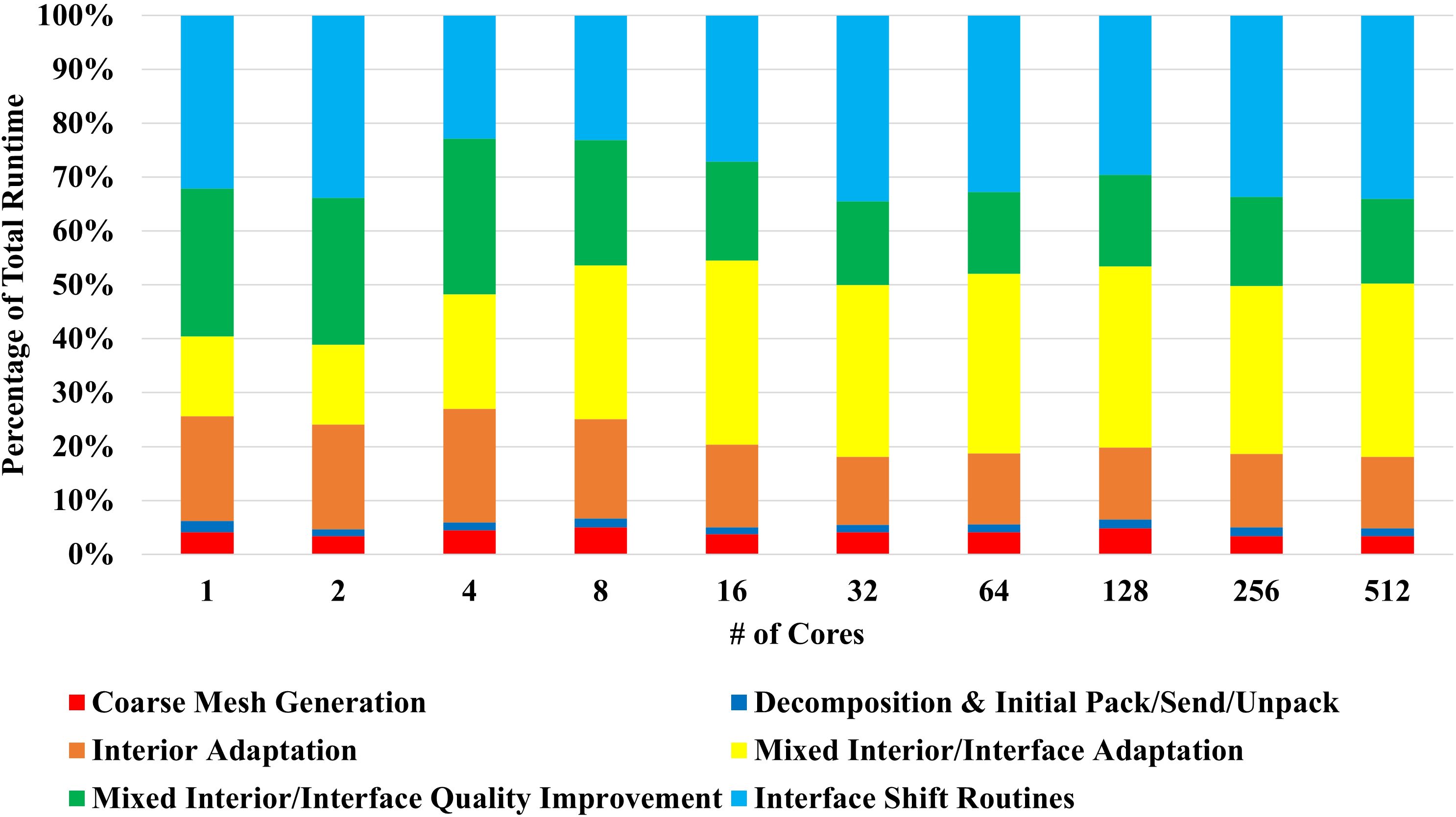}
\end{center}
\caption{A breakdown of end-to-end time for the A Posteriori DM\_CDT3D Method when adapting the delta wing geometry at 10 million complexity is shown.}
\label{fig:aposteriori_delta_10m_percentage_adaptation_breakdown}
\end{figure}

\begin{figure}[htbp]
\begin{center}
\includegraphics[width=0.85\textwidth]{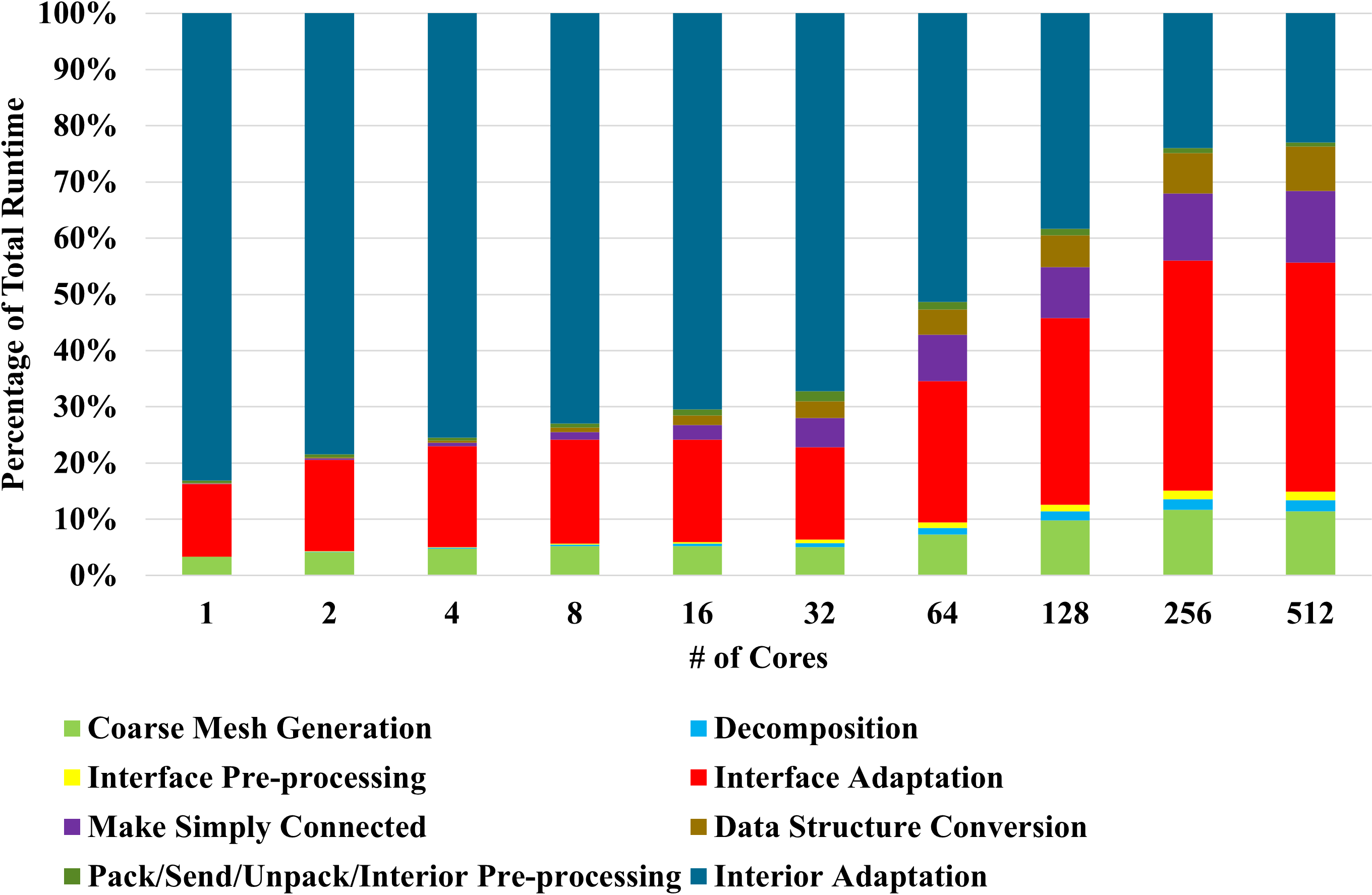}
\end{center}
\caption{A breakdown of end-to-end time for the A Priori DM\_CDT3D Method when adapting the delta wing geometry at 10 million complexity is shown.}
\label{fig:apriori_delta_10m_percentage_adaptation_breakdown}
\end{figure}

Table \ref{runtime_results_delta_10m} shows an end-to-end runtime comparison of the original shared memory CDT3D method \cite{EwCCDT3D}, the a posteriori distributed memory CDT3D approach \cite{Garner24EarlyDMCDT3D}, the presented a priori distributed memory CDT3D method, and \textit{refine} \cite{ParkRefine} when adapting the delta wing geometry at 10 million complexity on the Wahab supercomputer (the problem size remains fixed, i.e., strong scaling). Table \ref{mesh_sizes_delta_10m} shows the approximate size of the meshes generated by each method (consistent across all tested core configurations). Each CDT3D method generates meshes of about the same size (with a deviation of about a million tetrahedra) and are smaller than the meshes generated by \textit{refine} (by about 15 million tetrahedra). Each CDT3D method shows good scalability when utilizing 1-32 cores. The speedup for both DM\_CDT3D approaches improves by little between utilizing 256 and 512 cores. This is attributed to the communication overhead of the interface shift operations in the a posteriori approach (which contains some sequential components, see \cite{Garner24EarlyDMCDT3D} for a more detailed analysis of this method when adapting this same geometry). The a priori approach is limited by three aspects: (1) the sequential component of the algorithm (making subdomains simply connected after interface adaptation and converting the respective shared memory method data structures into distributed memory data structures for packing and distribution), (2) the fact that interface adaptation is limited to 32 cores since it occurs on one multicore node, and (3) load imbalance since PREMA's load balancing feature is turned off due to no overdecomposition (discussed in detail in section \ref{discussion}). Fig. \ref{fig:aposteriori_delta_10m_percentage_adaptation_breakdown} and Fig. \ref{fig:apriori_delta_10m_percentage_adaptation_breakdown} show performance breakdowns of each DM\_CDT3D approach. Overhead related to the shifting of interface data occupies about 30-35\% of the a posteriori DM\_CDT3D method's runtime. On the higher core configurations (greater than 32), all non-meshing operations of the a priori DM\_CDT3D method occupy approximately 15-25\% of end-to-end runtime. These include: decomposition, adaptation preprocessing, making the subdomains simply connected, converting the data structures, packing, sending, and unpacking data. This is an improvement over the 30-35\% of time occupied by non-meshing operations in the a posteriori DM\_CDT3D method. Additionally, this overhead becomes preferable as opposed to the overhead seen in methods where communication occupies more than 50\% of end-to-end time (as reported in the ECP study \cite{ECPMPIStudy}). \textit{refine} exhibits superlinear speedup (up to 32 cores) due to its optimizations that are designed for configurations of multiple cores (such as the re-ordering of points within subdomains for cache efficiency) \cite{TsolakisEvaluation, ParkRefine}. These optimizations cause significant overhead however when the method is executed with lower core configurations or executed sequentially. The a priori DM\_CDT3D approach exhibits better runtime performance than \textit{refine} when utilizing up to 256 cores while \textit{refine} shows slightly better performance when utilizing 512 cores.

\begin{table}[htb]
\caption{The runtime (approximately in \textbf{minutes}) of SM\_CDT3D, A Posteriori DM\_CDT3D, A Priori DM\_CDT3D, and \textit{refine} when adapting the delta wing geometry at 10 million complexity on the Wahab supercomputer is shown.}
\centering
\begin{tabular}{ccccccccccc}
\hline\hline
\multirow{2}{*}{Method} & \multicolumn{9}{c}{\# Cores} \\
 & $1$ & $2$ & $4$ & $8$ & $16$ & $32$ & $64$ & $128$ & $256$ & $512$\\\hline
SM\_CDT3D & $954$ & $522$ & $301$ & $158$ & $82$ & $42$ & $-$ & $-$ & $-$ & $-$ \\ 
A Posteriori DM\_CDT3D & $3378$ & $1387$ & $500$ & $269$ & $139$ & $103$ & $68$ & $54$ & $40$ & $38$\\
A Priori DM\_CDT3D & $1933$ & $771$ & $365$ & $176$ & $93$ & $55$ & $36$ & $28$ & $23$ & $22$\\
\textit{refine} & $12802$ & $5755$ & $2296$ & $970$ & $290$ & $136$ & $75$ & $48$ & $31$ & $21$\\
\hline\hline
\end{tabular}
\label{runtime_results_delta_10m}
\end{table}

\begin{table}[htb]
\caption{The approximate sizes of the meshes generated (consistent across all tested core configurations) by SM\_CDT3D, A Posteriori DM\_CDT3D, A Priori DM\_CDT3D, and \textit{refine} when adapting the delta wing geometry at 10 million complexity is shown. `M' means million.}
\centering
\begin{tabular}{ccc}
\hline\hline
Method & \# of Tetrahedra & \# of Vertices \\\hline
SM\_CDT3D & 95.65M & 16.08M \\ 
A Posteriori DM\_CDT3D & 96.6M & 16.2M \\
A Priori DM\_CDT3D & 97.66M & 17.2M \\
\textit{refine} & 112.1M & 19.1M \\
\hline\hline
\end{tabular}
\label{mesh_sizes_delta_10m}
\end{table}

\begin{figure}[htbp]
     \centering
     \begin{subfigure}[htb]{0.5\textwidth}
         \centering
         \includegraphics[width=\textwidth]{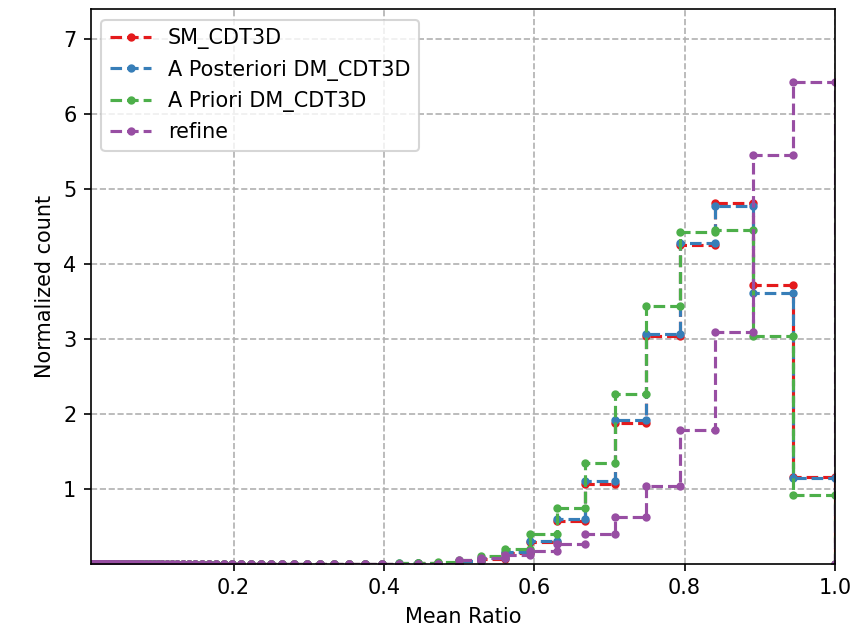}
     \end{subfigure}%
     \begin{subfigure}[htb]{0.5\textwidth}
         \centering
    \includegraphics[width=\textwidth]{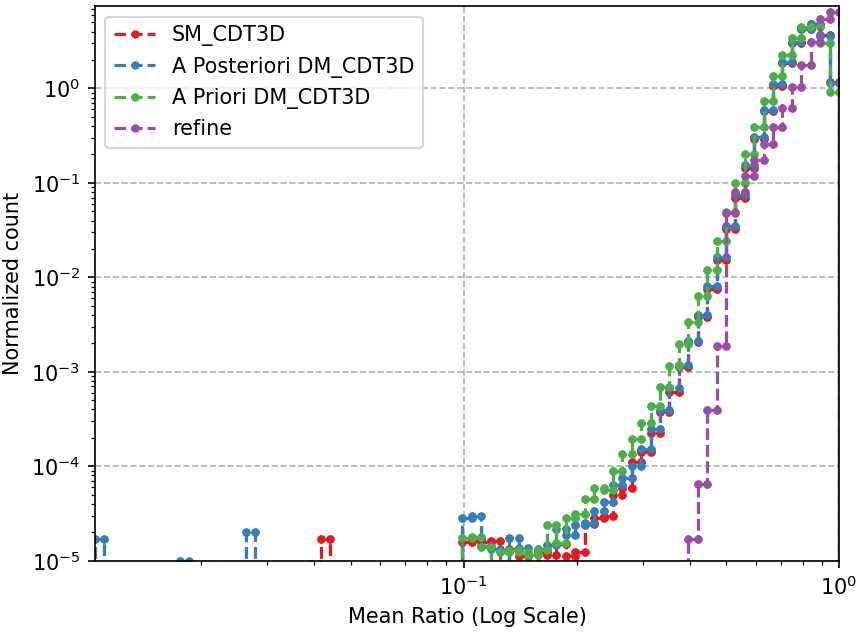}
     \end{subfigure}
        \caption{The mean ratios of elements generated by each method for the delta wing geometry at 10 million complexity are compared in linear and logarithmic scales.}
        \label{fig:delta_10m_MR}
\end{figure}

\begin{figure}[htbp]
     \centering
     \begin{subfigure}[htb]{0.5\textwidth}
         \centering
         \includegraphics[width=\textwidth]{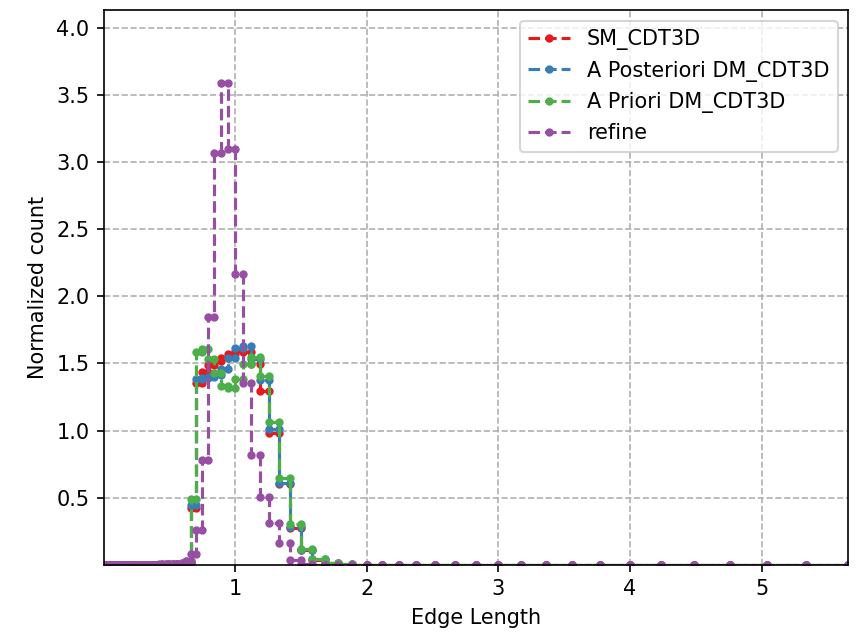}
     \end{subfigure}%
     \begin{subfigure}[htb]{0.5\textwidth}
         \centering
    \includegraphics[width=\textwidth]{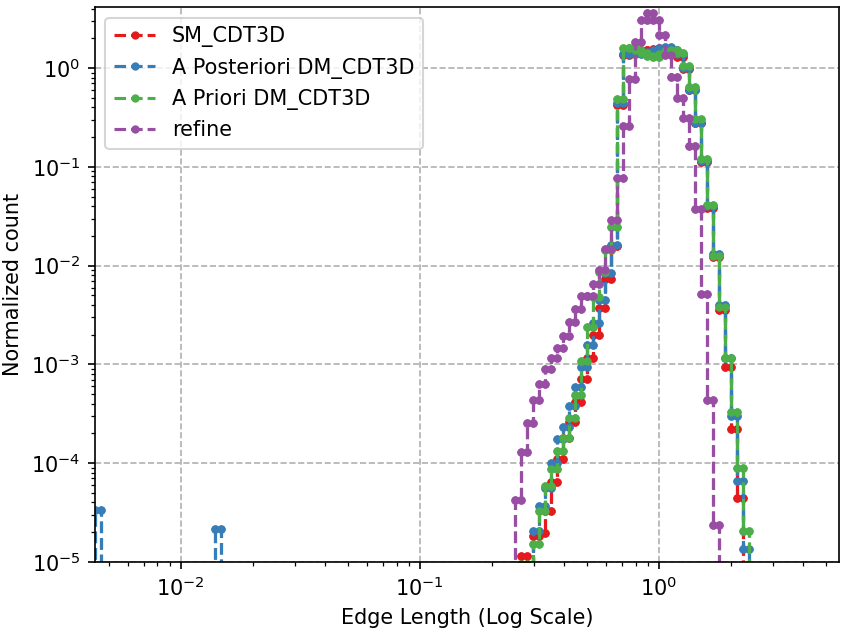}
     \end{subfigure}
        \caption{The edge lengths of elements generated by each method for the delta wing geometry at 10 million complexity are compared in linear and logarithmic scales.}
        \label{fig:delta_10m_EL}
\end{figure}

Metric conformity, characterized by element shape measure and edge length histograms, for the output meshes generated by each method are compared in Fig. \ref{fig:delta_10m_MR} and Fig. \ref{fig:delta_10m_EL}. It can be seen in both the linear and logarithmic scales that the meshes generated by both DM\_CDT3D methods exhibit good overall quality similar to that generated by the original shared memory CDT3D method. The a posteriori approach however generates meshes with a few elements of lower quality compared to the a priori approach. \textit{refine} exhibits the best lower bound on the mean ratio shape measure with a smaller deviation in edge length measure.

We now compare the performance of both the a priori DM\_CDT3D and \textit{refine} methods when adapting the delta wing geometry at 10 million complexity on the Anvil supercomputer. Table \ref{runtime_results_delta_10m_anvil} shows the end-to-end runtime comparison between the methods. Because we have observed that the runtime for these methods is expensive when executed utilizing small numbers of cores (and because access to the Anvil supercomputer was limited in terms of our awarded project allocation), we tested the methods with 48 CPU cores up to 1,536.

\begin{table}[htbp]
\caption[Performance for Delta Wing Adapted at 10 Million Complexity on Anvil]{The runtime (approximately in \textbf{minutes}) of the A Priori DM\_CDT3D and \textit{refine} methods when adapting the delta wing geometry at 10 million complexity on the Anvil supercomputer is shown.}
\centering
\begin{tabular}{ccccccc}
\hline\hline
\multirow{2}{*}{Method} & \multicolumn{6}{c}{\# Cores} \\
 & $48$ & $96$ & $192$ & $384$ & $768$ & $1536$\\\hline
A Priori DM\_CDT3D & $20$ & $15$ & $10$ & $7$ & $6$ & $6$\\
\textit{refine} & $47$ & $29$ & $24$ & $15$ & $-$ & $-$\\
\hline\hline
\end{tabular}
\label{runtime_results_delta_10m_anvil}
\end{table}

Overall, similar performance behavior is observed for both methods as their behavior when executed on Wahab. The a priori DM\_CDT3D method exhibits the optimal runtime, while speedup barely continues to improve when utilizing more than 384 cores for the method. This is again due to the same limitations as stated previously (with interface adaptation now limited to 96 cores). When utilizing larger shared memory nodes, DM\_CDT3D achieves better performance as opposed to utilizing smaller shared memory nodes such as those on Wahab. It should be noted that \textit{refine} crashes when executed with 768 cores and the method hangs during its first adaptation pass when executed with 1536 cores. The sizes of the meshes generated are approximately the same as when generated on Wahab. Additionally, the meshes generated by both methods on Anvil still have good quality (the same as in Figures \ref{fig:delta_10m_MR} and \ref{fig:delta_10m_EL}).

\subsubsection{End-to-end Performance \& Qualitative Results Adapted at 20 Million Complexity}
\begin{figure}[htbp]
\begin{center}
\includegraphics[width=0.95\textwidth]{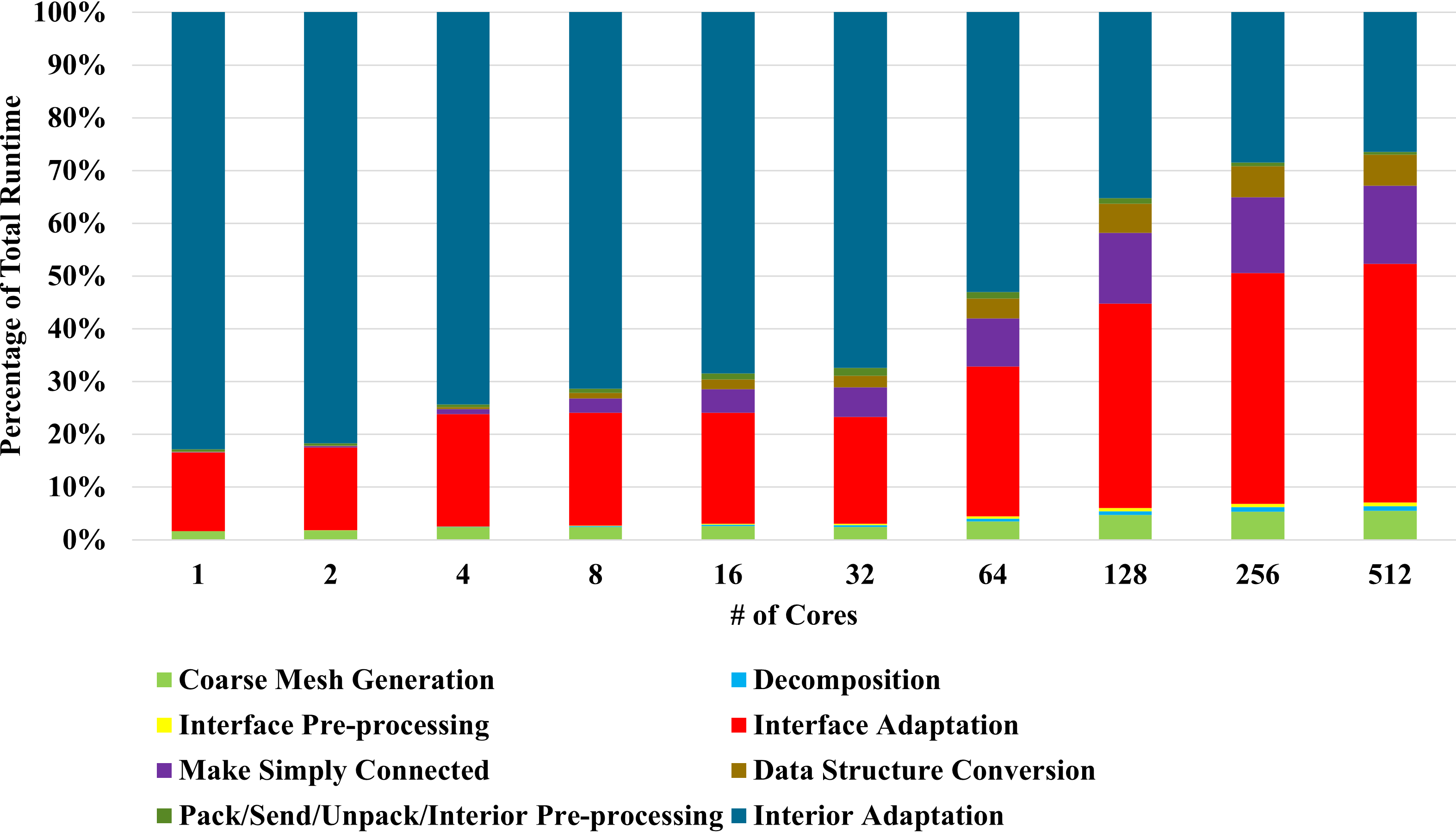}
\end{center}
\caption{A breakdown of end-to-end time for the A Priori DM\_CDT3D Method when adapting the delta wing geometry at 20 million complexity is shown.}
\label{fig:apriori_delta_20m_percentage_adaptation_breakdown}
\end{figure}

All of the remaining experiments were tested only on the Wahab supercomputer (again, due to limited access to the Anvil supercomputer). Since the original shared memory CDT3D method and \textit{refine} were also tested by adapting the delta wing geometry at 10 million complexity (up to 120 cores) in \cite{TsolakisEvaluation}, we now double the complexity while testing with more cores to further gauge each method's performance and compare them with the optimal DM\_CDT3D approach. Table \ref{runtime_results_delta_20m} shows the end-to-end runtime comparison of the methods when adapting the delta wing geometry at 20 million complexity (generating approximately 200 million elements, shown in table \ref{mesh_sizes_delta_20m}). The a posteriori DM\_CDT3D method is omitted since it is not the optimal approach between the two DM\_CDT3D approaches (as seen in the previous section and based on the data in \cite{Garner24EarlyDMCDT3D}). 
The same parameters were used for the methods with two exceptions for DM\_CDT3D. A PQR decomposition of 16 subdomains (4 across the x-axis, 2 across the y-axis, and 2 across the z-axis - 4x2x2) was used (as opposed to 8x1x2 when targeting 10 million complexity). Additionally, the interface adaptation phase utilizes the quality improvement module to ensure good final mesh quality (now needed when adapting the delta wing geometry at 20 million complexity). 
The CDT3D methods again produce smaller meshes than that of \textit{refine}. Both the shared memory and distributed memory CDT3D approaches show good scalability up to 32 cores again. It should be noted that \textit{refine} failed to adapt the mesh at 20 million complexity (it crashed) when executed sequentially or when utilizing 2 cores. It exhibits good scalability however when using 16 - 128 cores. DM\_CDT3D achieves the optimal runtime performance up to 256 cores again while \textit{refine} achieves better performance when using 512 cores on the Wahab supercomputer. Fig. \ref{fig:apriori_delta_20m_percentage_adaptation_breakdown} shows a breakdown of the end-to-end runtime for the phases of the a priori DM\_CDT3D algorithm when adapting the delta wing geometry at 20 million complexity. As expected, the percentage of time occupied by interface adaptation and interior adaptation remains about the same among configurations with 1-32 cores. This is because the same number of cores are utilized during each phase. When executing the method with more than 32 cores, the percentage of time occupied by interior adaptation decreases while interface adaptation increases. This is because the interior adaptation phase makes use of all the available cores while the interface adaptation phase is limited to the cores on one node. On the higher core configurations (greater than 32), all non-meshing operations again occupy approximately 15-25\% of end-to-end runtime. By the end of the interface adaptation phase, approximately 91 million tetrahedra are generated in each run (starting from a 12 million tetrahedra-sized coarse mesh) and the remaining 105 million are generated during interior adaptation. Although the interface adaptation phase generates about 40\% of the final mesh, the sum interior adaptation time occupies more than 80\% of end-to-end time (seen in the sequential run in Fig. \ref{fig:apriori_delta_20m_percentage_adaptation_breakdown}). This is because each subdomain's adaptation utilizes the full suite of operations from both the mesh adaptation and quality improvement modules of Fig. \ref{fig:cdt3d-anisotropic-pipeline} (unlocking/activating all elements, except the interfaces, before the post-refinement edge collapse operation and using the same parameter settings as the original shared memory CDT3D method to achieve metric conformity), as opposed to the interface adaptation phase which only utilizes a subset of operations (with less iterations of the aforementioned quality improvement operations). Some of these parameters (and their descriptions) utilized by the DM\_CDT3D method can be seen in table \ref{parameters} of the appendix.

\begin{table}[htbp]
\caption{The runtime (approximately in \textbf{minutes}) of SM\_CDT3D, A Priori DM\_CDT3D, and \textit{refine} when adapting the delta wing geometry at 20 million complexity on the Wahab supercomputer is shown.}
\centering
\begin{tabular}{ccccccccccc}
\hline\hline
\multirow{2}{*}{Method} & \multicolumn{9}{c}{\# Cores} \\
 & $1$ & $2$ & $4$ & $8$ & $16$ & $32$ & $64$ & $128$ & $256$ & $512$\\\hline
SM\_CDT3D & $2169$ & $1120$ & $611$ & $322$ & $164$ & $90$ & $-$ & $-$ & $-$ & $-$ \\
A Priori DM\_CDT3D & $4208$ & $2016$ & $803$ & $392$ & $188$ & $116$ & $77$ & $56$ & $50$ & $50$\\
\textit{refine} & $-$ & $-$ & $4879$ & $3853$ & $661$ & $320$ & $164$ & $80$ & $63$ & $46$\\
\hline\hline
\end{tabular}
\label{runtime_results_delta_20m}
\end{table}

\begin{table}[htbp]
\caption{The approximate sizes of the meshes generated (consistent across all tested core configurations) by SM\_CDT3D, A Posteriori DM\_CDT3D, A Priori DM\_CDT3D, and \textit{refine} when adapting the delta wing geometry at 20 million complexity is shown. `M' means million.}
\centering
\begin{tabular}{ccc}
\hline\hline
Method & \# of Tetrahedra & \# of Vertices \\\hline
SM\_CDT3D & 192.9M & 32.38M \\ 
A Priori DM\_CDT3D & 196.8M & 34.34M \\
\textit{refine} & 214.4M & 36.39M \\
\hline\hline
\end{tabular}
\label{mesh_sizes_delta_20m}
\end{table}

\begin{figure}[htbp]
     \centering
     \begin{subfigure}[htb]{0.5\textwidth}
         \centering
         \includegraphics[width=\textwidth]{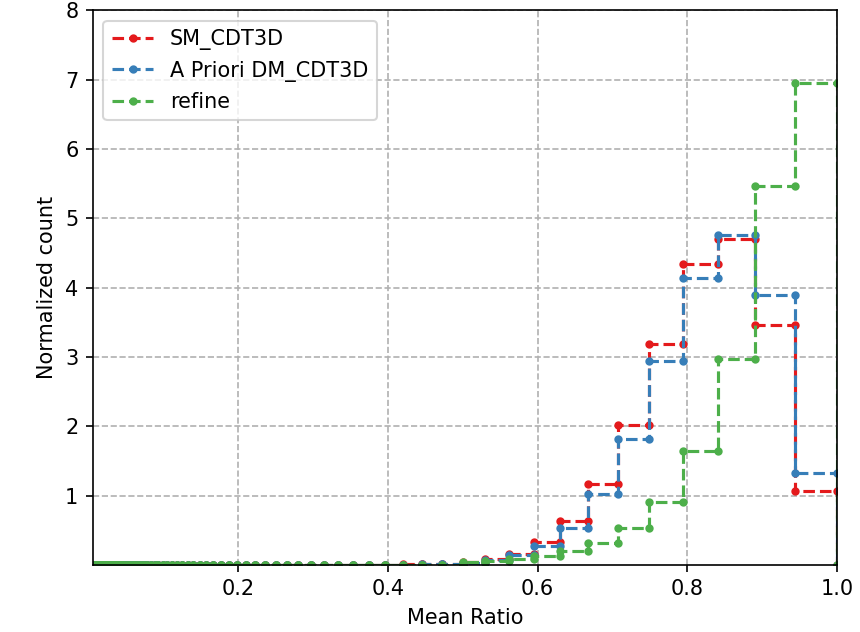}
     \end{subfigure}%
     \begin{subfigure}[htb]{0.5\textwidth}
         \centering
    \includegraphics[width=\textwidth]{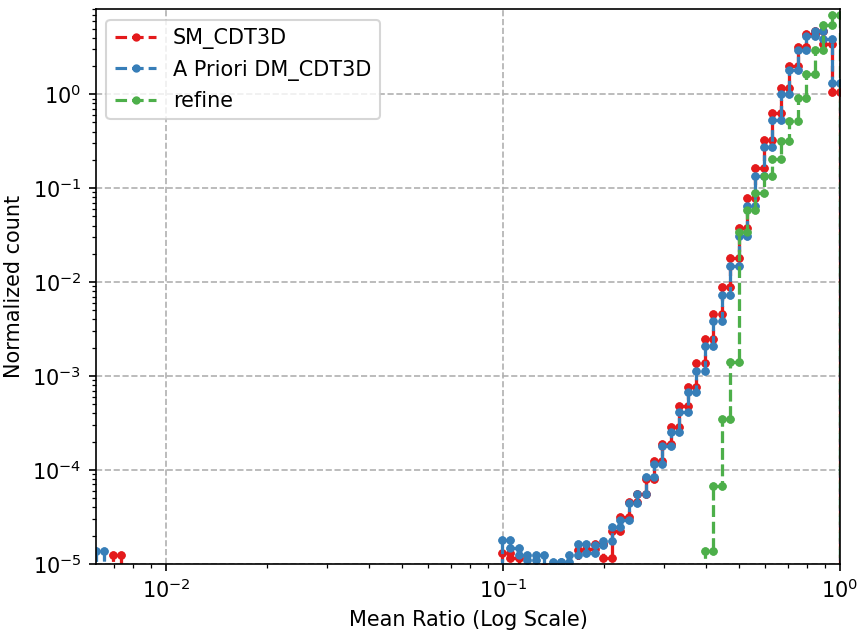}
     \end{subfigure}
        \caption{The mean ratios of elements generated by each method for the delta wing geometry at 20 million complexity are compared in linear and logarithmic scales.}
        \label{fig:delta_20m_MR}
\end{figure}

\begin{figure}[htbp]
     \centering
     \begin{subfigure}[htb]{0.5\textwidth}
         \centering
         \includegraphics[width=\textwidth]{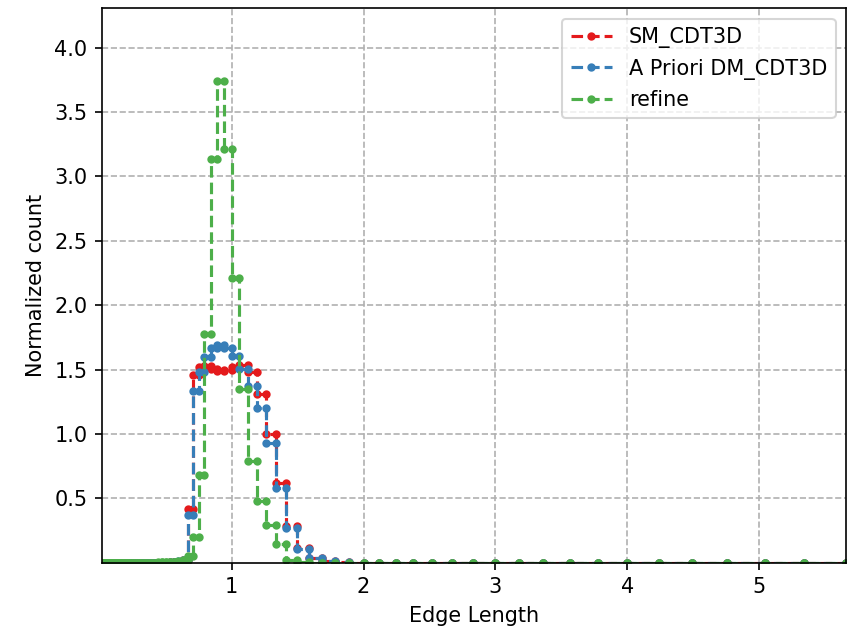}
     \end{subfigure}%
     \begin{subfigure}[htb]{0.5\textwidth}
         \centering
    \includegraphics[width=\textwidth]{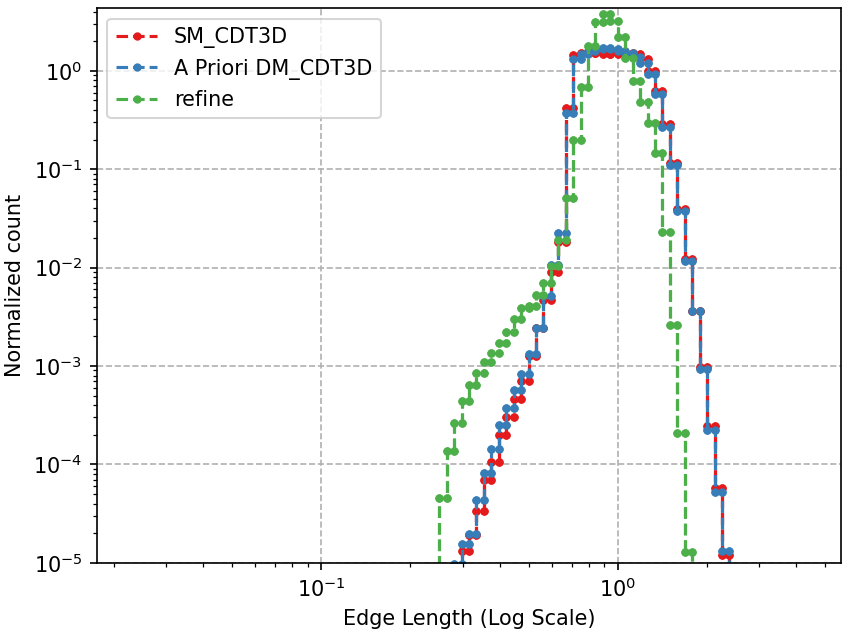}
     \end{subfigure}
        \caption{The edge lengths of elements generated by each method for the delta wing geometry at 20 million complexity are compared in linear and logarithmic scales.}
        \label{fig:delta_20m_EL}
\end{figure}

\begin{figure}[htbp]
     \centering
     \begin{subfigure}[htb]{0.5\textwidth}
         \centering
         \includegraphics[width=\textwidth]{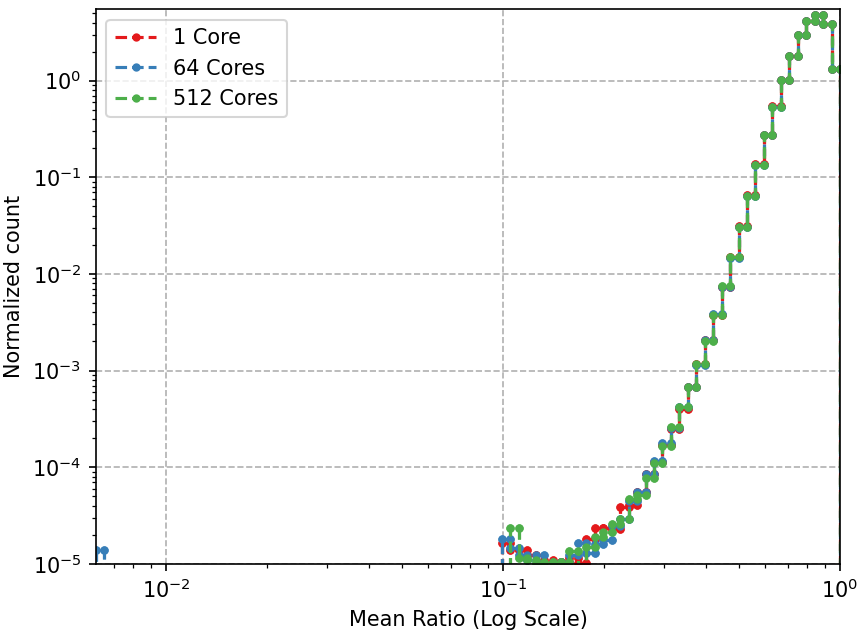}
     \end{subfigure}%
     \begin{subfigure}[htb]{0.5\textwidth}
         \centering
         \includegraphics[width=\textwidth]{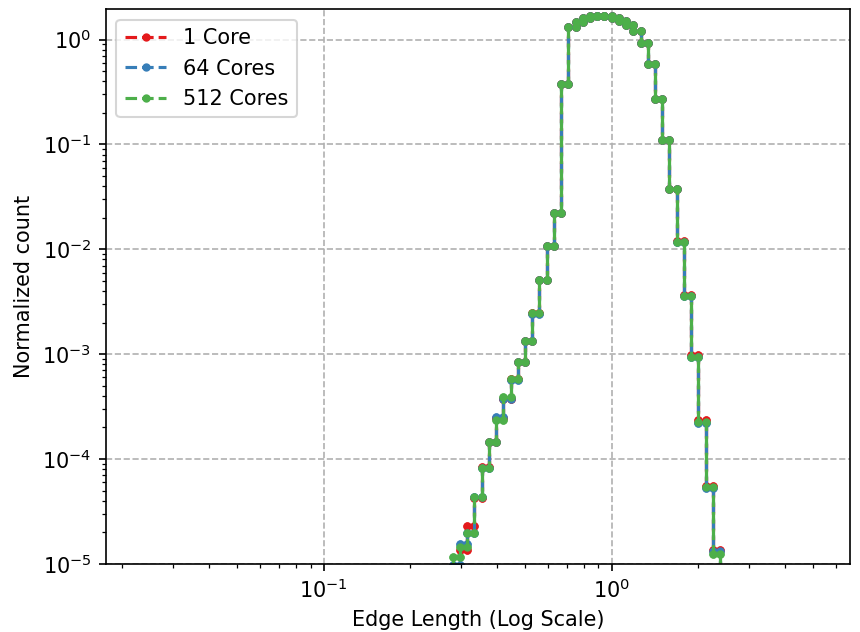}
     \end{subfigure}
        \caption{Shown is stability data of the A Priori DM\_CDT3D method for the delta wing geometry adapted at 20 million complexity.}
        \label{fig:delta_20m_Stability}
\end{figure}

The element shape measure and edge length histograms for the output meshes generated at 20 million complexity by each method are compared in Fig. \ref{fig:delta_20m_MR} and Fig. \ref{fig:delta_20m_EL}. The a priori DM\_CDT3D approach again generates a mesh of similar quality to that generated by SM\_CDT3D. The mesh generated by \textit{refine} exhibits the best overall quality. Fig. \ref{fig:delta_20m_Stability} provides a look at the stability and reproducibility (defined in section \ref{introduction}) of the a priori DM\_CDT3D method. When adapting the delta wing geometry at 20 million complexity, the method generates meshes of similar quality, achieving metric conformity regardless of the number of cores utilized. The meshes generated are not the same; therefore, the method meets the weak reproducibility requirement (since they are still of similar quality). 

\subsection{Cube Geometry} \label{cube_geometry}
\begin{figure}[!htb]
\begin{center}
\includegraphics[width=0.65\textwidth]{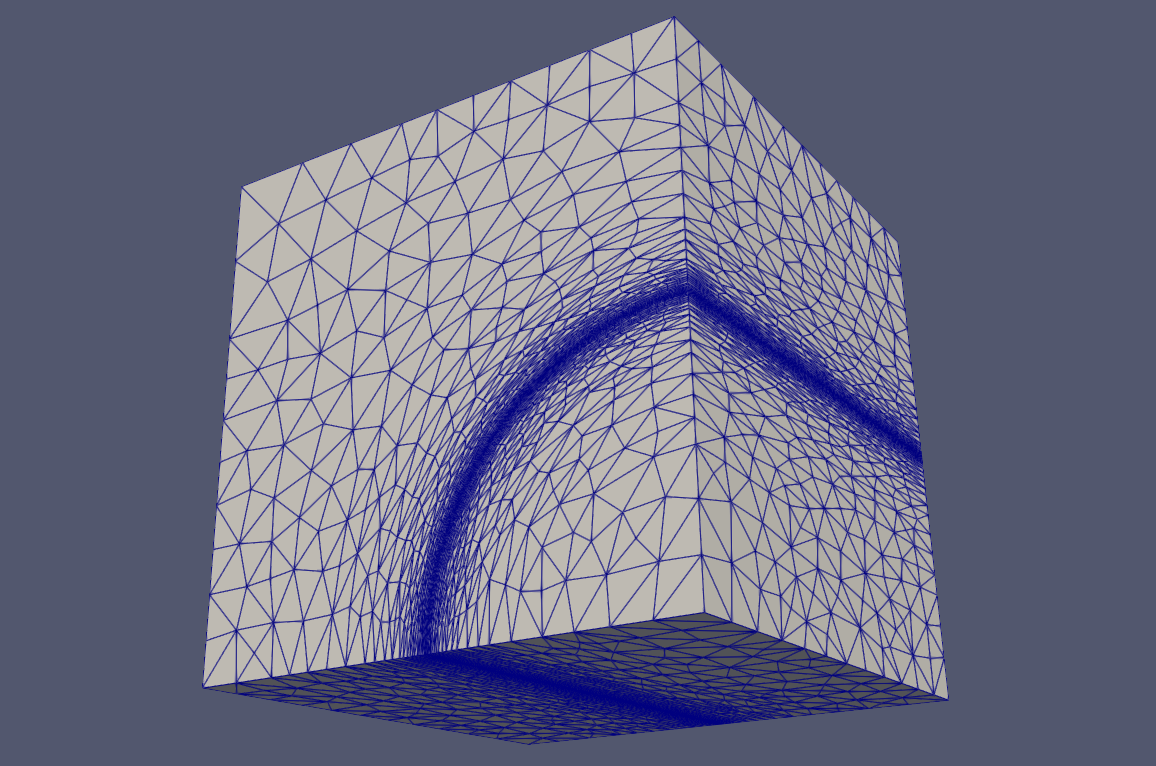}
\end{center}
\caption{Shown is a cube geometry that conforms to the polar-2 analytic metric field with a complexity of 8,000.}
\label{fig:cube}
\end{figure}

We test the methods using a case that was also utilized in the aforementioned shared memory CDT3D evaluation study \cite{TsolakisEvaluation} (where the metric field was scaled to a complexity of 500,000). We now scale the complexity to 100 million (approximately 1 billion elements will be generated) in order to gauge each method's performance when generating such a large mesh. Fig. \ref{fig:cube} shows the cube geometry where an analytically defined metric field \emph{M} (referred to as polar-2 in the first UGAWG benchmark \cite{UGAWG}) is used. This metric field represents a curved shear layer and is applied to the unit cube. The metric is specifically defined as:

\begin{equation}
M = \begin{bmatrix} 
    cos(t) & -sin(t) & 0 \\
    sin(t) & cos(t) & 0 \\
    0 & 0 & 1
\end{bmatrix}
\begin{bmatrix}
    {h_r^{-2}} & 0 & 0 \\
    0 & {h_t^{-2}} & 0 \\
    0 & 0 & {h_z^{-2}} \\
\end{bmatrix}
\begin{bmatrix} 
    cos(t) & sin(t) & 0 \\
    -sin(t) & cos(t) & 0 \\
    0 & 0 & 1
\end{bmatrix}
\end{equation}

where $r = \sqrt{x^2 + y^2}$, $t = atan2(y,x)$, $h_z = 0.1$, $h_0 = 0.001$, and $h_r = h_0 + 2(0.1 - h_0)|r - 0.5|$. The subscript $t$ is in the $\theta$ direction and subscript $r$ is the radial direction. The spacing in the tangential direction is defined by

\begin{equation}
    d = 10(0.6 - r) \quad \text{and} \quad 
    h_t = \begin{cases}
    0.1 & d < 0 \\
    d/40+0.1(1-d) & d \geq 0
    \end{cases}
\end{equation}

This polar distribution can be satisfied with high-quality elements by resolving curvature in the tangential direction near the layer. The initial grid conforms to the polar-2 metric with a complexity of 8,000.

Table \ref{runtime_results_cube_100m} shows the end-to-end runtime comparison of the methods when adapting the cube geometry at 100 million complexity. Table \ref{mesh_sizes_cube_100m} shows the approximate mesh sizes generated by each method. The parameters used in \cite{TsolakisEvaluation} for CDT3D regarding the cube case is utilized here. The same parameters that were used for DM\_CDT3D with the delta wing case adapted at 20 million complexity is used here with 3 exceptions. A PQR decomposition of 45 subdomains (3 across the x-axis, 3 across the y-axis, and 5 across the z-axis - 3x3x5) is used. The coarse mesh is adapted to a complexity at 1/15 of the target 100 million complexity (about 6,666,667). 3 layers of elements were marked as pseudo-active and their points were unlocked during interface adaptation preprocessing (as opposed to 5 for the delta wing cases). These parameters can be seen in Table \ref{parameters}. Both DM\_CDT3D and \textit{refine} fail to generate the mesh when using only one multicore node. This is because the amount of memory they each attempt to utilize exceeds the maximum available on one node. Although the shared memory CDT3D method successfully generates its meshes on one multicore node, DM\_CDT3D contains routines which allocate additional memory. For example, upon successful adaptation of a subdomain, qualitative statistics data are calculated and immediately packed and sent to the master process (to be merged with other qualitative statistics for a final report upon the program's termination). The routine for this quality calculation and the packing/unpacking of statistical data occurs regardless of whether or not one or more nodes are utilized (revealing the need to optimize DM\_CDT3D's memory usage according to the number of nodes utilized in future work). Both DM\_CDT3D and \textit{refine} successfully generate their meshes when utilizing more than one multicore node however (which is their intended usage). Both CDT3D methods generate smaller meshes (by about 100 million tetrahedra). While \textit{refine} exhibits excellent scalability, DM\_CDT3D again exhibits the optimal runtime performance (also when utilizing 512 cores). The shared memory CDT3D method takes approximately 8 hours (a full work-day) to generate about 1 billion elements. DM\_CDT3D reduces this to less than 4 hours (outperforming \textit{refine}) when utilizing 512 cores.

\begin{table}[htbp]
\caption{The runtime (approximately in \textbf{hours}) of SM\_CDT3D, A Priori DM\_CDT3D, and \textit{refine} when adapting the cube geometry at 100 million complexity on the Wahab supercomputer is shown.}
\centering
\begin{tabular}{ccccccccccc}
\hline\hline
\multirow{2}{*}{Method} & \multicolumn{9}{c}{\# Cores} \\
 & $1$ & $2$ & $4$ & $8$ & $16$ & $32$ & $64$ & $128$ & $256$ & $512$\\\hline
SM\_CDT3D & $183.58$ & $99.43$ & $56.2$ & $29.22$ & $14.7$ & $8.12$ & $-$ & $-$ & $-$ & $-$ \\
A Priori DM\_CDT3D & $-$ & $-$ & $-$ & $-$ & $-$ & $-$ & $6.32$ & $5.59$ & $3.75$ & $3.46$\\
\textit{refine} & $-$ & $-$ & $-$ & $-$ & $-$ & $-$ & $34.92$ & $16.68$ & $8.83$ & $5.98$\\
\hline\hline
\end{tabular}
\label{runtime_results_cube_100m}
\end{table}

\begin{table}[htbp]
\caption{The approximate sizes of the meshes generated (across all tested core configurations) by SM\_CDT3D, A Priori DM\_CDT3D, and \textit{refine} when adapting the cube geometry at 100 million complexity is shown. `M' means million and `B' means billion.}
\centering
\begin{tabular}{ccc}
\hline\hline
Method & \# of Tetrahedra & \# of Vertices \\\hline
SM\_CDT3D & 957.29M & 160.9M \\ 
A Priori DM\_CDT3D & 980.57M & 171.74M \\
\textit{refine} & 1.08B & 184.44M \\
\hline\hline
\end{tabular}
\label{mesh_sizes_cube_100m}
\end{table}

\begin{figure}[htbp]
     \centering
     \begin{subfigure}[htb]{0.5\textwidth}
         \centering
         \includegraphics[width=\textwidth]{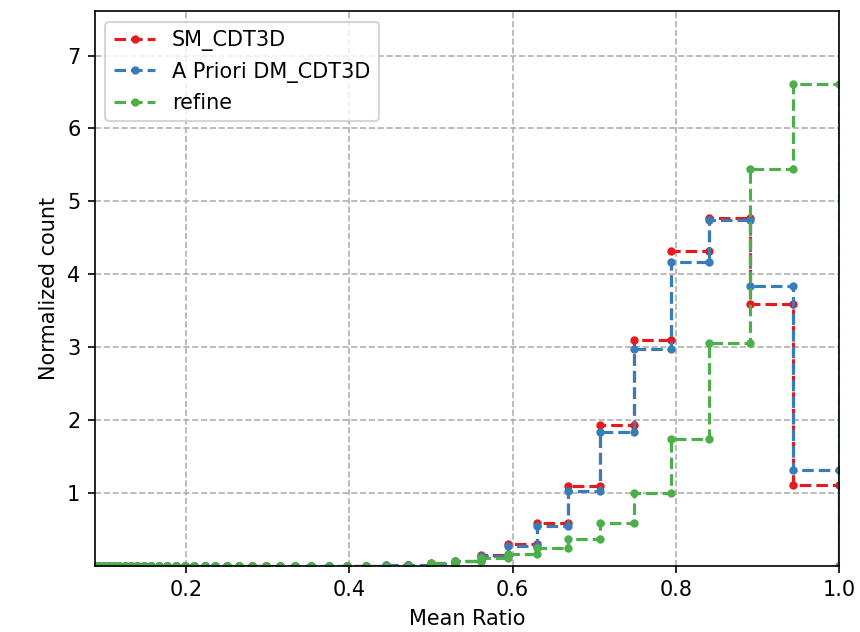}
     \end{subfigure}%
     \begin{subfigure}[htb]{0.5\textwidth}
         \centering
    \includegraphics[width=\textwidth]{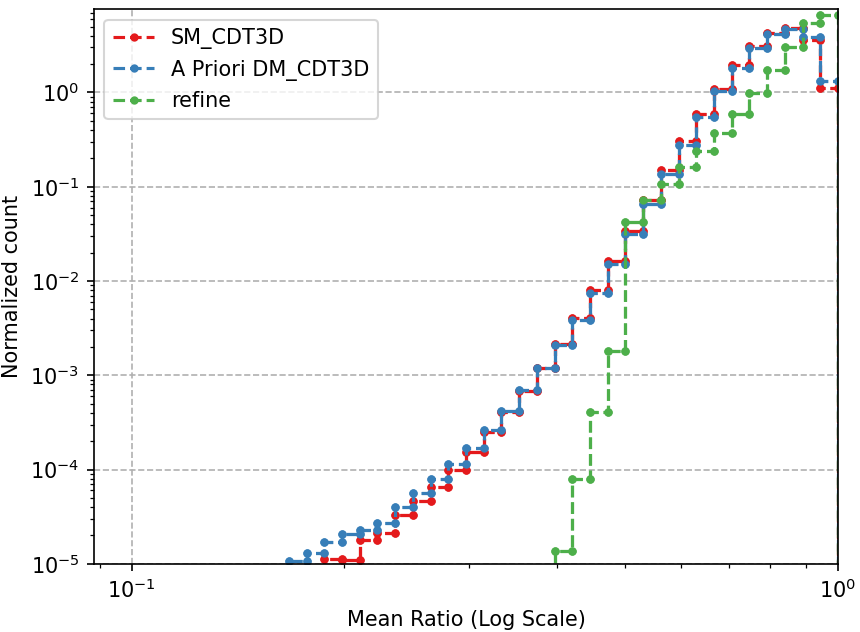}
     \end{subfigure}
        \caption{The mean ratios of elements generated by each method for the cube geometry at 100 million complexity are compared in linear and logarithmic scales.}
        \label{fig:cube_100m_MR}
\end{figure}

\begin{figure}[htbp]
     \centering
     \begin{subfigure}[htb]{0.5\textwidth}
         \centering
         \includegraphics[width=\textwidth]{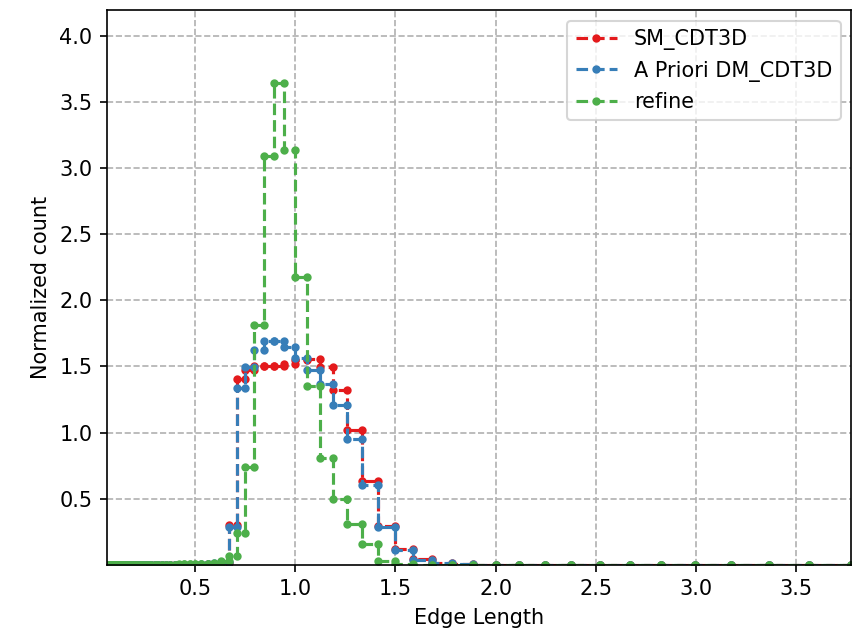}
     \end{subfigure}%
     \begin{subfigure}[htb]{0.5\textwidth}
         \centering
    \includegraphics[width=\textwidth]{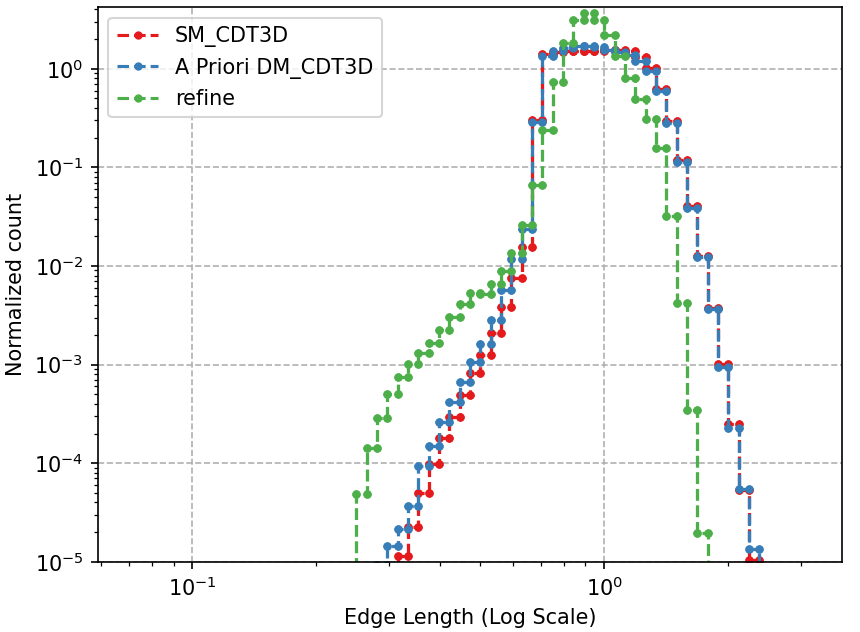}
     \end{subfigure}
        \caption{The edge lengths of elements generated by each method for the cube geometry at 100 million complexity are compared in linear and logarithmic scales.}
        \label{fig:cube_100m_EL}
\end{figure}

A percentage breakdown for DM\_CDT3D's end-to-end time when adapting the cube geometry is omitted since the same behavior is observed as with the delta wing geometry cases. The element shape measure and edge length histograms for the output meshes generated at 100 million complexity by each method are compared in Fig. \ref{fig:cube_100m_MR} and Fig. \ref{fig:cube_100m_EL}. The a priori DM\_CDT3D approach again generates a mesh of similar quality to that generated by SM\_CDT3D. The mesh generated by \textit{refine} again exhibits the best overall quality.

\section{Discussion and Future Work} \label{discussion}
As estimated in section \ref{related_work} when describing the drawbacks of collective communication techniques, \textit{refine} exhibits good scalability but its runtime performance is worse with lower core configurations likely due to its expensive communication overhead. The method only (slightly) outperforms DM\_CDT3D on the Wahab supercomputer when adapting the delta wing geometry using a configuration of 512 cores. \textit{refine} underperforms compared to the proposed method when utilizing core configurations of 256 or smaller on Wahab, and underperforms across all core configurations tested successfully (up to 384) on the Anvil supercomputer. On Anvil, \textit{refine} fails to generate meshes when utilizing 768 or 1,536 CPU cores. When executed successfully, \textit{refine} generates meshes of better quality compared to DM\_CDT3D for the same level of metric conformity but at the cost of generating 10-15\% more elements in each of the above cases. While \textit{refine}'s runtime can partially be attributed to this generation of additional elements, this tradeoff between mesh quality and runtime may not seem justifiable (compared to DM\_CDT3D's runtime) until executed with the configurations of 512 cores on Wahab. For example, \textit{refine} takes more than double the amount of time that DM\_CDT3D takes to generate the cube mesh when using 256 cores (while generating only about 10\% more elements compared to DM\_CDT3D). When generating approximately 1 billion elements for the same metric complexity, DM\_CDT3D outperforms \textit{refine} when using the same numbers of cores (across all configurations where the methods adapted the geometry successfully) on the Wahab supercomputer. As stated previously, DM\_CDT3D's scalability is constrained by the fact that its interface adaptation phase is limited to the number of CPU cores within a single node. \textit{refine} outperforms the proposed method when using 512 CPU cores (40 cores per node) to generate 100 million elements for the delta wing geometry on the Wahab supercomputer but DM\_CDT3D outperforms \textit{refine} across all tested core configurations on the Anvil supercomputer (96 cores utilized per node) for the same benchmark. When more cores are available per node, DM\_CDT3D exhibits better performance than \textit{refine}. This example (and the fact that DM\_CDT3D also exhibits better performance when generating a much larger mesh, e.g., 1 billion elements) exemplifies the conclusion made by the DoE's ECP study \cite{ECPMPIStudy, Klenk2017ECPStudy} that collective communication techniques must be avoided to achieve optimal performance when leveraging the concurrency offered by emerging HPC architectures. While DM\_CDT3D exhibits poor scalability (when executed on the distributed memory architectures used for our evaluation), it offers good end-user productivity compared to \textit{refine} when generating large meshes. For example, DM\_CDT3D generates approximately 65,600 elements per second (extending SM\_CDT3D's rate of about 35,722 when using 32 cores) on Wahab while \textit{refine} generates approximately 77,681 elements per second when using 512 cores to adapt the delta wing geometry at 20 million complexity. Using 384 cores, DM\_CDT3D exhibits an adaptation rate of about 232,523 (compared to \textit{refine}'s 124,555) when adapting the same geometry at 10 million complexity on the Anvil supercomputer. When adapting the cube geometry at 100 million complexity on Wahab, DM\_CDT3D has an adaptation rate of about 78,722 (compared to SM\_CDT3D's rate of 32,748 on 32 cores) while \textit{refine} has a rate of about 50,134 also when using 512 cores.

\begin{figure}[htbp]
\begin{center}
\includegraphics[width=0.80\textwidth]{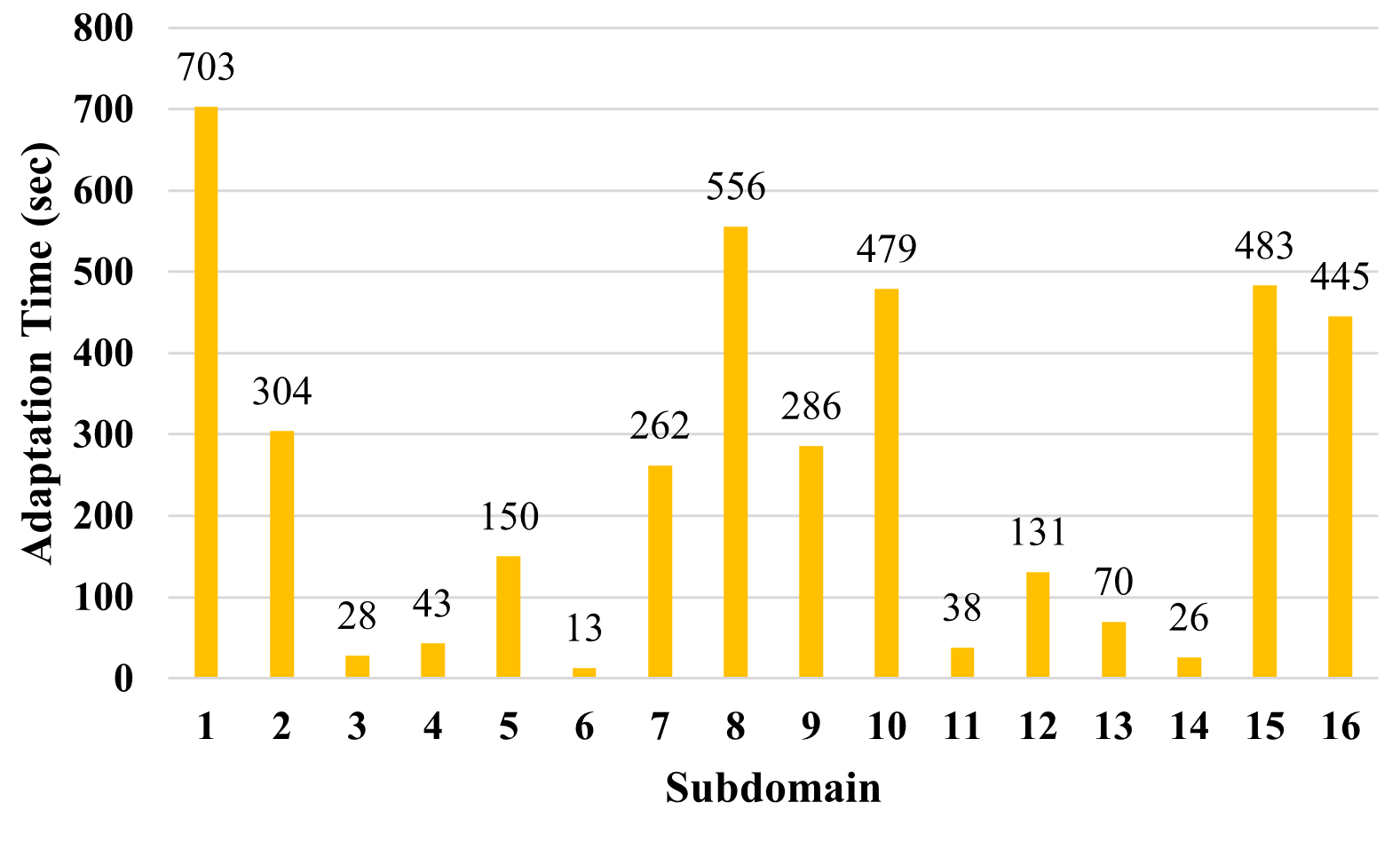}
\end{center}
\caption{Shown are the A Priori DM\_CDT3D subdomain adaptation times for the delta wing geometry at 20 million complexity with a 128-core (4 nodes) configuration.}
\label{fig:subdomain_adaptation_times}
\end{figure}

Fig. \ref{fig:subdomain_adaptation_times} shows the a priori DM\_CDT3D adaptation times of subdomains when adapting the delta wing geometry at 20 million complexity using a 128-core (4 nodes) configuration on Wahab. Recall that all configurations were tested with a 16-subdomain decomposition for the method when adapting this case. Table \ref{node_adaptation_times} shows the corresponding adaptation time per node. Although the total adaptation time is approximately the same per node with this 4-node run (indicating a good load balance), this is not the case when utilizing 512 cores (16 nodes) on Wahab. Each node is only assigned one subdomain, and Fig. \ref{fig:subdomain_adaptation_times} shows that some subdomains are adapted much quicker than others (leaving some nodes idle while others are still at work). This load imbalance can be alleviated by utilizing an overdecomposition (\# of subdomains $>>$ \# of processes) and PREMA's load balancing capabilities. The effectiveness of such an approach was demonstrated in \cite{Thomadakis22PREMA} and \cite{Thomadakis23TowardRuntime}. A caveat of the DM\_CDT3D method however is that if more subdomains are desired, then more time must be spent in generating the coarse mesh that must be dense enough for data decomposition. If the mesh is too coarse, either the interface adaptation phase would adapt the entire domain (due to the coarse pseudo-active/unlocked layers) or interface adaptation would be constrained (if less layers are unlocked to account for the coarse mesh) resulting in poor quality interface elements. Overdecomposition and coarse mesh complexity in turn impact the other parameters of the DM\_CDT3D method (i.e., what is needed to achieve good final mesh quality in the least amount of time, such as the number of layers unlocked/activated during each phase of preprocessing and parameters for the meshing operations themselves during each phase). Table \ref{parameters} (in the appendix) gives brief descriptions of the parameters utilized by the a priori DM\_CDT3D method.

\begin{table}[htb]
\caption{A Priori DM\_CDT3D statistics per node are shown for a 128-core (4 nodes) run when adapting the delta wing geometry at 20 million complexity on the Wahab supercomputer. Given a 16-subdomain decomposition, each node is assigned 4 subdomains. `M' means million and `Tets' means tetrahedra.}
\centering
\begin{tabular}{cccccccc}
\hline\hline
 & & & & \multicolumn{4}{c}{Subdomain Adaptation Times (sec)} \\
& \# Tets Assigned & Final \# Tets & \# Tets Generated & Min & Avg & Max & Total \\\hline
Node 1 & 20.1M & 49.8M & 29.7M & 28 & 269 & 703 & 1078 \\
Node 2 & 20.4M & 48.1M & 27.7M & 13 & 245 & 556 & 981 \\
Node 3 & 27.4M & 48M & 20.6M & 38 & 233 & 479 & 934 \\
Node 4 & 23.3M & 50.1M & 26.8M & 26 & 256 & 483 & 1024 \\
\hline\hline
\end{tabular}
\label{node_adaptation_times}
\end{table}

While the parameters used for DM\_CDT3D in our evaluation were determined after extensive testing, they still may not be the optimal parameters. The optimal settings for these parameters vary based on the geometry and the complexity to which it is adapted. Although it is the same geometry, adapting the delta wing at 20 million complexity versus 10 million complexity required a different data decomposition configuration and the use of additional meshing operations during interface adaptation to ensure good final mesh quality. Given the cost of utilizing HPC resources, it is not practical to consume large amounts of power over extended periods of time just to find the optimal parameters for every geometry. For example, \cite{FrontiersRegistrationParameters2024} compares registration methods for image-guided neurosurgery. These applications utilize medical image-to-mesh conversion techniques, and as such, contain numerous parameters that impact both performance and criteria designating if a mesh meets the requirements of the registration application. Using a single multicore node, terabytes of data were generated over a period of 5 months just to determine the optimal parameters for the test cases in \cite{FrontiersRegistrationParameters2024}. An early analytic model was developed to address a similar problem for applications built with an older version of PREMA \cite{BarkerModel2005}. It focused on optimizing performance by reducing communication overhead and predicting load balancing behavior for adaptive applications. Similar work regarding this regression problem will be addressed by building a machine learning model capable of determining the best parameters for the a priori DM\_CDT3D method that would produce the optimal mesh quality in the least amount of time. It should also be noted that the method of decomposition utilized currently in DM\_CDT3D (PQR) does not always generate equally sized subdomains in terms of the number of elements. When decomposing the coarse mesh using a 4x2x2 decomposition configuration (for the delta wing's 20 million complexity case), subdomains range in size from about 20 thousand to 2 million tetrahedra. While subdomains can change drastically in size due to the nature of adaptation, other methods of decomposition must also be tested for DM\_CDT3D in the future. Due to these issues, overdecomposition is not yet utilized and PREMA's load balancing feature is not yet utilized in DM\_CDT3D. Building a machine learning model by testing overdecomposition configurations with combinations of the remaining parameters is outside the scope of this paper and will be addressed in future work. 

While the distributed memory CDT3D method extends the performance of the original shared memory CDT3D method (by utilizing more than one multicore node), the data in Fig. \ref{fig:apriori_delta_10m_percentage_adaptation_breakdown} and Fig. \ref{fig:apriori_delta_20m_percentage_adaptation_breakdown} underline an important observation. The percentage of end-to-end runtime occupied by interface adaptation goes from about 20\% when using 32 cores to more than 40\% when using 256 cores. Again, this is because interface adaptation is limited to using only 32 cores within one multicore node when the method is executed on Wahab. When the method is executed on larger multicore nodes (on the Anvil supercomputer), overall performance improves because both the interface and interior (within the context of each individual subdomain) phases have access to more cores. To further improve scalability, the interface adaptation problem must be revisited to allow the method to leverage greater concurrency for interface adaptation than that within a single shared memory node. While PREMA's constructs were utilized extensively in the earlier a posteriori version of DM\_CDT3D, PREMA was used minimally in the a priori DM\_CDT3D approach due to the aforementioned issues. To further remove scalability limitations (specifically for interface adaptation and aside from load balancing with overdecomposition), a fully speculative approach would need to be implemented where interface elements are adapted simultaneously with interior elements (similar to \cite{NAVE2004191}). PREMA's asynchronous communication and computation capabilities would alleviate communication overhead in such an approach. Additionally, the sequential components (making subdomains simply connected and converting the data structures) must be parallelized to improve potential scalability. As specified previously, the presented method currently utilizes the default Pthread version of CDT3D as opposed to other backends offered by its tasking framework \cite{Tsolakis21TaskingFramework}. These other backends must be tested to gauge any difference in potential performance for the distributed memory CDT3D method. The DM\_CDT3D method's results and end-user productivity must also be verified through integration into a CFD simulation pipeline, similar to how the original shared memory CDT3D method was tested \cite{EwCCDT3D}.

Regarding the parallel mesh generation requirements described in section \ref{introduction}, the presented method satisfies the stability, code re-use, and reproducibility requirements. Stability was shown through the generated metric-conformed meshes (exhibiting weak reproducibility) that match the quality of the original shared memory CDT3D software. Code re-use was shown through the distributed method's utilization of operations that adhered to the speculative execution model. Moreover, some modifications were made to the shared memory code to ensure the efficient utilization of these operations (optimizing runtime as opposed to a naive, black box approach). The aforementioned challenges will be addressed to improve the method's scalability. Robustness will be addressed in the future, as additional work is needed to build upon CDT3D's use of the EGADS library \cite{EGADS} when processing CAD models in a distributed memory setting.


\section{Conclusion}
Presented is a distributed memory method for anisotropic mesh adaptation that is designed to meet the needs of large-scale, time-dependent CFD problems. Meshing functionality is separated from performance aspects so that the method can easily leverage the concurrency offered by emerging HPC architectures without requiring significant updates to its source code. Mesh adaptation is handled by a shared memory method called CDT3D. Several requirements regarding the distributed method's design are given based on lessons learned from re-designing some components of the shared memory method, enabling its integration into the distributed memory method and thus achieving good performance. All major operations within CDT3D are designed to adopt a speculative execution model, enabling the strict adaptation of either interior or subdomain interface elements so that each set of elements can be adapted in separate steps while maintaining mesh conformity. 

Results show that the distributed memory CDT3D method is able to produce meshes of comparable quality to those generated by the original shared memory CDT3D software and by the state-of-the-art method \textit{refine}. Given the costly overhead of collective communication techniques identified within the study of DoE's Exascale Computing Project \cite{ECPMPIStudy, Klenk2017ECPStudy} and seen in our evaluation (utilized by \textit{refine}), the distributed memory CDT3D method's emphasis on avoiding these techniques has proven beneficial. When tested on configurations up to 512 cores in a distributed memory architecture, the presented method exhibits the optimal runtime performance compared to \textit{refine} when generating a model geometry of approximately 1 billion elements. Furthermore, DM\_CDT3D showcases the performance (particularly on the Anvil supercomputer \cite{Anvil2022} when utilizing up to 1,536 cores) that may be achieved when efficiently leveraging the concurrency offered by both shared memory and distributed memory architectures (as opposed to designing an algorithm that focuses on only one or the other, e.g., the shared memory CDT3D or \textit{refine}). The next steps are to: (1) parallelize the sequential component of the proposed method's algorithm (i.e., making subdomains simply connected), (2) implement a machine learning model capable of providing the optimal parameter settings for the presented method (to also test overdecomposition configurations while utilizing PREMA's load balancing capabilities), enhancing potential performance and saving time by reducing the needed testing to find such parameters for large geometries and (3) integrate the presented method into a CFD simulation pipeline.
\newpage
\section*{Appendix}

\begin{table}[!htb]
\caption{Quantitative and qualitative statistics that show the affect of the pseudo-active modification and preemptive locking on CDT3D within DM\_CDT3D are shown. The Modified column is the experiment that includes these modifications for each subdomain's interior adaptation while respecting the order of operations throughout execution. The Naive column is the experiment that does not respect the order of operations and does not include the pseudo-active modification or preemptive locking for interior adaptation (default settings as a black box during both adaptation phases). `M' is million, `K' is thousand, and `Tets' means tetrahedra.}
\centering
\begin{tabular}{cccc}
\hline\hline
Adaptation Phase & Statistics & Naive & \textbf{Modified (Reduced By)} \\\hline
\multirow{7}{*}{Interface} & Adaptation Time (sec) & 321.09 & \textbf{75.4 (4.25x)} \\
& Start \# Tets & 12.1M & 12.1M \\
& Start \# Active Tets & 1.03M & 1.03M \\
& Start \# Vertices & 2.06M & 2.06M \\
& Start \# Unlocked Vertices & 193K & 193K \\
& End \# Tets & 18.1M & 18.1M \\
& End \# Vertices & 3.06M & 3.06M \\\hline
\multirow{7}{*}{Subdomain 1 Interior} & Adaptation Time (sec) & 1949.23 & \textbf{1220.99 (1.59x)} \\
& Start \# Tets & 9.04M & 9.06M \\
& Start \# Active Tets & 6.91M & 6.07M \\
& Start \# Vertices & 1.58M & 1.58M \\
& Start \# Unlocked Vertices & 1.48M & 1.15M \\
& End \# Tets & 48.7M & 48.8M \\
& End \# Vertices & 8.25M & 8.26M \\
\hline
\multirow{7}{*}{Subdomain 2 Interior} & Adaptation Time (sec) & 1937.07 & \textbf{1185.03 (1.63x)} \\
& Start \# Tets & 9.1M & 9.11M \\
& Start \# Active Tets & 6.92M & 6.07M \\
& Start \# Vertices & 1.58M & 1.58M \\
& Start \# Unlocked Vertices & 1.48M & 1.14M \\
& End \# Tets & 48.9M & 48.9M \\
& End \# Vertices & 8.24M & 8.25M \\
\hline\hline
\end{tabular}
\label{table_redundancy_full}
\end{table}

\begin{table}[!htb]
\caption{Some of the parameters used by the A Priori DM\_CDT3D method}
\centering
\scriptsize
\begin{tabular}{ccccc}
\hline\hline
Parameter & Delta-10M & Delta-20M & Cube-100M & Description \\\hline
PQR configuration & 8x1x2 & 4x2x2 & 3x3x5 & Determines data decomposition configuration \\\hline
\multirow{2}{*}{coarse mesh complexity} & \multirow{2}{*}{1.25M} & \multirow{2}{*}{1.25M} & \multirow{2}{*}{6.67M} & Complexity to which the coarse mesh is adapted \\ 
& & & & to ensure a dense enough mesh for data decomposition \\\hline
\multirow{2}{*}{Interface Adaptation Layers} & \multirow{2}{*}{5} & \multirow{2}{*}{5} & \multirow{2}{*}{3} & \# layers of tetrahedra (and their respective points) that are marked \\
& & & & as pseudo-active (unlocked) for interface adaptation \\\hline
\multirow{2}{*}{Interior Adaptation Layers} & \multirow{2}{*}{3} & \multirow{2}{*}{3} & \multirow{2}{*}{3} & \# layers of tetrahedra (and their respective points) that are marked \\
& & & & as pseudo-active (unlocked) for interior adaptation \\\hline
Interface - mdse & On & On & On & Pre-refinement edge collapse operation used to coarsen the input domain \\\hline
Interface - final\_mdse & Off & Off & Off & Post-refinement edge collapse operation used to remove small edges \\\hline
Interface - nqual & 0 & 1 & 1 & \# quality improvement iterations executed during the interface adaptation phase \\\hline
\multirow{2}{*}{Interface - nsmth} & \multirow{2}{*}{0} & \multirow{2}{*}{2} & \multirow{2}{*}{2} & \# vertex smoothing iterations executed during the quality improvement \\
& & & & module for the interface adaptation phase \\\hline
Interior - mdse & On & On & On & Pre-refinement edge collapse operation used to coarsen the input domain \\\hline
Interior - final\_mdse & On & On & On & Post-refinement edge collapse operation used to remove small edges \\\hline
Interior - nqual & 3 & 3 & 3 & \# quality improvement iterations executed for each subdomain's interior adaptation \\\hline
\multirow{2}{*}{Interior - nsmth} & \multirow{2}{*}{5} & \multirow{2}{*}{5} & \multirow{2}{*}{5} & \# vertex smoothing iterations executed during the quality improvement \\
& & & & module for each subdomain's interior adaptation \\
\hline\hline
\end{tabular}
\label{parameters}
\end{table}

\newpage
\section*{Acknowledgments}
This research was sponsored in part by the NASA Transformational Tools and Technologies Project (NNX15AU39A) of the Transformative Aeronautics Concepts Program under the Aeronautics Research Mission Directorate, the Richard T. Cheng Endowment, the Southern Regional Education Board (SREB) State Doctoral Scholar Fellowship, and the National Institute of General Medical Sciences of the National Institutes of Health under Award Number 1T32GM140911-03. The content is solely the authors’ responsibility and does not necessarily represent the official views of the National Institutes of Health. The authors would like to thank Dr. Christos Tsolakis for his assistance in the development of the a posteriori distributed memory CDT3D code and for his insight regarding potential causes of CDT3D's poor performance when utilized as a black box in the distributed memory method. The authors would also like to thank Dr. Mike Park for his management as part of NASA's Transformational Tools and Technologies Project (NNX15AU39A) of the Transformative Aeronautics Concepts Program, and for providing feedback on an earlier draft of this paper.

\bibliography{references}

\end{document}